\documentclass[twocolumn,tighten]{aastex631}

\shorttitle{Improved Timing of X-ray Isolated Neutron Stars}
\shortauthors{Bogdanov \& Ho}

\begin{document}

\title{THE ``MAGNIFICENT SEVEN'' X-RAY ISOLATED NEUTRON STARS REVISITED. \\ I. IMPROVED TIMING SOLUTIONS AND PULSE PROFILE ANALYSIS}

\correspondingauthor{Slavko Bogdanov}
\email{slavko@astro.columbia.edu}

\author[0000-0002-9870-2742]{Slavko Bogdanov}
\affiliation{Columbia Astrophysics Laboratory, Columbia University, 550 West 120th Street, New York, NY 10027, USA}

\author[0000-0002-6089-6836]{Wynn C.~G.~Ho}
\affil{Department of Physics and Astronomy, Haverford College, 370 Lancaster Avenue, Haverford, PA 19041, USA}

\begin{abstract}
We present the first systematic X-ray pulse timing analysis of the six members of the so-called ``Magnificent Seven'' nearby thermally emitting isolated neutron stars (XINS) with detected pulsations. Using the extensive collection of archival XMM-Newton, Chandra, and NICER observations spanning over two decades, we obtain the first firm  measurement of the spin-down rate for RX\,J2143.0+0654, while for the rest we improve upon previously published spin ephemerides and extend them by up to an additional decade. Five of the XINS follow steady spin-down with no indication of major anomalies in their long-term timing behavior; the notable exception is RX\,J0720.4$-$3125, for which, in addition to confirming the previously identified glitch,  we detect a second spin derivative. The high-quality folded X-ray pulse profiles produced with the updated timing solutions exhibit diverse and complex morphologies, as well as striking energy dependence. These peculiarities cannot be readily explained by blackbody-like isotropic emission and simple hot-spot configurations, hinting at the presence of complex multitemperature surface heat distributions and highly anisotropic radiation patterns, such as may arise from a strongly magnetized atmospheric layer.
\end{abstract}
\keywords{Pulsars: individual (RX\,J0420.0$-$5022, RX\,J0720.4$-$3125, RX\,J0806.4$-$4123, RX\,J1308.6$+$2127, RX\,J1856.5$-$3754, RX\,J2143.0+0654) --- Pulsars (1306) --- Neutron Stars (1108) --- X-ray astronomy (1810) --- X-ray sources (1822) }

\section{Introduction} \label{sec:intro}
\turnoffeditone
The R\"ontgensatellit (ROSAT) all-sky survey uncovered seven nearby ($D\lesssim 500$\,pc) thermally emitting neutron stars  (see, e.g., \citealt{2007Ap&SS.308..181H,2009ASSL..357..141T}, and references therein). These isolated objects, RX J0420.0$-$5022, RX J0720.4$-$3125, RX J0806.4$-$4123, RX J1308.6$+$2127, RX J1605.3$+$3249, RX J1856.5$-$3754, and RX J2143.0+0654, often collectively referred to as the ``Magnificent Seven'', exhibit very soft, seemingly purely blackbody-like spectra, with effective temperatures $k T_{\rm{eff}} \approx 40-100$\,eV, and X-ray luminosities $L_X\sim 10^{31}$\,erg\,s$^{-1}$. They have relatively long rotation periods, $P \approx 3-11$\,s, and spin-down rate measurements ($\dot{P}\equiv \rm{d}P/\rm{d}t$) suggest surface magnetic fields of $B_s \propto (P\dot{P})^{1/2}\sim 10^{13}$\,G and typical characteristic ages of $\tau_c=P/2\dot{P}\sim 1$\,Myr. They exhibit no radio emission despite deep searches \citep{2009ApJ...702..692K}. There are indications that these X-ray isolated neutron stars (XINS)\footnote{In the literature, these sources are sometimes referred to as X-ray dim isolated neutron stars (XDINS), even through they are not particularly dim in relative terms when compared against the current sample of X-ray detected Galactic neutron stars.} represent a significant fraction of all neutron stars (e.g., \citealt{2008MNRAS.391.2009K}) but their origin and evolutionary history are far from certain. 

XINS exhibit a number of peculiarities in their observed properties that defy straightforward explanations. For instance,  the confirmed optical and UV counterparts of all seven XINS \citep{2011ApJ...736..117K} reveal that the measured photometry exceeds the flux predictions in the Rayleigh-Jeans tail extrapolated from the blackbody-like X-ray spectra. Furthermore, although XINS are typically much less luminous than younger thermally cooling NSs, their X-ray luminosities are too large to be explained by passive cooling alone; they also exceed their spin-down luminosities ($\dot{E}\propto \dot{P}P^{-3}$), making them distinct from rotation-powered pulsars with comparable $P$ and $\dot{P}$ values. This implies that the observed surface thermal radiation does not come at the expense of the rotational energy reservoir of the star and requires an additional source, most probably release of internal heat due to magnetic field decay, which may explain the unexpectedly strong surface magnetic fields of XINS.  Regardless of the heating mechanism, the soft periodically modulated X-ray emission indicates the presence of a highly nonuniform heat distribution, which presumably traces the magnetic field topology at the surface and its vicinity. 
Thus, in principle, the observed surface X-ray emission harbors information about the magnetic field properties, including its surface distribution as well as its strength; the latter is also encoded in the beaming pattern of the surface atmospheric emission, which is strongly dependent on the magnetic field strength and field line orientation \citep[see, e.g.,][]{1992herm.book.....M,1994A&A...289..837P,2008ApJS..178..102H,2013MNRAS.434.2362P,2021ApJ...914..118D}. 

At present, a self-consistent model of the multiwavelength spectral energy distribution for all XINSs is lacking and their surface composition and thermal and magnetic maps remain unconstrained, except for possibly RX~J1856.5$-$3754 \citep{2007MNRAS.375..821H,2007MNRAS.380...71H}.  To rectify this situation, we have commenced an in-depth analysis of the voluminous archival X-ray data set of the current sample of XINS. Mapping the surface X-ray emission as a function of time and energy for these sources by applying realistic neutron star models that consider all practically important physics has the potential to reveal the underlying physical processes responsible for their peculiar properties. A crucial prerequisite for such an undertaking is the availability of reliable long-term phase-coherent timing solutions which permit the folding of all existing event data with sufficient time resolution to obtain energy-resolved pulse profiles with the best possible photon statistics.

In this paper, the first in a series, we present a comprehensive and uniform analysis of the six\footnote{Although based on all other indicators RX J1605.3$+$3249 is almost certainly an XINS, no X-ray pulsations have been detected from it to date (a claim of a 3.4\,s periodicity by \citealt{2014A&A...563A..50P} was not corroborated by deeper observations and is likely spurious). Thus, it is not included in this present work; see \citet{2019ApJ...880...74M} and \citet{2019A&A...623A..73P} for recent in-depth analyses of this object.} XINS that exhibit pulsed emission to obtain greatly improved long-term timing solutions and to examine the properties of their pulsed emission more closely. To accomplish this, we make use of the wealth of X-ray Multi-Mirror Mission (XMM-Newton) and Chandra X-ray Observatory exposures collected over the past two decades, as well as more recent Neutron Star Interior Composition Explorer (NICER) observations. Here, we focus exclusively on timing and pulse profile analyses; discussions of spectral analyses, including those made by previous works, will be presented in the second paper in the series. The third paper will focus on realistic modeling of the surface thermal radiation. The present work is organized as follows. Section~\ref{sec:data} summarizes the X-ray observations and reduction procedures used to extract the data products used in the analysis. In Section~\ref{sec:timing} we describe the timing procedure used in this work, and in Section \ref{sec:results} we lay out the resulting long-term timing analysis of the XINS, while Section~\ref{sec:profiles} presents an examination of the properties of the pulsed emission as a function of energy and time.  We offer a discussion in Section~\ref{sec:discussion} and conclusions in Section~\ref{sec:conclusion}.  

\begin{deluxetable*}{lRRRRRR}
\tabletypesize{\footnotesize}
\tablewidth{0pt}
\tablecolumns{7}
\tablecaption{Total filtered X-Ray Exposures (in Kiloseconds) Used in This Work \label{table:obs}}
\tablehead{\colhead{XINS\tablenotemark{a}} & \colhead{XMM-Newton} & \colhead{XMM-Newton} & \colhead{XMM-Newton} & \colhead{Chandra} & \colhead{Chandra} & \colhead{NICER} \\
\colhead{} & \colhead{EPIC pn} & \colhead{EPIC MOS1} & \colhead{EPIC MOS2} & \colhead{HRC} & \colhead{ACIS\tablenotemark{b}} & \colhead{XTI} }
\startdata
J0420.0$-$5022\tablenotemark{c}	&	172.1	& 109.8	&	111.8 &	\nodata 	& \nodata 	& \nodata	\\
J0720.4$-$3125	&	476.4	& 547.4 &	524.1 &	524.3	&	80.2 &	104.6 \\
J0806.4$-$4123	&	98.6	& 73.2	 &	72.1 &	29.8	&	18.1 &	171.8 \\
J1308.6+2127	&	130.8	& 180.6	 &	181.5 &	87.2	&	62.8 &	120.9 \\
J1856.5$-$3754	&	1986.2  & 2340.1 &	2431.7  &	1153.2	& \nodata	&	185.8 \\
J2143.0+0654	&	90.2    & 136.1 &	136.9 &	139.7	& 12.1	&	272.3 \\
\enddata
\tablenotetext{a}{See Tables~\ref{table:j0420_obs}-\ref{table:j2143_obs} in the Appendix for a detailed list of all observations for each target.}
\tablenotetext{b}{Only Chandra ACIS-S observations conducted in continuous clocking (CC3\_GRADED) mode are considered.}
\tablenotetext{c}{For RX J0420.0$-$5022, no usable Chandra or NICER data are available.}
\end{deluxetable*}

\vspace{-0.3cm}
\section{Observations and Data Analysis}
\label{sec:data}
The analysis presented herein makes use of archival X-ray observations of the six bright XINS with confirmed X-ray pulsations. For the purposes of this work, we selected only observations with sufficient time resolution to permit a pulse profile to be constructed. Thus, in the case of RX J0420.0$-$5022 we do not use XMM-Newton EPIC MOS1/2 full-frame (FF) mode observations due to the 2.6\,s readout time, which is 75\% of the neutron star spin period; likewise, we do not use Chandra ACIS-S data obtained in full imaging mode, which has a 3.2\,s readout time. 

The total filtered exposures for each XINS using each observatory/instrument combination used in this work  are summarized in Table~\ref{table:obs}, while a detailed list of the individual observations including observation identifiers, UTC date and time, detector mode and filter, and the effective exposure time are presented in Tables~\ref{table:j0420_obs}-\ref{table:j2143_obs} of the Appendix. The data reduction procedures employed for each telescope/instrument to arrive at the clean event lists used for timing  are described below.

\subsection{XMM-Newton EPIC}
In this work, we rely primarily on XMM-Newton European Photon Imaging Camera (EPIC) pn \citep{2001A&A...365L..18S} and MOS1/2 \citep{2001A&A...365L..27T} data sets spanning over two decades. We do not consider any Reflection Grating Spectrometer (RGS) event data due to the substantially lower sensitivity compared to the imaging data and the greatly reduced signal-to-noise ratio of the dispersed source emission.

The X-ray events were reduced and extracted using the Science Analysis Software (SAS) package version \texttt{xmmsas\_20211130\_0941-20.0.0} starting from the unprocessed (ODF) data products. After running the \texttt{epproc} and \texttt{emproc} pipelines for pn and MOS, respectively, the data were inspected for instances of strong background flaring which were excised by considering the count rate over the entire image in the 10--12 keV range for EPIC pn and $>$10 keV for EPIC MOS1/2 from light curves binned at 100\,s intervals. Segments of the light curve with count rates exceeding 0.4 and 0.35 counts s$^{-1}$ for pn and MOS1/2, respectively, were removed to produce a set of clean good time intervals.  For the imaging observations, the source events were extracted from circular regions of radius 32$''$, which encircle $\approx$90\% and $\approx$85\% of the total point source energy at $\sim$1.5 keV for the pn and MOS1/2, respectively, centered on the point source. For RX J1856.5$-$3754, there is a single 44\,ks EPIC pn observation (ObsID 0201590101) obtained in Fast Timing mode, which foregoes one imaging dimension to permit fast readout (29\,$\mu$s). The source events for this exposure were extracted from a rectangular region 13 pixels of width in the RAWX direction centered on row 37. The source event data were filtered using the recommended flag and pattern parameter values (PATTERN$\le$4 and \#XMMEA\_EP for EPIC-pn and PATTERN$\le$12 and \#XMMEA\_EM for EPIC MOS1/2) specified through the \texttt{evselect} tool in SAS. 

We note that a portion of the XMM-Newton data sets for RX\,J0720.4$-$3125 , RX\,J1308.6+2127, and RX J1856.5$-$3754 suffer from nonnegligible event pileup, as assessed by the \texttt{epatplot} task in SAS. However, although important for any detailed spectroscopic analyses, this effect does not significantly impact the pulse timing and cursory pulse profile analyses presented below.  

The timestamps of the filtered source events were translated from the Terrestrial Time (TT) standard to Barycentric Dynamical Time (TDB) using the \texttt{barycen} tool in SAS with the DE405 JPL solar system ephemeris.

\subsection{Chandra}

The Chandra X-ray Observatory data set of the XINS used in this work consists of Advanced CCD Imaging Spectrometer (ACIS) exposures conducted in continuous clocking (CC) mode and High Resolution Camera (HRC) exposures originally intended for high-resolution dispersed spectroscopy with the Low Energy Transmission Grating (LETG) placed in the optical path. The data analysis was carried out using CIAO version 4.15 \citep{2006SPIE.6270E..1VF} and the accompanying calibration database CALDB 4.10.4.  The barycentering of all Chandra ACIS and HRC source events was accomplished using the \texttt{axbary} task in CIAO assuming the DE405 JPL solar system ephemeris. 

\subsubsection{ACIS-S Continuous Clocking Observations}
A number of archival Chandra observations of XINS are available for which the back-illuminated ACIS-S3 chip was placed at the aim point. To avoid photon pileup that would be caused by high source count rates, the detector was configured in Continuous Clocking (CC3\_GRADED) mode, which affords a rapid (2.85\,ms) readout time by sacrificing one dimension of spatial imaging. Following recommended procedures, the source events were extracted from a rectangular region of width $5''$ along the imaging direction. Due to the bright nature of the sources under consideration and the small point spread function of Chandra, the background contribution to the source count rate is virtually negligible.

\subsubsection{HRC-S LETG Observations}
For the Chandra HRC-S observations with the LETG, we extracted the zeroth-order (i.e., undispersed) point-source emission using circular regions of radius $2''$ centered on the source position. In most instances, the first order dispersed spectrum is quite faint while also introducing higher background, and so is not used in the timing analysis. While the zeroth order provides no useful spectral information due to the poor intrinsic energy resolution of the HRC, because of the exceptionally soft spectra of all XINS we expect the vast majority of registered source events to have energies $\le$1\,keV.

The HRC event time tagging problem caused by a wiring error in the detector electronics\footnote{See \url{https://cxc.harvard.edu/cal/Hrc/timing_200304.html}  for further details.} identified early in the Chandra mission results in a degraded time resolution from the intrinsic 16 $\mu$s to $\approx$4 ms (determined mainly by the mean time between events over the entire detector). As this effective time resolution is still orders of magnitude smaller than the spin periods of the XINS, the HRC data are still suitable for the timing analyses that are the subject of this work.

\subsection{NICER XTI}
With the exception of RX\,J0420.0$-$5022, extensive NICER X-ray Timing Instrument \citep[XTI;][]{2016SPIE.9905E..1HG} observations have been conducted of the sample of XINS starting as early as 2017 June. To obtain filtered event lists suitable for timing analyses, we used the NICERDAS software that is part of HEASoft version 6.31, following the procedures recommended by the NICER team. Specifically, the cutoff rigidity COR\_SAX parameter (as originally defined for the BeppoSAX mission), which serves as an indicator of ambient in-orbit particle background levels (see, e.g., \citealt{2019ApJ...887L..25B}), was restricted to values $>$1.5. In addition, the undershoot rate, which is a measure of the severity of contamination due to optical photons at the lowest energies, was limited to a median value of $\le$200 during each good time interval. The rate of event overshoots, which are background contamination events caused by high-energy charged particles that deposit a large amount of charge on the detector, was restricted to values $\le$1.5. The Potsdam planetary geomagnetic activity index\footnote{See \url{https://heasarc.gsfc.nasa.gov/docs/nicer/analysis_threads/geomag/}.} $k_p$, which is an estimate of fluctuation disturbances to the local magnetic fields, was limited to values $\le$5. The filtered event arrival times were barycentered using the \texttt{barycorr} tool in HEASoft assuming the DE405 JPL solar system ephemeris.

\begin{deluxetable*}{lccccc}
\tabletypesize{\footnotesize}
\tablewidth{0pt}
\tablecolumns{7}
\tablecaption{Phase-coherent Timing Solutions for Five XINS \label{tab:timing}}
\tablehead{\colhead{Parameter\tablenotemark{a}}	&	\colhead{J0420.0$-$5022}	&	\colhead{J0806.4$-$4123}	&	\colhead{J1308.6$+$2127}	&	\colhead{J1856.5$-$3754}	&	\colhead{J2143.0$+$0654}}
\startdata		 						
\hline
    \multicolumn{6}{c}{\textit{Assumed quantities}} \\                
\hline
R.A. (J2000.0)\tablenotemark{b}	&	$04^{\rm h}20^{\rm m}01.\!^{\rm s}95$	&	$08^{\rm h}06^{\rm m}23.\!^{\rm s}40$	&	$13^{\rm h}08^{\rm m}48.\!^{\rm s}27$	&	$18^{\rm h}56^{\rm m}35.\!^{\rm s}795$	&	$21^{\rm h}43^{\rm m}03.\!^{\rm s}40$	\\
Decl. (J2000.0)\tablenotemark{b}	&	$-50^{\circ}22^{\prime}48\farcs1$	&	$-41^{\circ}22^{\prime}30\farcs9$	&	$+21^{\circ}27^{\prime}06\farcs78$	&	$-37^{\circ}54^{\prime}35\farcs54$	&	$+06^{\circ}54^{\prime}17\farcs5$	\\			
$\mu_{\alpha}\cos\delta$ (mas\,yr$^{-1}$)\tablenotemark{b}	&	\nodata	&	\nodata	&	$-207(20)$	&	$+326.6(5)$	&	\nodata	\\
$\mu_{\delta}$ (mas\,yr$^{-1}$)\tablenotemark{b}	&	\nodata &	\nodata	&	$+84(20)$	&	$-61.9(4)$	&	\nodata	\\		
Position epoch (MJD) &	52,590 &	52,326	& 51,719	&	52,868 &	54,388	\\	
\hline
    \multicolumn{6}{c}{\textit{Measured quantities}} \\                
\hline									
Epoch (MJD TDB)\tablenotemark{c}	&	55,285.49844	&	54,771.32	&	53,415.68	&	52,372.68301	&	53,156.650258	\\
$\nu$ (Hz) &		$0.28960290947^{(+6)}_{(-5)}$ 	& $0.08794776394^{(+2)}_{(-2)}$	&	 $0.096969489591^{(+7)}_{(-7)}$ 	&	$0.14173936875^{(+2)}_{(-2)}$ &  $0.10606446051^{(+9)}_{(-14)}$ \\
$\dot{\nu}$ ($10^{-15}$\,Hz\,s$^{-1}$) &	$-2.4393^{+0.0005}_{-0.0005}$  	&	$-0.0818^{+0.0001}_{-0.0001}$ 	&	$-1.05427^{+0.00003}_{-0.00003}$	& 	$-0.60373^{+0.00008}_{-0.00007}$ & $-0.4663^{+0.0006}_{-0.0004}$ \\
Range of dates (MJD)	&	52,638--58,625	&	51,856--59,982	&	52,274--60,077		&	51,613-60,036 &	53,156--60,184	\\
H-test significance ($\sigma$)\tablenotemark{d} &	$12.09$	&	$29.11$	&	$139.56$	&	$35.13$	&	$27.01$	\\
\hline
    \multicolumn{6}{c}{\textit{Derived quantities}} \\                
\hline
$P$ (s)	&	$3.4530039833^{(+7)}_{(-6)}$	&	$11.370385729^{(+2)}_{(-2)}$	&	$10.3125220543^{(+7)}_{(-8)}$		&	$7.0552028617^{(+8)}_{(-11)}$	&	$9.428228789^{(+9)}_{(-13)}$	\\
$\dot{P}$ ($10^{-14}$\,s\,s$^{-1}$)	&	$2.9085^{+0.0005}_{-0.0006}$	&	$1.058^{+0.002}_{-0.002}$			&	$11.2119^{+0.0003}_{-0.0003}$	&	$3.0051^{+0.0004}_{-0.0004}$	&	$4.145^{+0.005}_{-0.004}$	\\
$\dot{E}$ ($10^{30}$\,erg\,s$^{-1}$)\tablenotemark{e}	&	$27.9$	&	$0.3$	&	$4.0$	&	$3.4$	&	$2.0$	\\
$\tau_c$ (Myr)\tablenotemark{f}	&	1.9	&	17.0	&	1.5	&	3.7	&	3.6		\\
$B_s$ ($10^{13}$\,G)\tablenotemark{g}	&	1.0	&	1.1	&	3.4	&	1.5	&	2.0		\\
\enddata
\tablenotetext{a}{The numbers quoted for the fitted parameters correspond to the 50$^{\rm th}$ percentile value of the posterior distribution and the $-/+$ values correspond to the 16$^{\rm th}$ ($-$) and 84$^{\rm th}$ ($+$) percentiles of the posterior credible intervals. Numbers quoted in parentheses give the uncertainty in the last digit of the nominal value. }
\tablenotetext{b}{Astrometric parameters measured from optical or X-ray imaging observations and used for barycentering of the event data are from \citet{2004AandA...424..635H}, \citet{2002ApJ...579L..29K}, \citet{2010ApJ...724..669W}, and \citet{2009AandA...499..267S}.}
\tablenotetext{c}{Reference epoch for measured and derived spin parameters quoted in the table.}
\tablenotetext{d}{Single-trial H-test pulse detection significance of the barycentered and stacked event data sets folded at the highest-likelihood timing solution.}
\tablenotetext{e}{Spin-down luminosity $\dot{E}=-4\pi^2I\nu\dot{\nu}$, assuming a neutron star with moment of inertia $I=10^{45}$\,g\,cm$^2$.}
\tablenotetext{f}{Pulsar characteristic age $\tau_c\equiv -\dot{\nu}/2\nu$.}
\tablenotetext{g}{Surface equatorial magnetic field strength $B_{\rm surf}=3.2\times10^{19}(-\dot{\nu}/\nu^3)^{1/2}$\,G under the assumption of vacuum dipole magnetic braking for a neutron star with $R=10$ km and $M=1.4$\,M$_{\odot}$.}

\end{deluxetable*}

\section{Pulse Timing Procedure}
\label{sec:timing}
To enable the timing and pulse profile analyses presented here, we first barycentered the XMM-Newton, Chandra, and NICER data sets assuming the best known position for each source and the JPL DE405 solar system ephemeris using the appropriate software tool for each observatory as described in Section~\ref{sec:data}. For RX\,J0720$-$3125, RX\,J1308.6$+$2127, and RX\,J1856.5$-$3754, which have measured proper motions, we accounted for the change in position over time by applying the barycentering for each observation using the proper-motion-corrected position at that epoch. In practice, due to the slow spins of the XINS, this position adjustment has a small effect on the corrected event timestamps and thus on the outcomes of the timing analyses. Moreover, setting the position and proper motion parameters free in the fit does not lead to an improvement in the timing solutions nor does it result in improved constraints on the astrometric parameters. 

Initially, we attempted to fold the extracted source event data at the most recent published timing solutions for each XINS. However, since some of these ephemerides are over a decade old, they typically did not accurately predict the spin of the neutron star for the more recent data sets, as evidenced by phase drift and smearing of the accumulated profile. In light of this, we proceeded to carry out a formal timing analysis for the six XINS to obtain more up-to-date and precise timing solutions. We conducted this analysis with an unbinned event-based Bayesian likelihood evaluation approach as implemented in the \texttt{event\_optimize} code in the PINT pulsar timing software package \citep{2021ApJ...911...45L}, which has been thoroughly tested and is used extensively for timing of Fermi Large Area Telescope $\gamma$-ray pulsars (see, e.g., \citealt{2022Sci...376..521F,Smith_2023}). In this technique, the photon event data are not grouped into time of arrival (TOA) measurements but are instead evaluated collectively as an unbinned ensemble against a timing model (with likelihood evaluation based on the H-test; \citealt{1989A&A...221..180D}; \citealt{2010A&A...517L...9D}) using a smooth pulse profile template constructed from one or more Gaussians (as necessary to reproduce the main features of the pulses). An initial profile template was constructed from the observation with the strongest pulse detection, which was most often the longest XMM-Newton exposure available. For parameter set sampling, the code uses the popular \texttt{emcee} affine-invariant Markov chain Monte Carlo (MCMC) ensemble sampler \citep{2013PASP..125..306F}. For every inference run, we considered 200 sampler walkers and at least 1000 steps per chain to ensure convergence, with the first 100 steps used for burn-in. The photon based likelihood sampling technique is particularly beneficial for NICER XTI event data since it allows the inclusion of all short exposures ($\lesssim$1\,ks) that would otherwise be discarded in a conventional timing analysis as they do not provide sufficient photon statistics to produce a useful TOA measurement. In addition, for the fainter XINS such as RX\,J0420.0$-$5022 and RX\,J2143.0+0654, this approach is more advantageous due to the limited photon statistics that are obtained from the typical XMM and Chandra exposures available for these targets. The same holds true for RX\,J1856.5$-$3754, which, although the brightest in terms of total source count rate, is difficult to time due to the exceptionally low pulsed fraction.  One drawback of this event-based likelihood procedure is that it is substantially more computationally intensive than a conventional TOA analysis in terms of processor and random access memory requirements, with computational time and memory usage scaling proportionally with the number of photons considered in the analysis. 

As free parameters in the timing model we consider the spin frequency ($\nu$), the first frequency derivative ($\dot{\nu}$), and the arbitrary pulse phase offset relative to the reference Modified Julian Date (MJD). For each target we also explored the possibility of a change in the spin-down rate by introducing a second frequency derivative ($\ddot{\nu}$).  In most cases, the 3$\sigma$ credible intervals obtained for $\ddot{\nu}$ overlap with zero, suggesting that for the currently available data the addition of this parameter is not warranted. This is also indicated by the fact that including $\ddot{\nu}$ does not result in an improvement in the pulsed significance as determined by the H-test.  As described later, for RX J0720.4$-$3125 the timing model required the introduction of additional glitch parameters as well as a second frequency derivative ($\ddot{\nu}$).  For all six XINS, the latest previously published ephemerides were used as the initial guess of the free parameters for the sampler.

\section{Timing Results}
\label{sec:results}
The results of the timing analysis for five of the XINS are compiled in Table~\ref{tab:timing}, and for RX J0702.4$-$3125 in Table~\ref{tab:j0720_timing}. We quote measured parameters in frequency space ($\nu$ in Hz and $\dot{\nu}$ in Hz\,s$^{-1}$) but for convenience we also list the corresponding conversions to period space (with $P$ in s and $\dot{P}$ in s\,s$^{-1}$).   For the event-based likelihood analysis results, we quote the customary 50$^{\rm th}$ percentile and the 16$^{\rm th}$ and 84$^{\rm th}$ percentiles credible intervals as the lower and upper uncertainties, respectively, of the posterior distribution. Figures~\ref{fig:triangle} and \ref{fig:j0720_triangle} in the Appendix show the posterior distributions of the timing parameters in the form of corner plots.  The reference epochs (in MJD) used for the quoted spin measurements correspond to the previously published values as listed in the ATNF pulsar catalog \citep{2005AJ....129.1993M}, except for RX J0702.4$-$3125, for which the reference epoch of MJD 56,598 corresponds to the midpoint of the postglitch interval (see Section~\ref{sec:j0720_timing} for details). 

The final timing solutions in the form of Tempo-style parameter (.par) files are provided  as supplemental material to this paper; they contain the values of the fitted parameters corresponding to the highest likelihood sample found in the inference analysis for each XINS. In what follows, for each target we provide initial discovery and background information, a summary of previous work specific to pulse timing, and present the updated timing results.

\subsection{RX J0420.0$-$5022}
RX J0420.0$-$5022 (1RXS J042003.1$-$502300) was identified as a neutron star candidate by \citet{1999A&A...351L..53H}. Follow-up ROSAT observations showed a remarkably soft thermal spectrum and yielded an improved position, both of which strengthened its classification as a neutron star. While initially a 22.7\,s period was suggested, more sensitive observations with XMM-Newton found a period of 3.45\,s instead, the shortest among the XINS \citep{2004AandA...424..635H}.  The same observations also showed that RX J0420.0$-$5022 was the coolest INS, with blackbody $kT \simeq 45$ eV; it is also the faintest in terms of observed photon flux in the soft X-ray band.  The first coherent timing solution for this XINS was derived by \citet{2011ApJ...740L..30K}, who obtained a spin-down rate of $\dot{\nu}=(-2.3\pm 0.2)\times 10^{-15}$ Hz s$^{-1}$ based on the 2010--2011 XMM-Newton observing campaign and $\dot{\nu}=(-2.314\pm0.008) \times 10^{-15}$ Hz s$^{-1}$ when observations dating back to 2002 were included. However, for the latter they noted a number of possible alternative timing solutions due to cycle count ambiguities, of which the quoted one was judged the best based on the $\chi^2$ value of the TOA fit. The spin parameters imply $B_s=1\times10^{13}$ G, $\tau_c=1.9$\,Myr, and by far the highest spin down luminosity among the XINS of $\dot{E}=3\times10^{31}$\,erg\,s$^{-1}$. 

For the present timing analysis, we consider XMM-Newton data spanning the date range 2002 December 30 to 2019 May 22.  The only available archival Chandra ACIS-S observations were obtained in full frame imaging mode, which has a readout time of 3.2\,s, rendering them unusable for a timing analysis of this 3.45\,s neutron star. No NICER XTI observations of RX J0420.0$-$5022 have been conducted to date, presumably due to its relatively faint nature. For the barycentering and timing analysis, we assume the source position obtained by \citet{2004AandA...424..635H} based on Chandra imaging observations.

Despite the eight year gap in data, from our event based likelihood analysis we obtain a fairly precise spin-down rate measurement of  $\dot{\nu}=(-2.4393\pm0.0005) \times 10^{-15}$\,Hz\,s$^{-1}$. This value is consistent with the 2010--2011 $\dot{\nu}$ value from \citet{2011ApJ...740L..30K} but not with their longer baseline (2002--2011) $\dot{\nu}$, suggesting that their chosen solution was not the correct one. This is expected since RX J0420.0$-$5022 has the shortest spin period among the XINS by a wide margin and is thus most susceptible to pulse cycle count ambiguities.  We note that a number of alternative, equally spaced solutions are also apparent in the $\nu-\dot{\nu}$ plane from our analysis, which likely arise due to the large gaps in data coverage from 2003--2010 and 2011--2019.  Nevertheless, the quoted $\dot{\nu}$ is strongly favored  based on the strongest pulse detection significance determined from the H-test ($12.09\sigma$ single trial).

\begin{figure}
    \includegraphics[,width=0.46\textwidth]{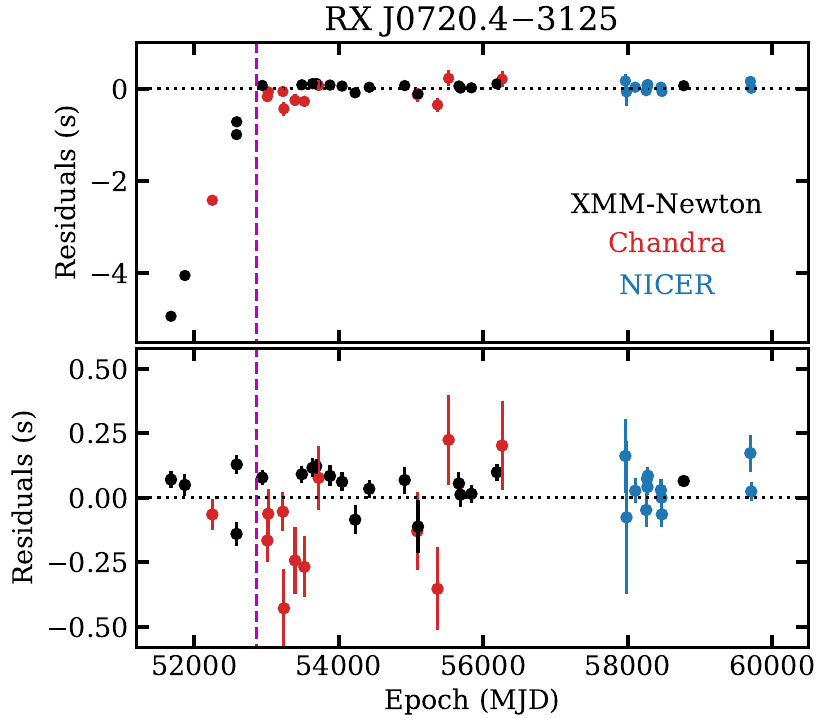}
    \caption{Timing residuals expressed in seconds for RX J0720.4$-$3125 from a TOA analysis in Tempo2 using XMM-Newton (black), Chandra (red), and NICER (blue) observations. The top panel shows the best fit timing solution of all observations after MJD 52866 (marked by the vertical dashed line) without a glitch to illustrate the sudden change in spin behavior. The bottom panel shows the best fit with a glitch included with an assumed epoch of MJD 52886. See Table~\ref{tab:j0720_timing} for the timing model used and text for further details.}
    \label{fig:j0720_residuals}
\end{figure}

\begin{deluxetable}{lC}
\tabletypesize{\footnotesize}
\tablewidth{0pt}
\tablecolumns{7}
\tablecaption{Timing Solution for RX J0720.4$-$3125 \label{tab:j0720_timing}}
\tablehead{\colhead{}	&	\colhead{Parameter\tablenotemark{a}}}
\startdata									
\hline
    \multicolumn{2}{c}{\textit{Assumed quantities}} \\                
\hline 		
R.A. (J2000.0)\tablenotemark{b}	&	$07^{\rm h}20^{\rm m}24.\!^{\rm s}9620$			\\
Decl. (J2000.0)\tablenotemark{b}	&	$-31^{\circ}25^{\prime}50\farcs08$	\\				
$\mu_{\alpha}\cos\delta$ (mas\,yr$^{-1}$)\tablenotemark{b}	&	$-93.9(12)$	\\
$\mu_{\delta}$ (mas\,yr$^{-1}$)\tablenotemark{b}	&	$+52.8(13)$	\\		
Position epoch (MJD) & $52,645.1$ \\
Glitch epoch (MJD)  &	52,866	 \\
\hline
    \multicolumn{2}{c}{\textit{Measured quantities}} \\                
\hline									
Epoch (MJD TDB)	&	56,598.0	\\
$\nu$ (Hz) &		0.1191733514^{(+1)}_{(-2)} 	\\
$\dot{\nu}$ ($10^{-15}$\,Hz\,s$^{-1}$) &	-1.0156^{+0.0004}_{-0.0004} \\
$\ddot{\nu}$ ($10^{-27}$\,Hz\,s$^{-2}$) &		5.9^{+0.9}_{-0.8}	\\
$\Delta\Phi$ ($\phi/2\pi$)\tablenotemark{c}	&	 0.03^{+0.002}_{-0.002}		\\
$\Delta\nu$ ($10^{-9}$\,Hz)\tablenotemark{c} &		5.02^{+0.05}_{-0.05}	\\
$\Delta\dot{\nu}$ ($10^{-17}$\,Hz\,s$^{-1}$)\tablenotemark{c}   &		-1.20^{+0.04}_{-0.04} \\
Range of dates (MJD)	& 51,575-59,720 \\	
H-test single trial significance ($\sigma$) &	204.73 \\
\hline
    \multicolumn{2}{c}{\textit{Derived quantities}} \\                
\hline
$P$ (s)	&		8.391137685^{(+8)}_{(-11)}	\\
$\dot{P}$ ($10^{-14}$\,s\,s$^{-1}$)	&	7.151^{+0.003}_{-0.003}	\\
$\ddot{P}$ ($10^{-25}$\,s\,s$^{-2}$)	& -4.1^{+0.6}_{-0.6} \\
$\dot{E}$ ($10^{30}$\,erg\,s$^{-1}$)	& 4.8	\\
$\tau_c$ (Myr)	&	1.9 \\
$B_s$ ($10^{13}$\,G)	& 2.5	 \\
\enddata
\tablenotetext{a}{The numbers quoted for the fitted parameters correspond to the 50$^{\rm th}$ percentile  of the posterior distribution with the 16$^{\rm th}$ ($-$) and 84$^{\rm th}$ ($+$) percentiles given in parentheses.}
\tablenotetext{b}{Astrometric parameters measured from optical imaging observations \citep{2007ApJ...660.1428K}.}
\tablenotetext{c}{$\Delta\Phi$ $:=$ glitch permanent spin phase increment; $\Delta\nu$ $:=$ permanent frequency increment; $\Delta\dot{\nu}$ $:=$ permanent first frequency derivative increment.}
\end{deluxetable}

\subsection{RX J0720.4$-$3125}
\label{sec:j0720_timing}
RX J0720.4$-$3125 (1RXS J072025.1$-$312554) is a XINS with a spin period of $P=8.39$\,s, discovered by \citet{1997A&A...326..662H} over the course of the ROSAT All-Sky Survey. After multiple attempts to constrain the spin-down rate \citep{2002ApJ...570L..79K,2002MNRAS.334..345Z,2004MNRAS.351.1099C}, a fully phase-connected long-term timing solution was finally obtained by \citet{2005ApJ...628L..45K} using Chandra and XMM-Newton observations. A spin-down rate of $\dot{P}=6.98\times 10^{-14}$ s s$^{-1}$ was measured, which implies a characteristic age of $\tau_c =2$\,Myr and a dipole magnetic field strength of $B_s=2.4\times10^{13}$\,G. 

\citet[][see also \citealt{2006A&A...451L..17H,2007ApJ...659L.149V,2009A&A...498..811H}]{2010A&A...521A..11H} used additional observations to extend the timing solution but identified residuals suggestive of long term variations in the spin behavior that could not be accounted for solely by a $\dot{P}$. The interpretations offered for this behavior included free precession of the neutron star or a glitch. However, a follow-up analysis \citep{2012MNRAS.423.1194H} concluded that the apparent timing irregularities are best explained by a glitch that occurred at some point during the gap in observations spanning the MJD interval $52,866 \pm 73$. \citet{2007ApJ...659L.149V} reported the first frequency derivative to be $\dot{\nu}=(-1.04\pm0.03)\times10^{-15}$\,Hz\,s$^{-1}$ and the glitch permanent frequency increment to be $\Delta\nu=(4.1\pm1.2)\times10^{-9}$ Hz, while the glitch permanent frequency first derivative increment was not well constrained with $\Delta\dot{\nu}=(-4\pm3)\times10^{-17}$\,Hz\,s$^{-1}$. 

RX J0720.4$-$3125 has been observed extensively over the past two decades (though with a sizable gap in coverage between 2012 and 2017), with a total on-source exposure second only to RX J1856.5$-$3754 (see Table~\ref{table:obs}). 
XMM-Newton has targeted it in 21 separate observations for which suitable on-source data is available, spanning from 2000 May 13 to 2019 October 16, for total effective net exposures of 476.4, 547.4, and 524.1\,ks for pn, MOS1, and MOS2, respectively.  Archival Chandra observations consisting of
21 HRC-S LETG and 11 ACIS-S Continuous Clocking mode exposures that cover the period between 2000 February 1 and 2012 November 28 are also used in this analysis. Finally, there are 58 mostly brief NICER XTI exposures that add up to 104.6\,ks of clean on-source time accumulated from 2017 June 30 to 2022 May 21. Collectively, the usable event data from the three observatories cover a period of 22 years, providing an additional decade of timing baseline coverage compared to the last published ephemeris. For the astrometric parameters, we adopt the coordinates and proper motion values measured by \citet{2007ApJ...660.1428K} using optical observations. 

As in prior published analyses of RX J0720.4$-$3125, to account for the anomalous long-term spin behavior we introduce three glitch parameters: the glitch spin phase increment $\Delta\Phi$, the permanent frequency increment $\Delta\nu$, and the permanent frequency first derivative increment $\Delta\dot{\nu}$. For illustrative purposes, Figure~\ref{fig:j0720_residuals} shows the results of a TOA based timing analysis of RX J0720.4$-$3125 that considers a steady spin down model based on the most recent XMM-Newton and NICER data set that is propagated backwards in time. A clear discontinuity in the model is apparent around MJD 52886 (which we fix as the presumed epoch of the glitch), consistent with the glitch interpretation (see top panel of Fig.~\ref{fig:j0720_residuals}). Moreover, we find that even after accounting for the glitch, there are broad residuals that suggest the presence of a second frequency derivative $\ddot{\nu}$, which we also include as a free parameter.

The updated timing solution for RX J0720.4$-$3125 using the event-based likelihood technique is presented in Table~\ref{tab:j0720_timing}, while the posterior distributions of the free parameters are shown in Figure~\ref{fig:j0720_triangle}. The resulting values of  $\dot{\nu}=(-1.0156 \pm 0.0004)\times 10^{-15}$\,Hz\,s$^{-1}$, $\Delta\nu=(5.02\pm0.05)\times10^{-9}$ Hz, and $\Delta\dot{\nu}=(-1.20\pm0.04)\times10^{-17}$\,Hz\,s$^{-1}$ are in agreement with the values reported by \citet{2007ApJ...659L.149V} but with substantially improved precision, owing to the greatly extended time span of the current archival data set. We also obtain a fairly well constrained value of the second frequency derivative $\ddot{\nu}=5.9^{+0.9}_{-0.8}\times10^{-27}$\,Hz\,s$^{-2}$, making RX J0720.4$-$3125 the only XINS thus far with a significant $\ddot{\nu}$. With this improved timing model we obtain a stacked pulse profile with a 204.73$\sigma$ single-trial significance, by far the highest among the XINS.

\subsection{RX J0806.4$-$4123}
The soft ROSAT source RX J0806.4$-$4123 (1RXS J080623.0$-$412233) was selected as a likely nearby neutron star by \citet{1998AN....319...97H}. Exploratory observations with XMM-Newton established the presence of soft blackbody-like emission and identified a periodicity at $P=11.37$\,s \citep{2002A&A...391..571H}.  Using additional observations, \citet{2009ApJ...705..798K} were only able to obtain a marginally significant spin-down rate of $\dot{\nu}=(-4.3\pm2.3)\times10^{-16}$ Hz s$^{-1}$ based on XMM-Newton data from 2008 and 2009. Very recently, the spin-down rate was measured to be $(-7.3\pm1.2)\times10^{-17}$ Hz s$^{-1}$ based solely on the NICER XTI data set covering the period from 2019 to 2023 \citep{Posselt24}.

RX J0806.4$-$4123 has the least amount of devoted on-source exposure among the XINS.
Here, we mainly make use of XMM-Newton observations spanning the time period from 2000 November 8 to 2019 May 16 that yield net exposures of 98.6, 73.2, and 72.1 ks for pn, MOS1, and MOS2, respectively. Two Chandra observations of RX J0806.4$-$4123 have been conducted so far, one 18.1 ks ACIS-S Continuous Clocking mode exposure from 2010 March 21 and one 29.8 ks HRC-I LETG exposure from 2018 September 1, which we include in the timing analysis.  Finally, we consider NICER XTI data collected between 2019 March 11 and 2023 February 7 for a total clean on-source exposure of 104.6 ks. All observations are compiled in Table~\ref{table:j0806_obs}. For the barycentering of event data and timing analysis, the source sky coordinates measured by \citet{2004AandA...424..635H} from Chandra ACIS-S observations were considered. 

The event-based likelihood timing analysis results in a spin frequency derivative $\dot{\nu}=(-8.18 \pm 0.01) \times 10^{-17}$\,Hz\,s$^{-1}$, in full agreement with that found independently and using alternative methods by \citet{Posselt24} but with substantially reduced uncertainties due to the wider time span of the data set (2000--2023). 
The inferred $\dot{\nu}$ value is the smallest among the XINS, which implies a low spin-down luminosity of only $\dot{E}=3\times10^{29}$ erg s$^{-1}$ and a characteristic age of $\tau_c=17$\,Myr, an order of magnitude larger than the rest of the XINS.  The combined XMM-Newton, Chandra, and NICER data yield a pulse detection significance of 29.11$\sigma$ based on the H-test in the optimal 0.27--0.89 keV energy range.

\subsection{RX J1308.6+2127}
\label{sec:J1308}
RX J1308.6+2127 (also cataloged as RBS\,1223 and RXS J130848.6+212708) was identified as a strong  neutron star candidate among the set of ROSAT bright sources by \citet{1999A&A...341L..51S}. A periodicity of 5.16\,s was initially reported by \citet{2002A&A...381...98H}, which later investigations identified to actually correspond to half of the true neutron star spin period of $P=10.31$\,s \citep{2004AdSpR..33..638H,2005A&A...441..597S}. RX J1308.6+2127 is notable in that it is the only XINS to exhibit two clearly separated pulses per rotation, which suggest the presence of at least two distinct X-ray emitting regions on the stellar surface. The first and only published coherent timing solution was obtained by \citet{2005ApJ...635L..65K}, who arrived at a frequency rate of change value of $\dot{\nu}=(1.053\pm0.003)\times10^{-15}$ Hz s$^{-1}$. 

\citet{2011A&A...534A..74H} have used XMM-Newton observations of RX J1308.6+2127 to place constraints on the emission geometry, surface chemical composition, magnetic field strength, and neutron star compactness. More recently, \citet{2017MNRAS.468.2975B} claimed the detection of a narrow phase-dependent feature in the X-ray spectrum with an energy of $\sim$740 eV and an equivalent width of $\sim$15\,eV. The purported feature was detected only in $\sim$1/5 of the phase cycle, and appears to be present for the entire time span covered by the observations examined (2001 December to 2007 June). The strong dependence on the pulsar rotation and the narrow width was speculated to be due to resonant cyclotron absorption/scattering in a confined high magnetic field structure close to the stellar surface.

For our updated timing analysis of RX J1308.6+2127 we include a more recent XMM-Newton observation from 2019 December 16 (ObsID 0844140101), a single 87.2 HRC-S LETG observation (ObsID 4595) from 2005 April 12, and eight ACIS-S CC mode observations obtained between 2006 February 12 and 2006 August 3. To this we add a collection of NICER XTI exposures covering the period 2017 November 14 to 2018 June 26 with a combined clean effective exposure of 120.9\,ks. In the timing analysis, the right ascension and declination for the source were assumed to correspond to the Chandra position reported in \citet{2002ApJ...579L..29K}.

The median value we obtain for the rate of change of frequency for RX J1308.6+2127 is $\dot{\nu}=(-1.05427\pm 0.00003)\times10^{-15}$ Hz s$^{-1}$, in excellent agreement with the \citet{2005ApJ...635L..65K} timing solution but with a $\approx$100-fold improvement in precision, despite the decade-long gap in X-ray observations of this source (2007--2017). By folding the entire collection of event data at the newly derived ephemeris we obtain a maximum pulse detection significance of 139.56$\sigma$ as determined by the H-test in the 0.30--1.21\,keV band.

\subsection{RX J1856.5$-$3754}

RX J1856.5$-$3754 (1RXS J185635.1$-$375433) is regarded as the canonical XINS and is the X-ray brightest. It was the first of the XINS to be identified as a strong neutron star candidate based on its soft spectrum and the lack of a bright optical counterpart \citep{1996Natur.379..233W}. It is one of the nearest known NSs of any variety, with a parallax distance of only $D=123^{+11}_{-15}$\,pc \citep{2010ApJ...724..669W}. Despite these facts, it was the sixth XINS to have pulsations detected \citep{2007ApJ...657L.101T};  although exceptionally bright in terms of photon flux, the thermal radiation features an extraordinarily low pulsed fraction of $\sim$1\%, necessitating deep XMM-Newton observations to uncover the $7.055$\,s periodicity.  This property can be plausibly interpreted as a thermally emitting region that is nearly aligned with the spin axis and/or a close alignment between the spin axis and the observer line of sight \citep{2007MNRAS.380...71H}. 

Previously, \citet{2008ApJ...673L.163V} conducted a timing analysis, which produced only an upper limit on the spin-down rate. A definitive measurement was obtained in a follow-up study by \citet{2009ApJ...692L..62K}. Most recently, \citet{2022MNRAS.516.4932D} presented an updated timing solution for RX J1856.5$-$3754, with $\dot{\nu}=-6.042(4) \times 10^{-16}$ Hz s$^{-1}$ but using XMM-Newton EPIC-pn exposures only up to and including 2022 April 3 (ObsID 0810842001) and a portion of presently available NICER observations through 2019 June 18 (ObsID 2614010139). 

There exist a large number of archival X-ray observations of RX\,J1856.5$-$3754 because it is observed with great regularity as an instrument calibration target. This has resulted in a bountiful archival data set with $\approx$2\,Ms of XMM-Newton, $\approx$1.6 Ms of Chandra, and 186\,ks of NICER XTI exposures accumulated over the course of the past two decades.
Here, we make use of EPIC pn as well as MOS1/2 exposures and include two additional XMM-Newton observations (ObsIDs 0810842201 and 0810842301) from 2022 September 24 and 2023 April 1 (see Table~\ref{table:j1856_obs} in Appendix) that were not used in \citet{2022MNRAS.516.4932D}. We also analyze an extended set of NICER XTI observations up to 2022 October 10 that also include 33 calibration exposures acquired at off-axis angles\footnote{These observations were carried out at the same 1.47$\arcmin$ offset as NICER observations conducted for the millisecond pulsar PSR J0437$-$4715 to calibrate the off-axis response of the XTI (see \citealt{2019ApJ...887L..25B} for further details). This offset results in only an $\approx$8\% reduction  in the effective area at all energies due to vignetting at the position of RX\,J1856.5$-$3754.}  that are usable for a timing analysis. Finally, we consider a number of Chandra ACIS-S Continous Clocking mode exposures as well as the zeroth order point source counts from the extensive set of Chandra HRC-S LETG observations, which were also not considered in \citet{2022MNRAS.516.4932D}. For the astrometric parameters of RX J1856.5$-$3754, we use the measurements obtained by \citet{2010ApJ...724..669W}.

The event-based likelihood timing analysis for RX\,J1856.5$-$3754 results in $\dot{\nu}=(-6.0373^{+0.0008}_{-0.0007}) \times 10^{-16}$ Hz s$^{-1}$, which is fully consistent with the measurement from \citet{2022MNRAS.516.4932D} within uncertainties, although with significantly improved precision owing to the use of a larger data set that is also extended by an additional year. Despite the order of magnitude more exposure than the other XINS, resulting in $\approx 2\times10^7$ counts collected for this source, the pulse detection significance of $35.13\sigma$ from the H-test (in the optimal energy band 0.15--0.74\,keV) is substantially lower than those for RX\,J0720.4$-$3125 and RX\,J1308.6+2127 because of the exceptionally low pulse amplitude.

\begin{figure*}
    \centering
    \includegraphics[trim={0.5cm 0.5cm 0.5cm 0},clip,width=0.49\textwidth]{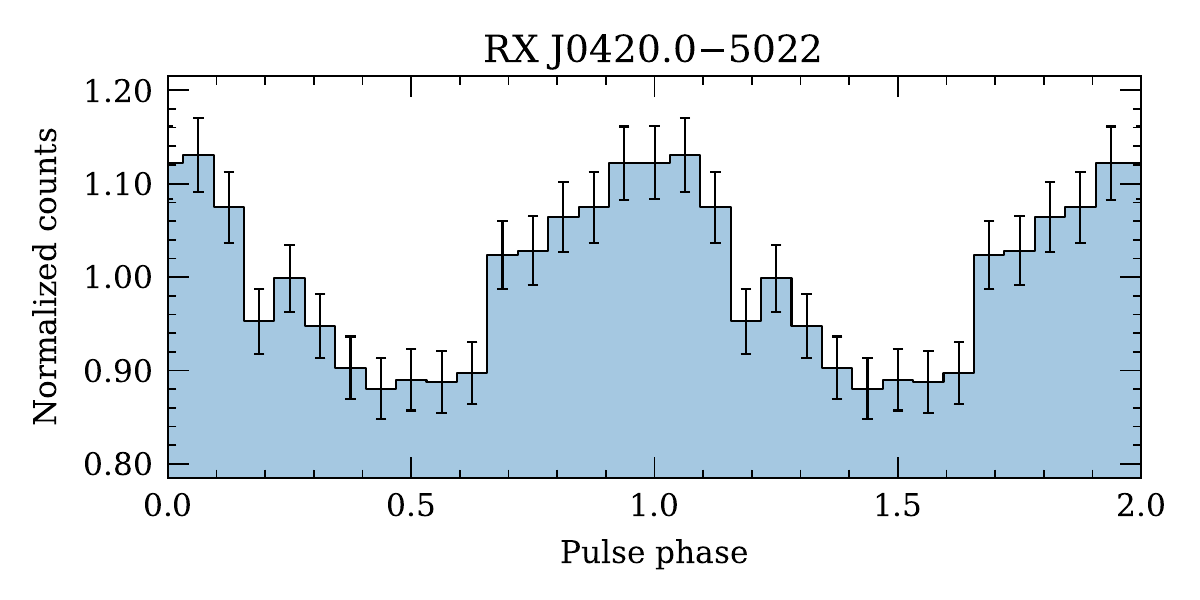}
    \includegraphics[trim={0.5cm 0.5cm 0.5cm 0},clip,width=0.49\textwidth]{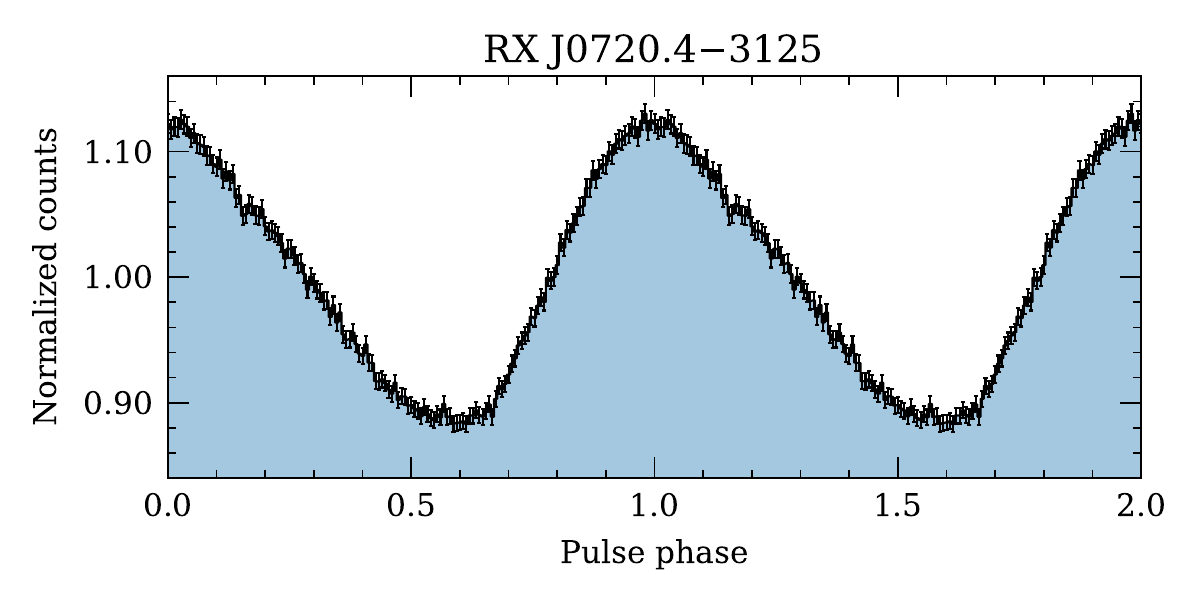}
    \includegraphics[trim={0.5cm 0.5cm 0.5cm 0},clip,width=0.49\textwidth]{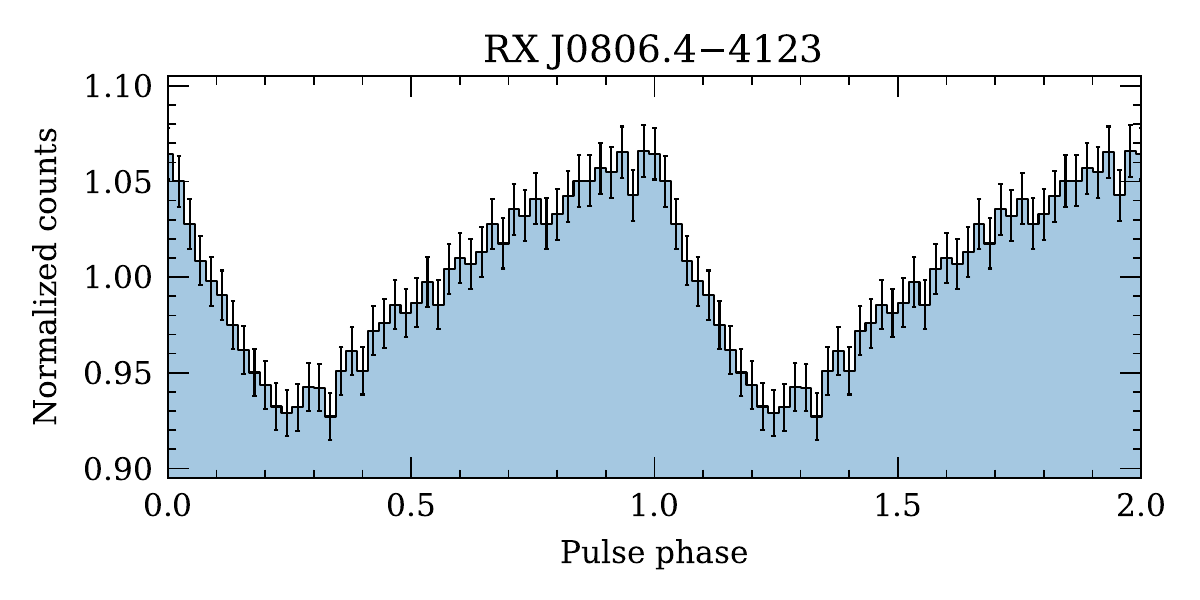}
    \includegraphics[trim={0.5cm 0.5cm 0.5cm 0},clip,width=0.49\textwidth]{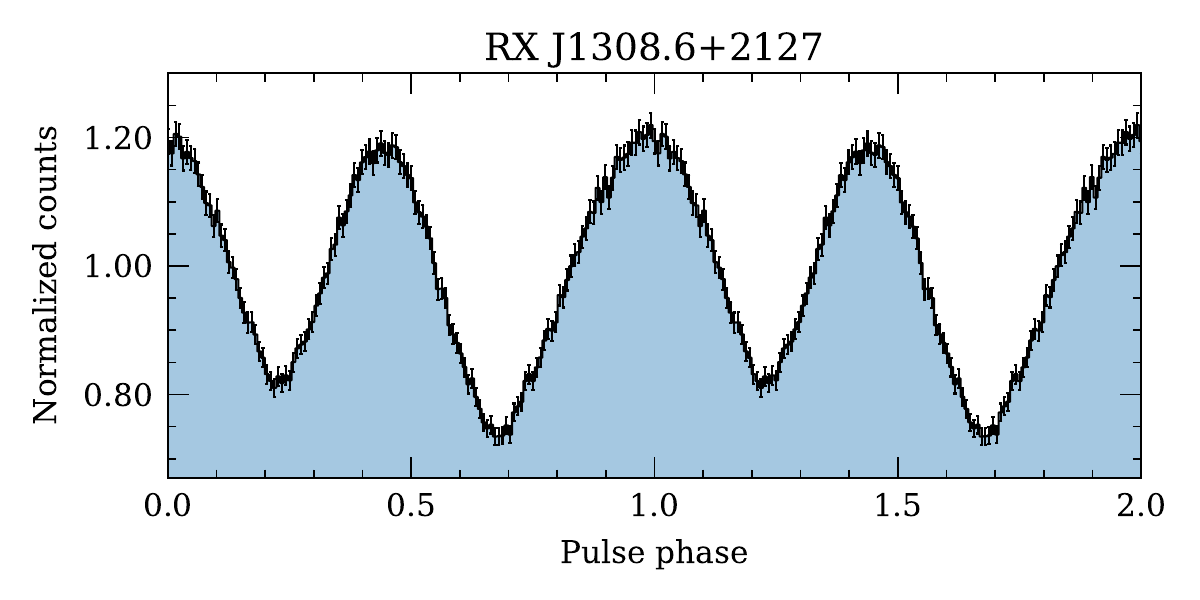}
    \includegraphics[trim={0.5cm 0.5cm 0.5cm 0},clip,width=0.49\textwidth]{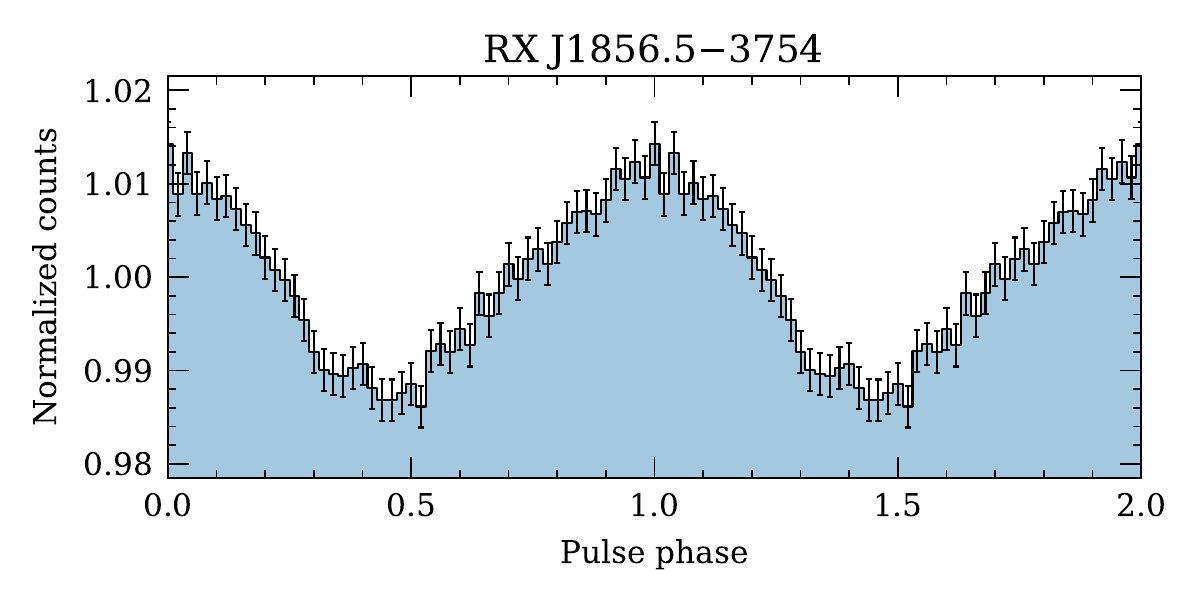}
    \includegraphics[trim={0.5cm 0.5cm 0.5cm 0},clip,width=0.49\textwidth]{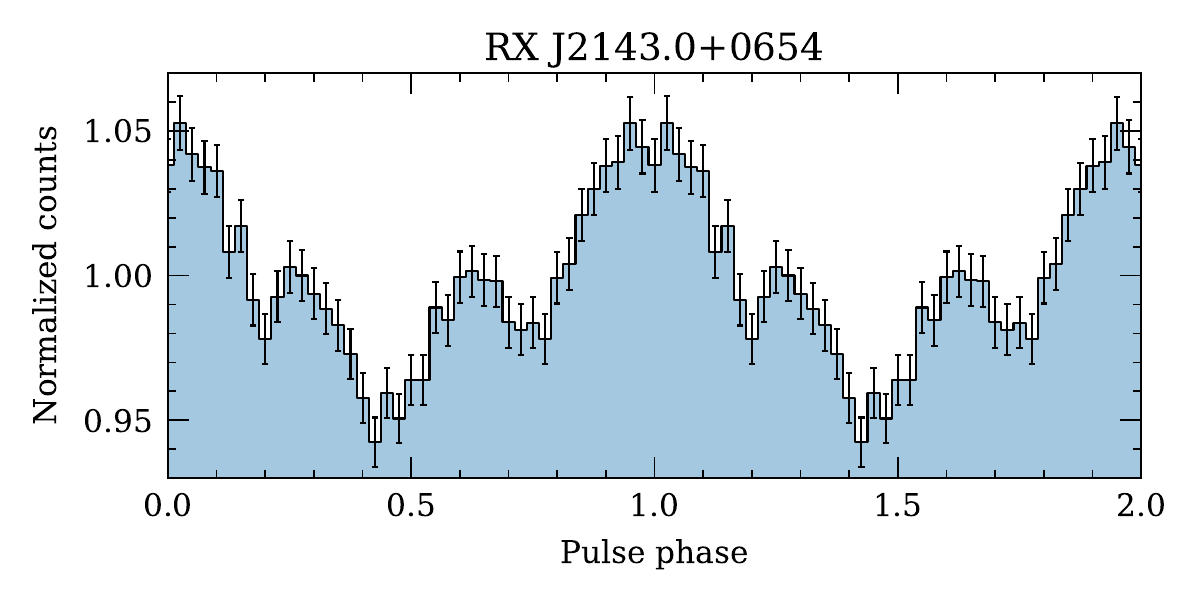}
    \caption{Normalized pulse profiles of the six XINS obtained by folding and stacking all XMM-Newton EPIC pn and MOS1/2, Chandra HRC-S and ACIS, and NICER (where available) X-ray event data at the timing solutions reported in Tables~\ref{tab:timing} and \ref{tab:j0720_timing}. See text for the energy range used for each profile. A rotational cycle is repeated for clarity. }
    \label{fig:profiles}
\end{figure*}

\begin{figure*}
    \centering
    \includegraphics[trim={0.0cm 0.4cm 0.0cm 0.3cm},clip,width=0.40\textwidth]{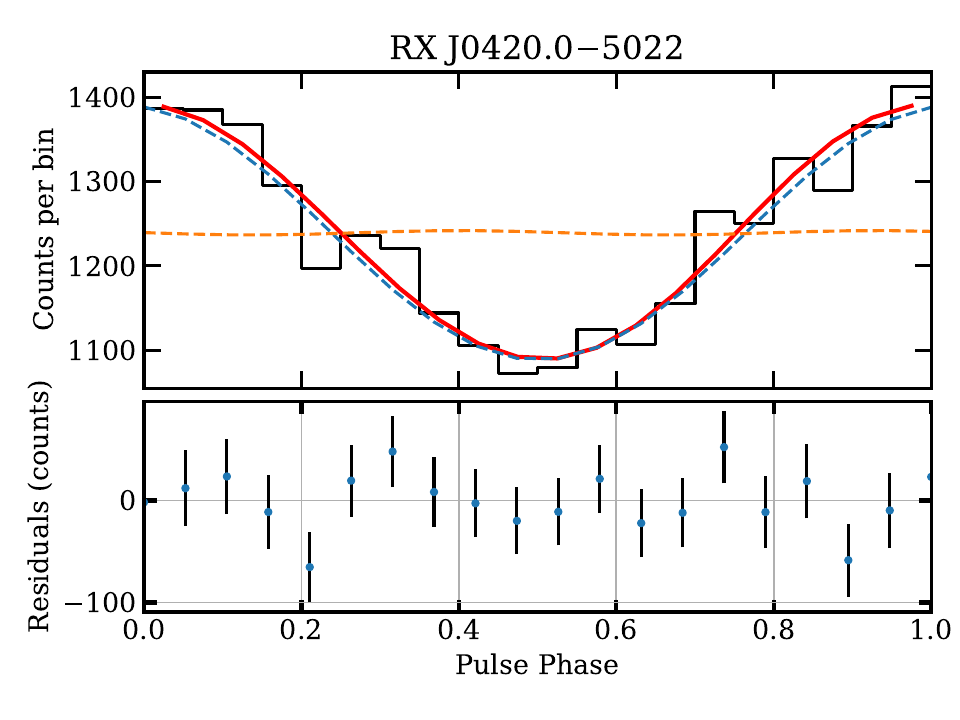}~~~
    \includegraphics[trim={0.0cm 0.4cm 0.0cm 0.3cm},clip,width=0.40\textwidth]{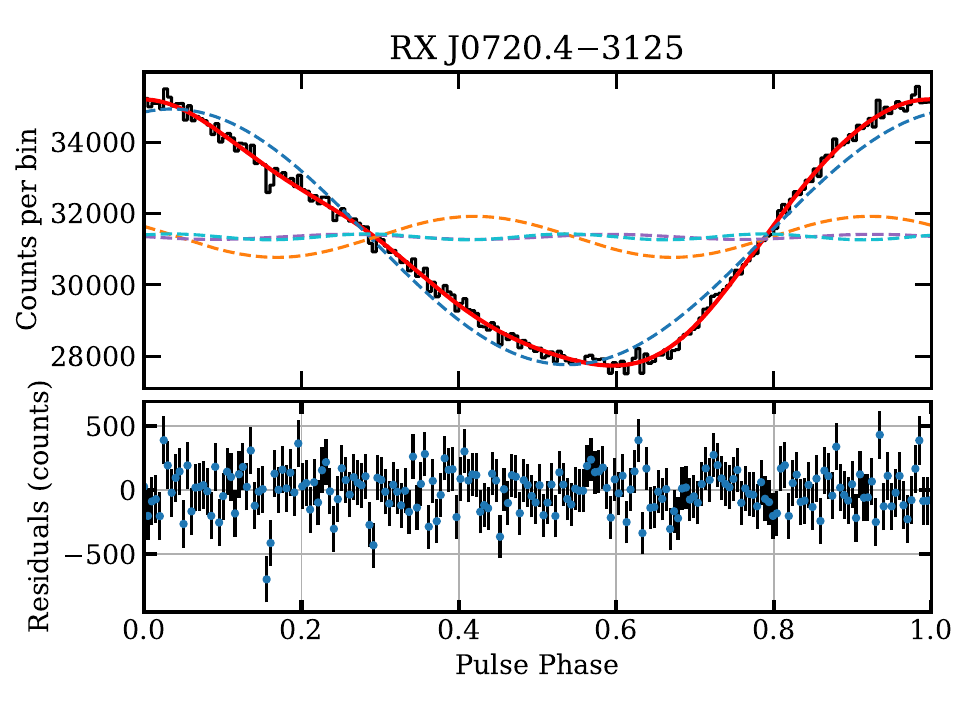}
    \includegraphics[trim={0.0cm 0.4cm 0.0cm 0.3cm},clip,width=0.40\textwidth]{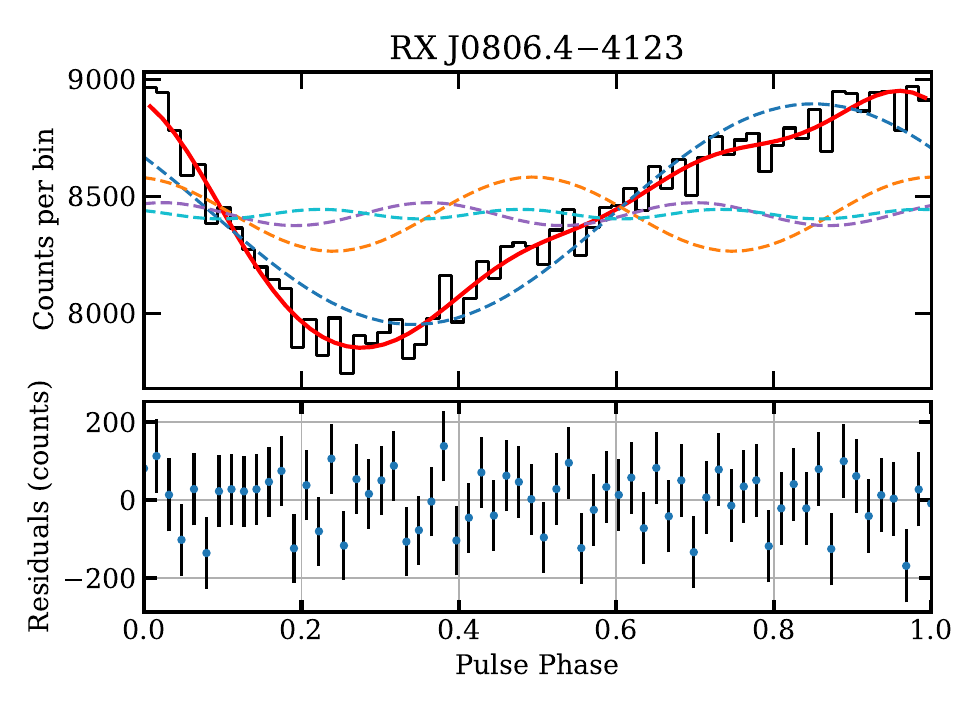}~~~
    \includegraphics[trim={0.0cm 0.4cm 0.0cm 0.3cm},clip,width=0.40\textwidth]{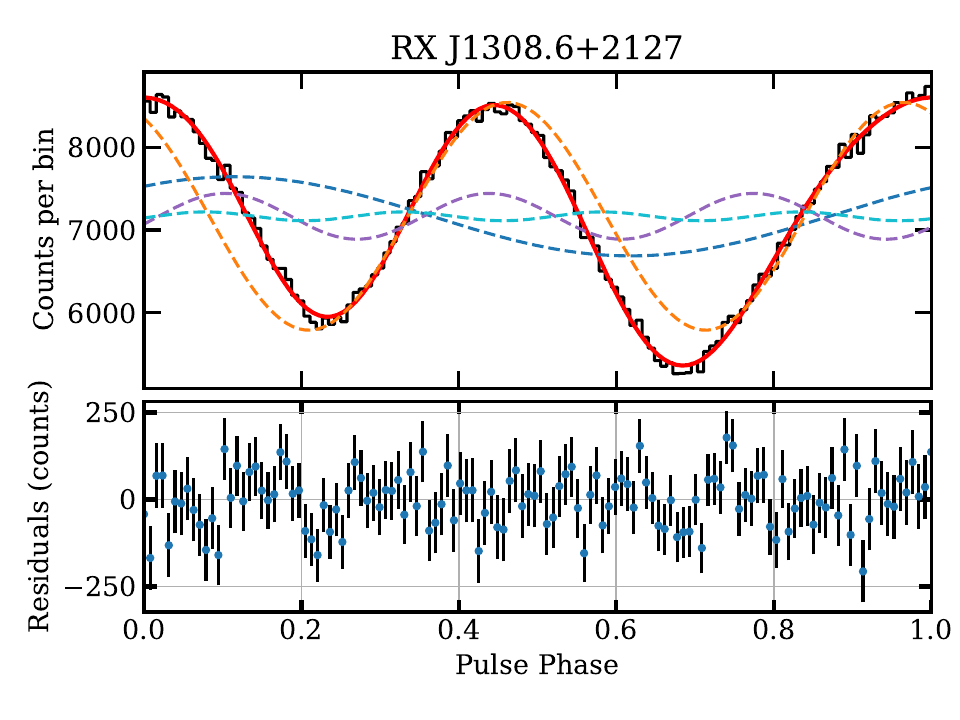}
    \includegraphics[trim={0.0cm 0.4cm 0.0cm 0.3cm},clip,width=0.40\textwidth]{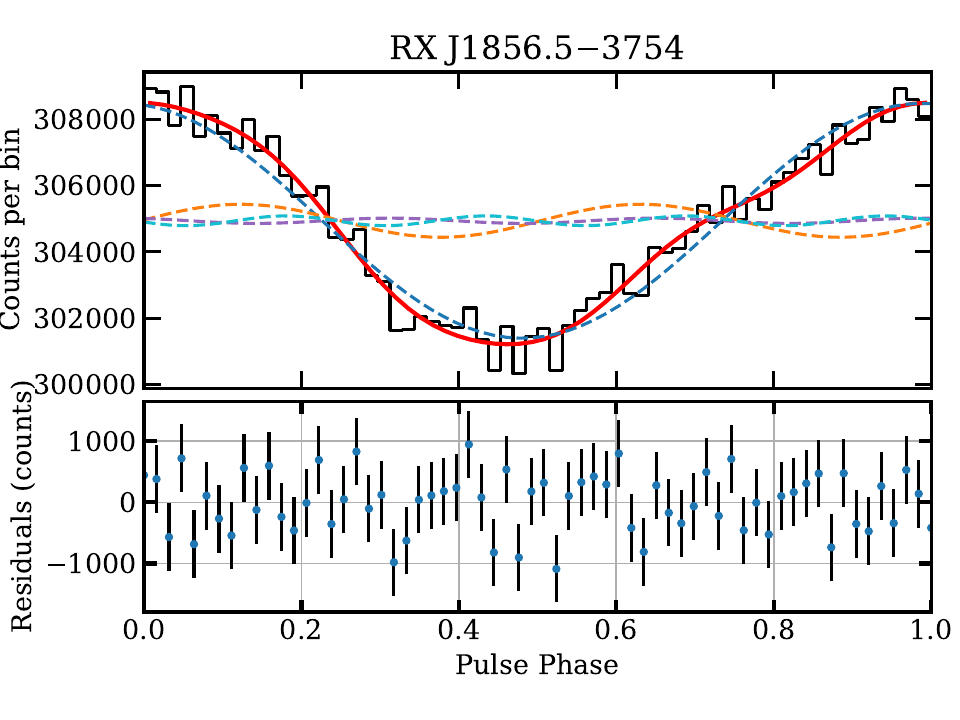}~~~
    \includegraphics[trim={0.0cm 0.4cm 0.0cm 0.3cm},clip,width=0.40\textwidth]{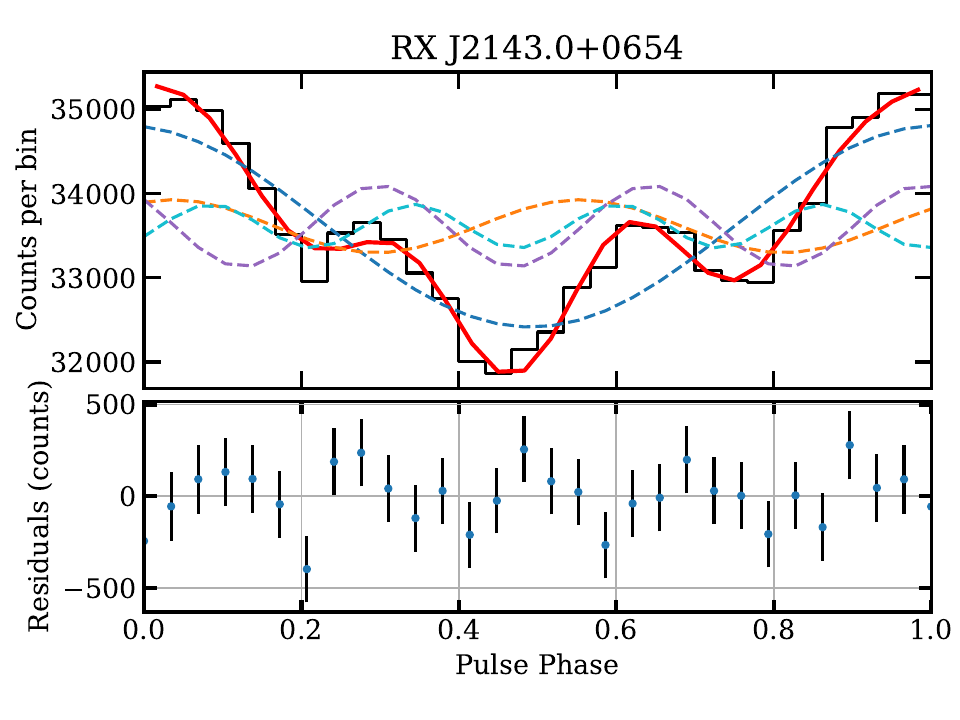}
    \caption{The folded profiles from Figure~\ref{fig:profiles} but with the most significant Fourier components superposed. Error bars are omited to improve clarity and a single rotational cycle is shown.  The solid red lines show the best fit with a model of the profile constructed from the empirical Fourier coefficients of the set of photon phases. The dashed lines show the sinusoids corresponding to the first harmonically related components of the fit (blue, orange, purple, and cyan for the first, second, third, and fourth harmonic, respectively).  The bottom panel for each pulsar shows the residuals from the fit.}
    \label{fig:harms}
\end{figure*}

\begin{figure*}
    \centering
    \includegraphics[trim={0.5cm 0.5cm 0.5cm 0},clip,width=0.40\textwidth]{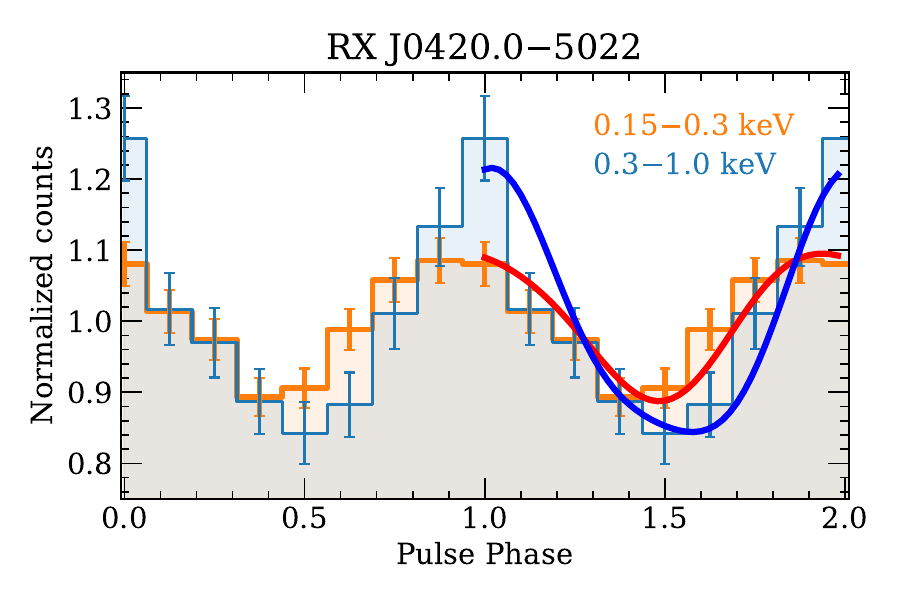}~~~~
    \includegraphics[trim={0.5cm 0.5cm 0.5cm 0},clip,width=0.40\textwidth]{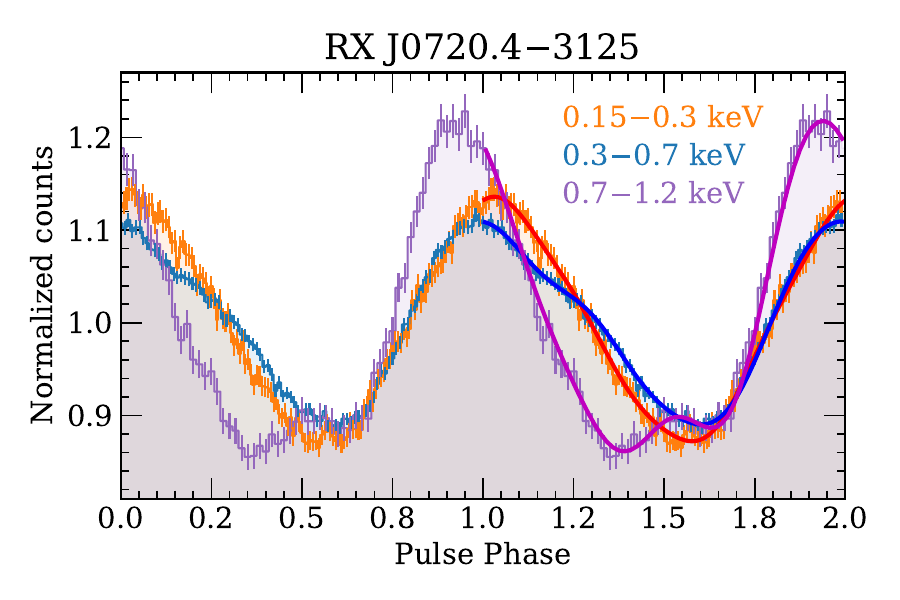}
    \includegraphics[trim={0.5cm 0.5cm 0.5cm 0},clip,width=0.40\textwidth]{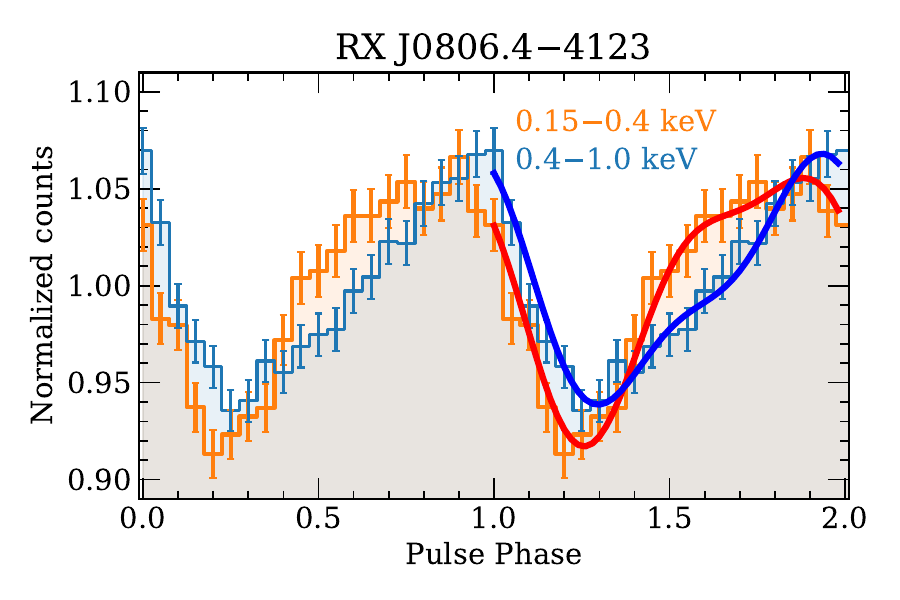}~~~~
    \includegraphics[trim={0.5cm 0.5cm 0.5cm 0},clip,width=0.40\textwidth]{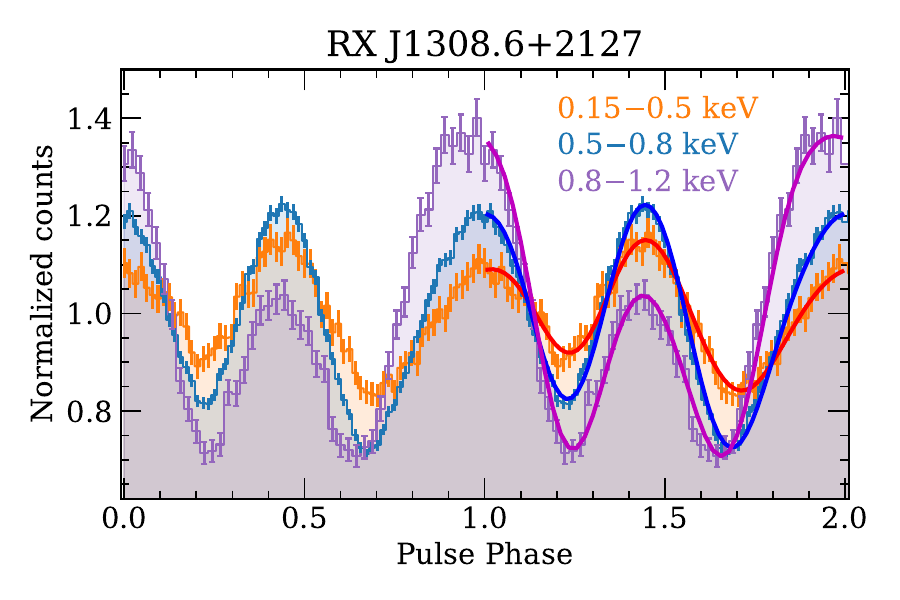}
    \includegraphics[trim={0.5cm 0.5cm 0.5cm 0},clip,width=0.40\textwidth]{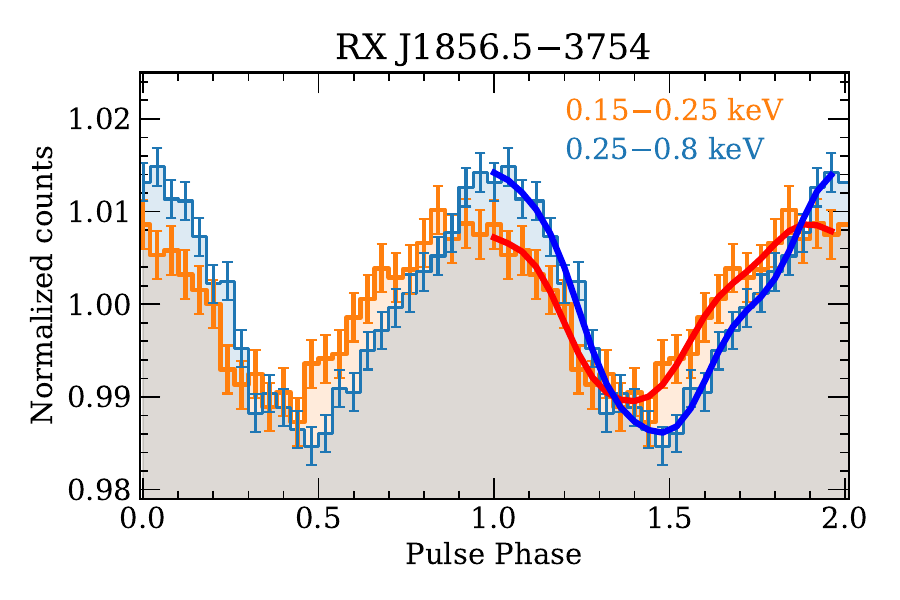}~~~~
    \includegraphics[trim={0.5cm 0.5cm 0.5cm 0},clip,width=0.40\textwidth]{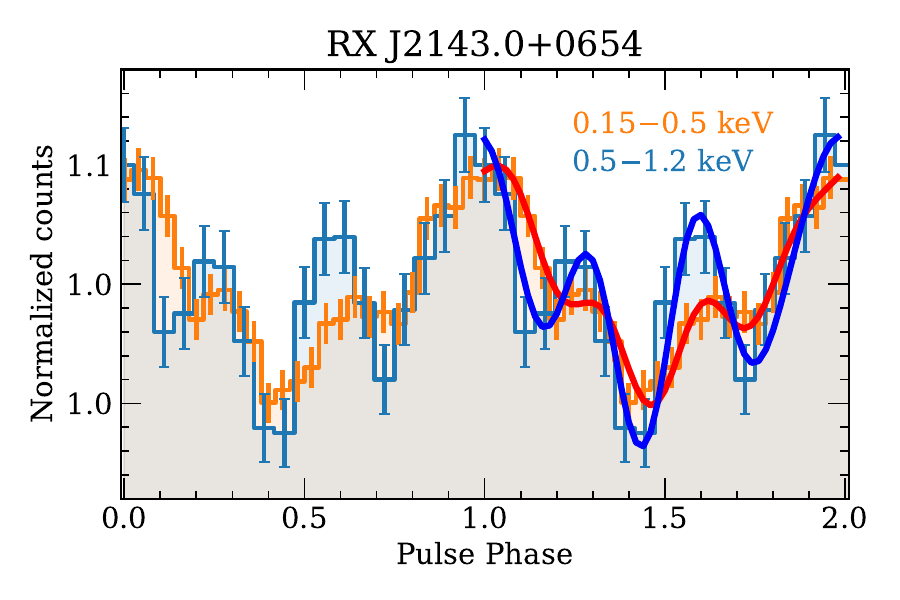}
    \caption{Normalized pulse profiles of the six XINS split into multiple energy bands. For RX J0720.4$-$3125 and RX J1308.6+2127, profiles for three energy intervals are shown in orange, blue and purple for the soft, mid, and hard bands, respectively. For the rest, the binned profile data are split into two energy ranges with the softer and harder bands corresponding to the orange and blue curves, respectively. The superposed solid red, blue, and purple lines in the second rotational cycle show the best fit with a representation of the profile constructed from the empirical Fourier coefficients of the set of photon phases for the same energy range.}
    \label{fig:profiles_multi}
\end{figure*}

\begin{figure}
    \centering
    \includegraphics[width=0.42\textwidth]{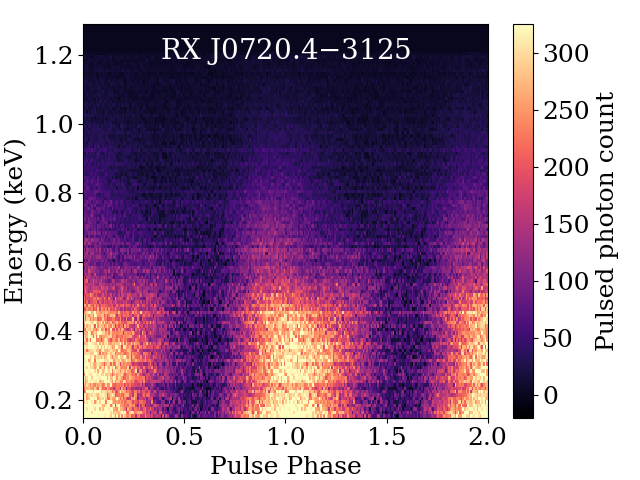}
    \includegraphics[width=0.42\textwidth]{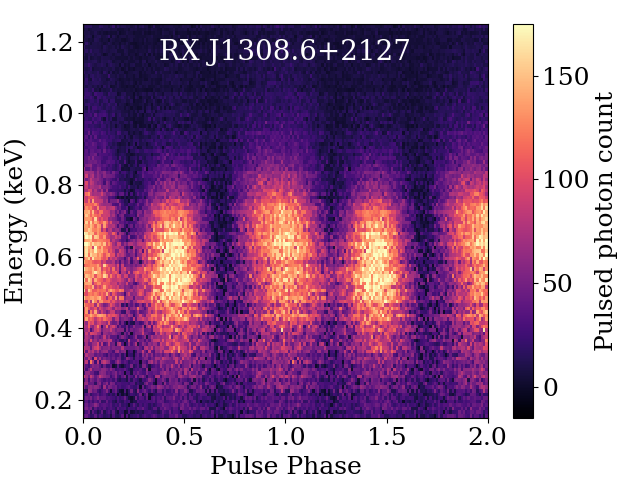}
    \includegraphics[width=0.42\textwidth]{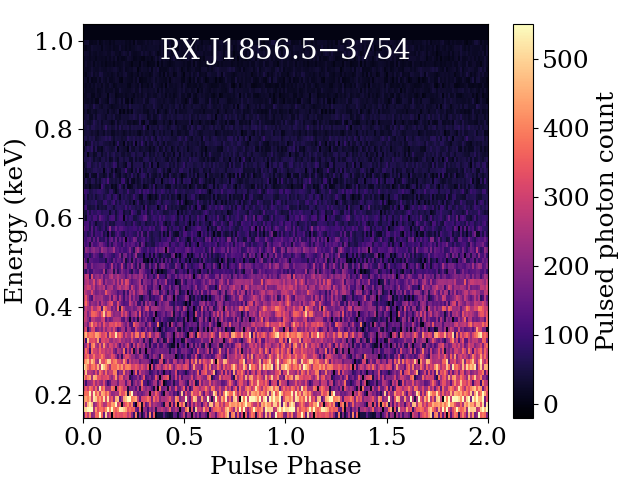}
    \caption{Two-dimensional histograms of pulsed counts versus rotational phase and photon energy for RX J0720.4$-$3125, RX J1308.6+2127, and RX J1856$-$3754 (from top to bottom, respectively). The color bar shows the number of pulsed counts in each pixel with counts increasing from darker to lighter colors. The horizontal striping is an artifact produced by instrumental edge features.  Two pulse phase cycles are shown for clarity.}
    \label{fig:xins_2d}
\end{figure}

\subsection{RX J2143.0+0654}
RX\,J2143.0+0654 (also known as RBS\,1774 and 1RXS J214303.7+065419) was identified as a likely neutron star candidate by \citet{2001A&A...378L...5Z}. Its true nature was established by \citet{2005ApJ...627..397Z}, who uncovered a spin period of 9.437\,s and a thermal, blackbody-like spectrum. Follow up X-ray timing observations by \citet{2009ApJ...692L..62K} yielded only a marginally significant spin-down rate of  $\dot{\nu}= {(-4.6\pm 2.0) \times 10^{-16}}$ Hz s$^{-1}$. 

For the updated timing analysis we consider XMM-Newton observations covering the period 2004 May 31 to 2019 May 22, though with an 11 year gap in coverage between 2008 and 2019. The total filtered on-source exposures are 90.2, 136.1 and 136.9\,ks for pn, MOS1, and MOS2, respectively. The Chandra data set for RX\,J2143.0+0654 consists of a single 12.1\,ks ACIS-S CC mode observation from 2009 December 24 and four HRC-S LETG observations from 2018 August 18 to 2018 September 14 that total 139.7\,ks. The 29 NICER XTI observations targeting this source were carried out between 2022 July 26 and 2023 August 28, yielding a substantial boost in event data with 272.3\,ks of clean exposure. For the barycentering and timing analysis, we take the position published in \citet{2009AandA...499..267S} of the probable optical counterpart of RX J2143.0+0654.  

With the greatly extended timing baseline obtained by including additional Chandra, XMM-Newton, and NICER observations spanning 2018 August to 2023 August, we obtain the first solid spin-down measurement $\dot{\nu}=(-4.663^{+0.006}_{-0.004}) \times 10^{-16}$ Hz s$^{-1}$ or $\dot{P}=(4.145^{+0.005}_{-0.004})\times10^{-14}$\,s\,s$^{-1}$  for RX\,J2143.0+0654, consistent with the initial constraint from \citet{2009ApJ...692L..62K}. The posterior distribution is broad, asymmetric, and multipeaked (see Fig.~\ref{fig:triangle} in the Appendix), which is presumably caused by the large gap in X-ray observations from 2009 to 2018. The current spin-down rate measurement implies a surface dipole field of $B_s=2\times10^{13}$\,G, characteristic age of $\tau_c=3.6$\,Myr, and a spin-down luminosity of $\dot{E}=2\times10^{30}$\,erg\,s$^{-1}$, in line with the rest of the XINS sample.  Folding the combined event list from all observatories results in an H-test pulse detection significance of 27.01$\sigma$ in the 0.15--1.18\,keV band.

\section{Pulse Profile Analysis}
\label{sec:profiles}
With the updated timing solutions in hand, we can now construct high signal-to-noise pulse profiles and examine their morphology and energy dependence in greater detail than previously possible. Figure~\ref{fig:profiles} shows a compilation of the normalized (counts per bin divided by the mean across all bins) broadband pulse profiles obtained by folding and stacking the event data from all telescopes considered in the timing analysis.  We note that this stacking results in an average of the instrument response curves as a function of energy of the different telescopes, weighted based on the relative effective areas and exposure times. 
The energy range used for each XINS is the interval that was found to produce the maximum pulse detection significance:  0.16--1.15, 0.15--1.21, 0.27--0.89, 0.30--1.21, 0.15--0.74, and 0.15--1.18\,keV, for the six XINS in alphanumerical order, respectively.

For some of the XINS,  the high fidelity of the updated  profiles reveal interesting morphology. RX J0720.4$-$3125, RX J0806.4$-$4123, and RX J1856.5$-$3754 show a single, albeit clearly asymmetric, peak per rotation period. The pulse from RX J1856.5$-$3754 is characterized by a shallow leading edge and a steeper trailing edge.  RX J0720.4$-$3125 has a clearly asymmetric ``sawtooth'' pulse, with a rapid rise and a more gradual decline, while 
RX J0806.4$-$4123 exhibits an even more exaggerated triangular wave morphology, with a slower rise and a very fast decline of the pulse. RX J1308.6+2127 has two clearly separated peaks per rotation period, with comparable amplitude but with one noticably broader than the other.
The current data set folded at the improved timing solution for J2143.0$+$0654 shows the most complex broad band profile among the XINS, with a main broad peak flanked by two weaker and narrower pulses.  
In the case of RX J0420.0$-$5022, due to the paucity of pulsed photons the detailed structure of the pulse is not clearly evident. It appears to show a single pulse per rotation, but limited information can be extracted from the present data set.

\subsection{Harmonic Decomposition}
\label{sec:harmonics}
The complexity of the pulse shapes can be judged more quantitatively using the harmonic content of the periodic signal in the Fourier domain. In particular, the presence of significant harmonics in the power spectrum beyond the fundamental frequency provides an indication of how strongly the pulse shape deviates from a pure sinusoid. The total profiles for the six XINS with their Fourier components superposed are shown in Figure~\ref{fig:harms}. Due to the limited data set, the profile RX J0420.0$-$5002 is adequately described by a single broad sine wave. RX J0720.4$-$3125 and RX J1856.5$-$3754 have a significant second harmonic (i.e., first overtone), which arise due to the pronounced skewness of the single pulse. RX J0806.4$-$4123 requires the first four Fourier components to fully describe the highly asymmetric pulse shape. In RX J1308.6+2127, the second harmonic  is predictably more prominent than the fundamental, as expected from a signal with two pulses per cycle, but additional higher harmonics are also present due to the differing widths of the two pulses. Despite the limited photon statistics for RX J2143.0+0654, its power spectrum is the richest in harmonic content due to the multiple features of the pulse profile.

\subsection{Energy Dependence}
In Figure~\ref{fig:profiles_multi} we show the profiles for the six XINS but split into different energy ranges. For RX J0720.4$-$3125 and RX J1308.6+2127, the data are divided into three, while for the rest into two adjoining bands. The energy ranges were selected through a visual inspection of the folded data in a number of trial ranges; the bands selected were chosen because they exhibited the largest discrepancies between the profiles in the different energy ranges. The Chandra HRC events are excluded from these energy-resolved profiles as they provide no reliable spectral information.  The wealth of pulsed source counts collected for RX J0720.4$-$3125, RX J1308.6+2127, and RX J1856.5$-$3754 allows us to examine the energy dependence of the pulsed emission in even greater detail. For this purpose, in Figure~\ref{fig:xins_2d} we present two-dimensional histograms of pulsed counts as a function of both pulse phase and photon energy for these three XINS. To obtain the pulsed counts in every energy band considered we subtracted the DC (i.e.,~unpulsed) component, with the level of this DC component taken from the pulse phase bin with the minimum number of counts in each energy interval considered.

One of the outliers again is RX J0720.4$-$3125, which appears to undergo a dramatic transformation of the profile around 0.6\,keV, with a narrower and offset pulse at higher energies (see Figure~\ref{fig:xins_2d}). This behavior is immediately apparent in Figure~\ref{fig:profiles_multi}, where the X-ray profile is shown in three energy bands, 0.15--0.3, 0.3--0.7, and 0.7--1.2 keV. In \citet{2015ApJ...807L..20B}, this peculiarity was interpreted as being due to an absorption feature based on a more limited data set (with only a single XMM-Newton observation used to draw that conclusion). An alternative interpretation for this feature is the presence of two distinct and widely separated X-ray emitting regions on the stellar surface with grossly different areas and temperatures, analogous to those of the Puppis A central compact object RX J0822$-$4300  (see \citealt{2009ApJ...695L..35G}) and the so-called ``Three Musketeers'' middle-aged rotation-powered pulsars \citep[e.g.,][]{2005ApJ...623.1051D}.

Similar behavior is observed in RX J1308.6+2127, where the relative strengths of the two distinct pulses, which have comparable amplitudes across the entire soft band, change substantially above $\approx$0.8\,keV. The broader pulse, in particular, also shifts ahead by $\approx$0.05 rotational cycles at higher energies, while the narrower pulse, which is slightly stronger in the softest band (0.15--0.5\,keV), greatly diminishes in amplitude above $\approx$0.8\,keV.  

The other XINS show more subtle but no less peculiar changes in their pulsed emission as a function of energy. For RX J0806.4$-$4123, J1856.5$-$3754, and perhaps RX J0420.0$-$5022, there is compelling (albeit presently marginal) evidence for a phase lag of the emission at higher energies as well as a possible narrowing of the pulse. In RX J2143.0+0654, there are hints that the relative strengths of the three pulses may change with energy and undergo phase shifts as a function of energy. However, in all these instances, the photon statistics of the pulsed emission are not sufficiently high for definitive conclusions to be drawn.

\subsection{Hard Pulsed Emission}
There have been recent reports of evidence for faint non-thermal components at higher energies in some XINS \citep{2020ApJ...904...42D,2022MNRAS.516.4932D}. Hard X-ray radiation from neutron stars can be produced by a variety of mechanisms. Extended or compact, unresolved non-thermal radiation can be produced by a wind nebula or a bow shock driven by a particle outflow from the neutron star as it encounters the ambient medium \citep{2005AdSpR..35.1116G,2015SSRv..191..391K}. A hard tail can also be caused by inverse Compton scattering of the surface thermal radiation by a population of intervening energetic particles; for low optical depth this results in a power-law tail \citep{1986PASJ...38..819N}. Finally, in rotation-powered neutron stars (especially young, energetic ones), pulsed emission with a power-law spectrum can arise from particle acceleration in the pulsar magnetosphere. The possible presence of a hard tail in the spectra of XINS has led to speculations that this high-energy excess could be due to exotic processes such as emission of axions from the core of the neutron star \citep{2021PhRvL.126b1102B}.

To test for one of these possibilities, namely pulsed non-thermal emission, we consider only events in the range 1.5--5\,keV, where the contribution of the soft thermal radiation is expected to be negligible and the non-source background to not be dominant. We only use XMM-Newton data since the NICER XTI background at higher energies is fairly high due to the lack of imaging capabilities and the Chandra exposures are substantially less sensitive.  We find that in the 1.5--5\,keV band none of the six XINS exhibit statistically significant pulsations, with H-test significances of 1.15$\sigma$,
2.47$\sigma$,
2.25$\sigma$,
2.49$\sigma$,
0.51$\sigma$, and
1.54$\sigma$, 
for RX J0420.0$-$5022, RX J0720.4$-$3125, RX J0806.4$-$4123, RX J1308.6$+$2127, RX J1856$-$3752, and RX J2143.0+0654, respectively. These non-detections translate into 3$\sigma$ upper limits on the pulsed fractions of the hard emission above 1.5\,keV of  6.1\%, 2.1\%, 4.5\%, 6.0\%, 1.5\%, and 5.9\%, respectively.

We defer an in-depth spectroscopic investigation of the apparent excess at high energies for all XINS to a subsequent paper.

\section{Discussion}
\label{sec:discussion}

\begin{figure}
    \centering
    \includegraphics[trim={0 1.5cm 0 0},clip,width=0.45\textwidth]{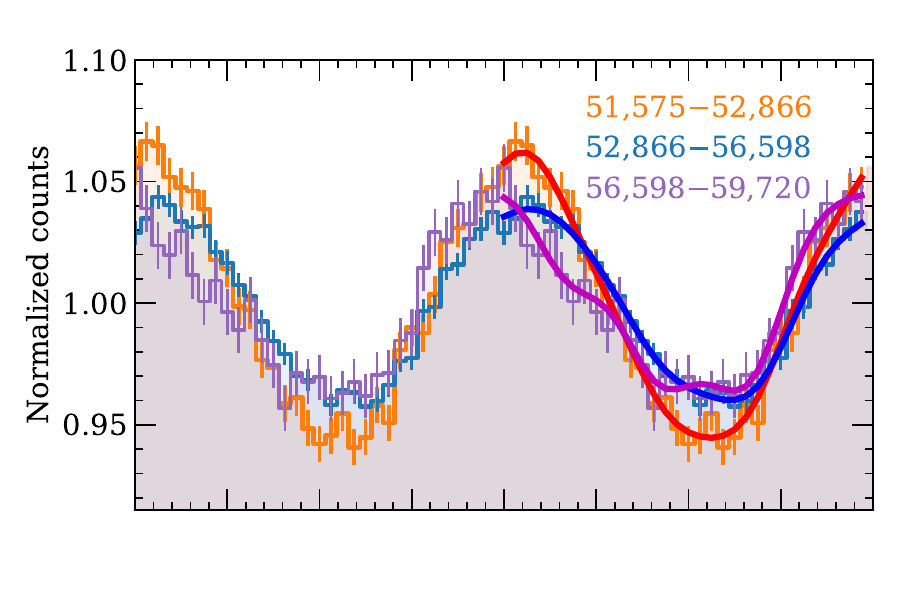}
    \includegraphics[trim={0 0.35cm 0 0.9cm},clip,width=0.45\textwidth]{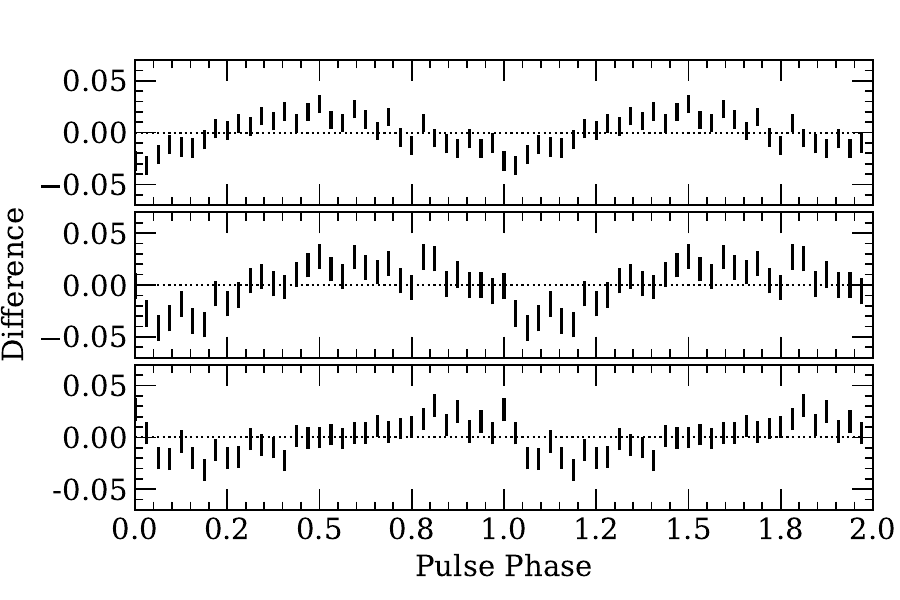}
    \caption{Top panel: XMM-Newton EPIC pn+MOS1/2 pulse profiles of RX J0720.4$-$3125 from three MJD ranges: one including all data prior to the glitch (orange) and the other two using events from the first (blue) and second (purple) halves of the post-glitch data set. The solid red, blue, and purple lines show the model based on the empirical Fourier coefficients of the phase folded event data. Bottom three panels: Differences in normalized counts between the profiles from the second and first (upper panel), third and first (middle panel), and third and second (lower panel) MJD intervals from the top panel.}
    \label{fig:j0720_variability}
\end{figure}

The pulsed thermal radiation observed from the XINS presumably arises due to changing aspect of hot surface emitting regions as the star rotates. The peculiar features of the observed XINS profiles, their Fourier domain properties, and their variation with photon energy are not easily reproducible by one or more hot spots on the surface that radiate a simple blackbody spectrum with equal intensity in all directions. Therefore, any deviations of the profile from a sinusoidal shape and large changes as a function of energy would arise due to a combination of: i) complicated hot spot arrangements; ii) non-isotropic surface radiation, such as that produced by an atmospheric layer; and/or iii) phase and energy dependent attenuation, by processes such as cyclotron absorption \citep[e.g.,][]{2012ApJ...751...15S}. We note that sharpened and/or distorted pulse shapes could also be produced by gravitational light bending for ultra-compact neutron stars ($R_{\rm NS} \lesssim8$ km), although such small stellar radii are not favored by current constraints \citep{PhysRevLett.121.161101,2021ApJ...918L..28M,2021ApJ...918L..27R}. Moreover, for the comparatively slow spin periods under consideration here ($3-11$\,s), asymmetries in the profile introduced by Relativistic Doppler boosting and aberration of the surface emission induced by the rotation of the star are insignificant \citep[see, e.g.,][]{2014ApJ...791...78A}. 

The XINS dipole magnetic field strengths deduced from the spin-down rate are $\sim$$10^{13}$\,G, which falls in a regime where the emergent radiation from a surface atmospheric layer is expected to be strongly affected by the field \citep{2014PhyU...57..735P}. This manifests in a  highly anisotropic radiation pattern, with a ``pencil'' beam in the direction of the magnetic field and a fan beam that peaks at intermediate angles \citep{1992herm.book.....M,1994A&A...289..837P}. The detailed beaming pattern has a strong energy dependence and is determined in part by the magnetic field strength and orientation, as well as on the chemical composition of the topmost emitting layer of the atmosphere \citep{2007MNRAS.377..905M,2008ApJS..178..102H}. The observed properties of the pulsed emission from the six XINS can, in principle, be explained by emission from such a magnetized surface layer \citep[see, e.g.,][]{2007MNRAS.375..821H,2007MNRAS.380...71H}. RX J2143.0+0654, for instance, has a broad main peak flanked by two weaker and much narrower peaks, which is an unusual feature for thermal radiation, but could in principle be a consequence of such a pencil+fan angular dependence of the surface radiation. However, a formal analysis using a realistic neutron star model is required to verify this assertion. Such an investigation will be presented in a subsequent paper in this series.

\subsection{The Curious Case of RX J0720.4$-$3125}
\label{sec:j0720_spin}
The presence of a significant second spin derivative in RX J0720.4$-$3125 allows us to compute a pulsar braking index as $n=\nu\ddot{\nu}/\dot{\nu}^2$.  In the idealized model of a rotation-powered neutron star as a magnetic dipole spinning down in a vacuum, pulsar electrodynamics predicts a value of $n=3$ \citep{1969Natur.221..454G}. More realistic treatments of pulsar magnetospheres that incorporate wind outflows, magnetic field obliquity, and field evolution can produce values less than 3 \citep[e.g.,][]{1988MNRAS.234P..57B,1997MNRAS.288.1049M,1999ApJ...525L.125H,2015MNRAS.452..845H}. 
However, the braking index for RX J0720.4$-$3125 based on the inferred $\nu$, $\dot{\nu}$, and $\ddot{\nu}$ is an unphysical value of $n=682$. This implies that the measured $\ddot{\nu}$ is not due to magnetic braking and could instead be due to timing noise \citep[see, e.g.,][and references therein]{2020MNRAS.494.2012P}. An additional indicator of timing noise evident in Figure~\ref{fig:j0720_residuals} is the significant scatter of the TOAs around the best fit model even after the inclusion of $\ddot{\nu}$.

While previous investigations have found that most of the XINS show no evidence of long-term variability down to levels of a few percent (e.g., \citealt{2009ApJ...705..798K}; \citealt{2012A&A...541A..66S}; \citealt{2022MNRAS.516.4932D}), RX J0720.4$-$3125 has been reported to undergo spectral and profile changes on timescales of years  \citep{2004A&A...415L..31D, 2012MNRAS.423.1194H,2015ApJ...807L..20B}. Figure~\ref{fig:j0720_variability} shows the XMM-Newton profile in the 0.15--1.2 keV range split into three MJD ranges: the first covering the interval prior to the assumed glitch epoch (MJD 51,575), the second after the glitch and up to the midpoint of the post glitch data span (MJD 56,598), and the third from this midpoint onward. It is apparent that the three profiles are not identical, with systematic differences in pulse amplitude, shape, and peak phase (bottom three panels Fig.~\ref{fig:j0720_variability}), which cannot be explained merely by statistical fluctuations. The difference in peak phase is consistent with the level of scatter observed in the timing residuals shown in Figure~\ref{fig:j0720_residuals}. Thus, the likely explanation for the apparent $\ddot{\nu}$ and the scatter in the timing residuals is this gradual change in the pulsed emission over time, though the underlying process responsible for this atypical behavior remains to be understood.  A thorough analysis of the spectral and profile variations in RX J0720.4$-$3125 that may shed more light onto this mystery is deferred to a subsequent work.

Based on an analysis of the set of XMM-Newton EPIC pn Full Frame mode observations, \citet{2017A&A...601A.108H} proposed that the true spin period of RX J0720.4$-$3125 may actually be twice that of the commonly reported value. To verify this claim, we examined the power spectra of the deepest (65.1 ks) XMM-Newton observation from 16 October 2019 (ObsID 0852980201, obtained in Small Window mode) and the collection of NICER XTI exposures. We find no evidence for significant signal power around 0.0596 Hz, even when dividing the event data in the energy bands considered in \citet{2017A&A...601A.108H} (see in particular their Fig.~5). We also carried out a timing analysis of the NICER XTI data alone restricted to events above 0.7 keV, where there is an apparent difference between the two pulses per rotation for the presumed 0.0596\,Hz spin frequency. The NICER XTI data was chosen as it provides a uniform, high quality data set far away from the glitch and its span is such that a $\ddot{\nu}$ is not important. The highest likelihood solution found ($\nu=0.0596$ Hz, $\dot{\nu}=-5.78\times10^{-16}$ Hz\,s$^{-1}$) results in an H-test significance that is lower than the best solution for the nominal spin at 0.119 Hz, implying that there is no significant contribution to the signal power from the pulsar at a frequency of 0.0596\,Hz. This is confirmed by harmonic decomposition of the folded data, which reveals that the Fourier component of the putative fundamental frequency 0.0596\,Hz is not required to accurately reproduce the observed pulses.  Moreover, at twice the nominal spin period, the two pulses per rotation are statistically indistinguishable and separated by exactly 0.5 in spin phase in all energy ranges considered. For reference, in the case of RX J1308.6+2127 (see Section~\ref{sec:J1308}), the two pulses are not half a rotational cycle apart and their peak-to-peak separation changes depending on the choice of energy range (see Fig.~\ref{fig:profiles_multi}).
Based on these findings, we conclude that the commonly reported spin of $P=8.39$\,s is in fact the true rotation period of the RX J0720.4$-$3125 neutron star.

\section{Conclusions}
\label{sec:conclusion}
We have presented a comprehensive timing reanalysis of all available X-ray data of the six nearby XINS with confirmed pulsations.  In specific cases, the resulting timing solutions obtained provide orders of magnitude improvements in precision of the spin frequency and spin-down rate measurements compared to previous published results, while also extending the baseline to span over two decades.  For RX J2143.0+0654, which previously had only a marginal measurement on its spin down, we obtained the first firm measurements of $\dot{\nu}$. Five of the XINS are well behaved in the sense that their long-term timing behavior is consistent with steady spin-down, with no clear evidence for measurable timing noise, higher order spin frequency derivatives, or glitches. The exception is RX J0720.4$-$3125, which is already known to have undergone a glitch around MJD 52,886, and exhibits broad residuals that can be accommodated by a second spin derivative ($\ddot{\nu}$). The erratic spin behavior is possibly related to the previously identified long-term variations in the X-ray spectrum and pulse profile of this source, though the underlying physics responsible for these irregularities remains unknown.

The final timing solutions resulting from the analysis presented herein are included as supplementary products to this paper in the form of Tempo-style pulsar parameter (.par) files. The updated spin ephemerides allowed us to produce folded X-ray pulse profiles that reveal previously unseen details. The pulsed emission from the XINS shows surprising complexity, with pronounced asymmetries in the otherwise broad pulses expected from thermally emitting neutron stars.  Moreover, the profiles show peculiar energy dependence, with significant shape changes, phase shifts in the pulse peak and, in some instances, enhancements in pulsed fraction at higher energies.  These features are difficult to reconcile with a simple model of an isotropic blackbody-like radiation from small hot spots on the stellar surface. Thus, it may be necessary to invoke more sophisticated models that include the detailed beaming patterns of magnetic atmospheres and more complicated multi-temperature hot regions on the neutron star surface. 

In a follow up paper, we will present a detailed investigation of the spectra (both phase averaged and phase resolved) and their variation with time to gain additional insight into the surface temperature distribution, chemical composition, and magnetization of the stellar surface. Ultimately, we intend to employ detailed modeling methods of the extensive X-ray data sets to fully map the surface temperature distribution on the stellar surface for each XINS. 

In the meantime, continued monitoring of the XINS sample is needed to maintain phase coherence of these timing solutions and to monitor for glitches and other timing anomalies. For RX J0420.0$-$5022, RX J0806.4$-$4123, and RX J2143.0+0654, in particular, additional data is necessary to improve the photon statistics of their  pulse profiles, which would lead to tighter constraints on their surface emission properties. When coupled with the prospect of additional XINS discoveries with eROSITA \citep[see, e.g,][]{2023A&A...674A.155K,2024arXiv240117290K}, this would provide new opportunities to glean insight into the properties of these puzzling objects and lead to an improved understanding of the Galactic neutron star population and uncertain aspects of neutron star evolution.\\

We thank Deven Bhakta for implementing parallel processing in the PINT \texttt{event\_optimize} code, and Paul Ray and Scott Ransom for helpful discussions. This work was funded in part by NASA Astrophysics Data Analysis Program (ADAP) grant No.~80NSSC20K0884.  We acknowledge use of computing resources from Columbia University's Shared Research Computing Facility project, which is supported by NIH Research Facility Improvement Grant 1G20RR030893-01, and associated funds from the New York State Empire State Development, Division of Science Technology and Innovation (NYSTAR) Contract C090171, both awarded April 15, 2010. The work presented is based in part on observations with \textit{XMM-Newton}, an ESA Science Mission with instruments and contributions directly funded by ESA Member states and NASA, and on observations from the NICER mission, which is funded by NASA. The XMM-Newton Science Analysis Software (SAS) is developed and maintained by the Science Operations Centre at the European Space Astronomy Centre and the Survey Science Centre at the University of Leicester. This research has made use of data obtained from the Chandra Data Archive, contained in the Chandra Data Collection (CDC) 224\dataset[doi:10.25574/cdc.224]{https://doi.org/10.25574/cdc.224}, and software provided by the Chandra X-ray Center (CXC) in the application package CIAO, as well as data products and software provided by the High Energy Astrophysics Science Archive Research Center (HEASARC), which is a service of the Astrophysics Science Division at NASA/GSFC and the High Energy Astrophysics Division of the Smithsonian Astrophysical Observatory.  We acknowledge extensive use of NASA's Astrophysics Data System (ADS) bibliographic services and the ArXiv. 

\facilities{CXO, NICER, XMM}
\software{Astropy \citep{2022ApJ...935..167A}, CIAO \citep{2006SPIE.6270E..1VF}, emcee \citep{2013PASP..125..306F}, HEASoft \citep{2014ascl.soft08004N}, Matplotlib \citep{Hunter:2007}, PINT \citep{2021ApJ...911...45L}, SAS \citep{2004ASPC..314..759G}, Tempo2 \citep{2006MNRAS.369..655H}}

\appendix
\section{Logs of X-Ray Observations}
In Tables~\ref{table:j0420_obs}-\ref{table:j2143_obs} we list the XMM-Newton, Chandra, and NICER observations used for the timing analysis presented in Section~\ref{sec:timing} for each target. Each table provides the numeric observation identifier (ObsID), the calendar date and UTC start time, the instrument, imaging mode/filter used (where applicable), and the net exposure time after applying the data filtering procedures described in Section~\ref{sec:data}. For a number of XMM-Newton observations, one or more of the detectors have no scientifically useful data due to the detector not being active or the filter wheel being closed, and are thus not listed.

\section{Timing Solution Posterior Distributions}
\begin{figure*}[t!]
    \centering
    \includegraphics[trim={0.45cm 0.6cm 0.5cm 0.5cm},clip,width=0.41\textwidth]{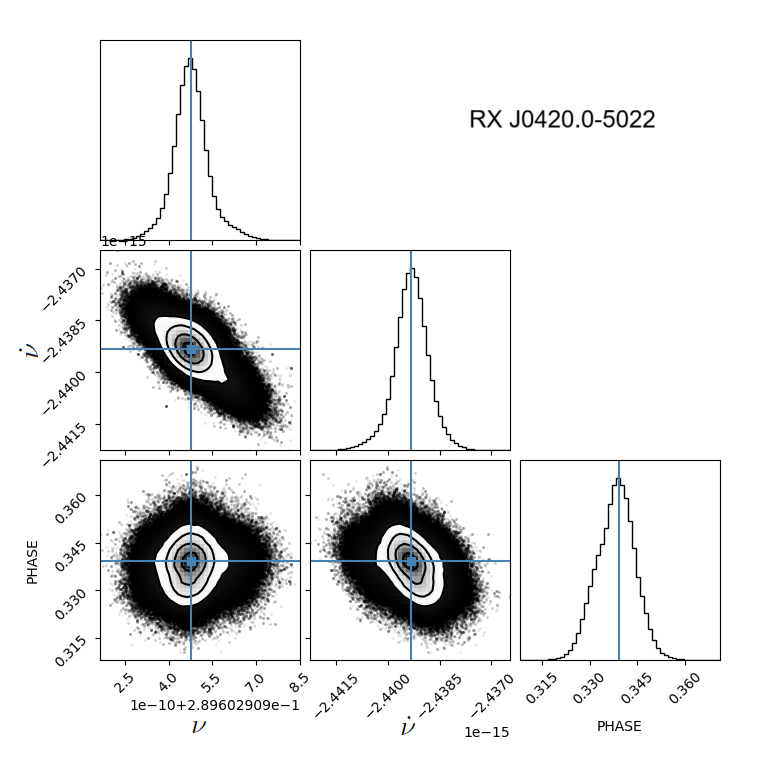}
    \includegraphics[trim={0.45cm 0.6cm 0.5cm 0.5cm},clip,width=0.41\textwidth]{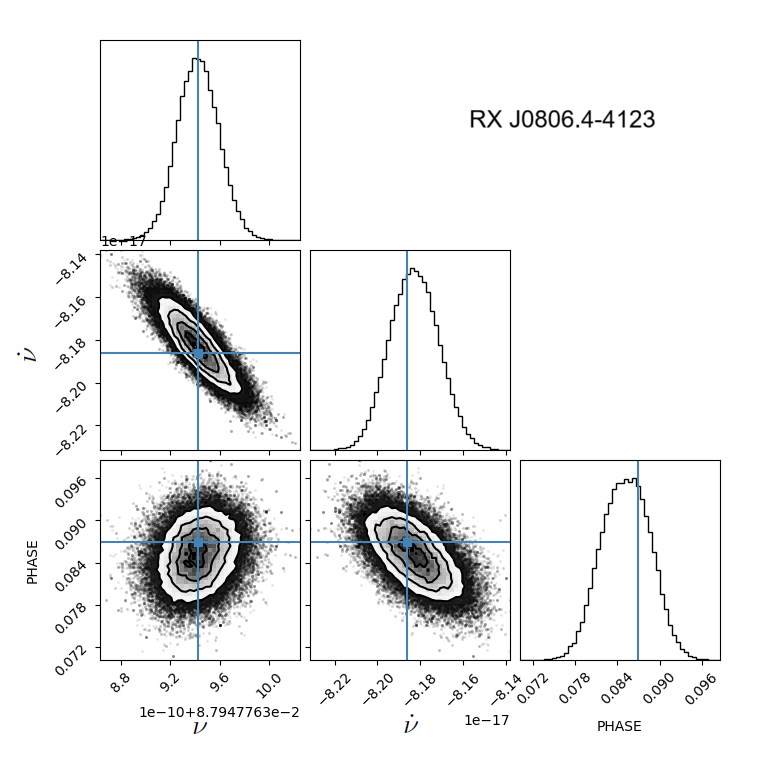}
    \includegraphics[trim={0.45cm 0.5cm 0.5cm 0.5cm},clip,width=0.41\textwidth]{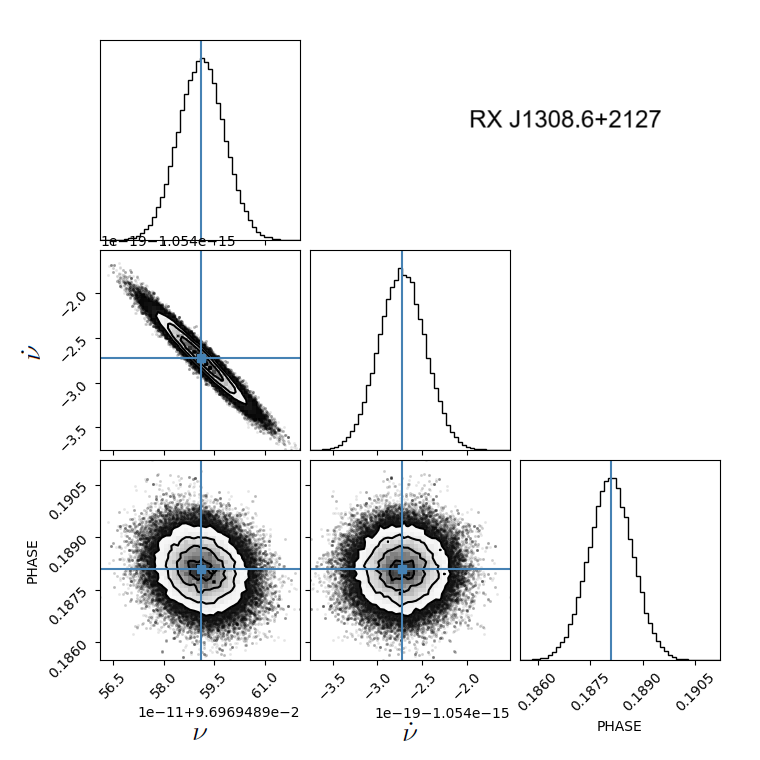}
    \includegraphics[trim={0.45cm 0.6cm 0.5cm 0.5cm},clip,width=0.41\textwidth]{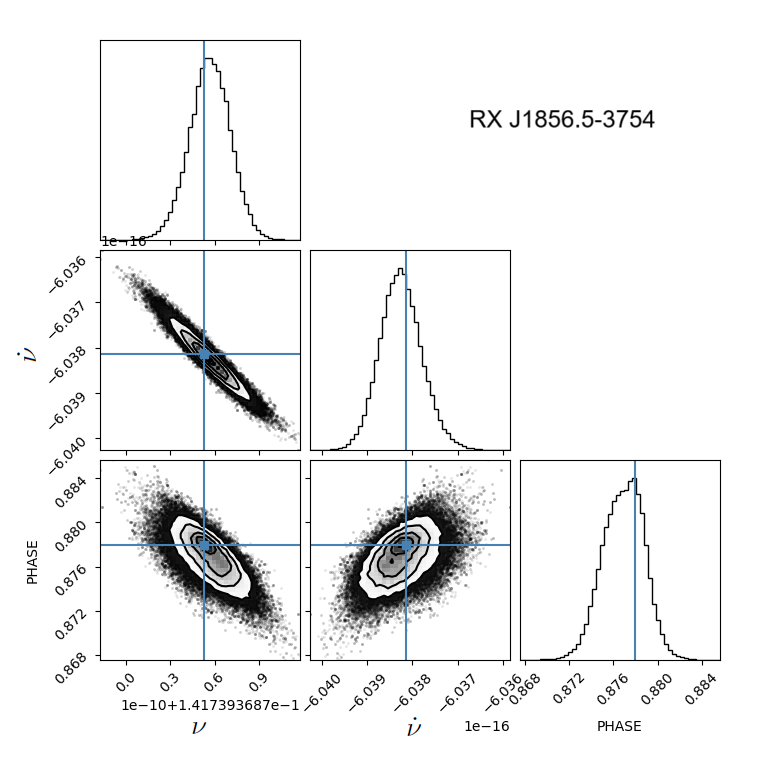}    
    \includegraphics[trim={0.45cm 0.6cm 0.5cm 0.5cm},clip,width=0.41\textwidth]{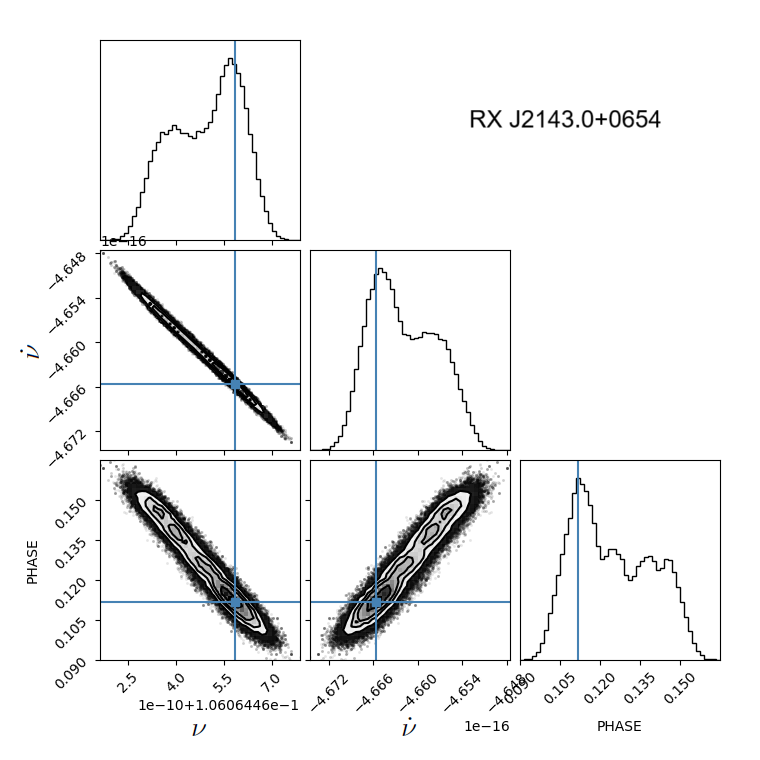}
    \caption{Frequency ($\nu$), frequency derivative ($\dot{\nu}$), and pulse phase posterior distributions inferred for RX J0420.0$-$5022 (top left), RX J0806.4$-$4123 (top right), RX J1308.6+2127 (middle left), RX J1856.5$-$3754 (middle right), and RX J2143.0+0654 (bottom) from the event-based likelihood analysis described in Section~\ref{sec:timing}. The contours correspond to the 0.5, 1, 1.5, and 2$\sigma$ credible interval levels. The blue square marks the sample with the maximum likelihood that was used to fold the event data and produce the pulse profiles shown throughout the paper. The median value and credible intervals are reported in Table~\ref{tab:timing}.}
    \label{fig:triangle}
\end{figure*}

\begin{figure*}
    \centering
    \includegraphics[trim={0.45cm 0.1cm 0.4cm 0.2cm},clip,width=0.999\textwidth]{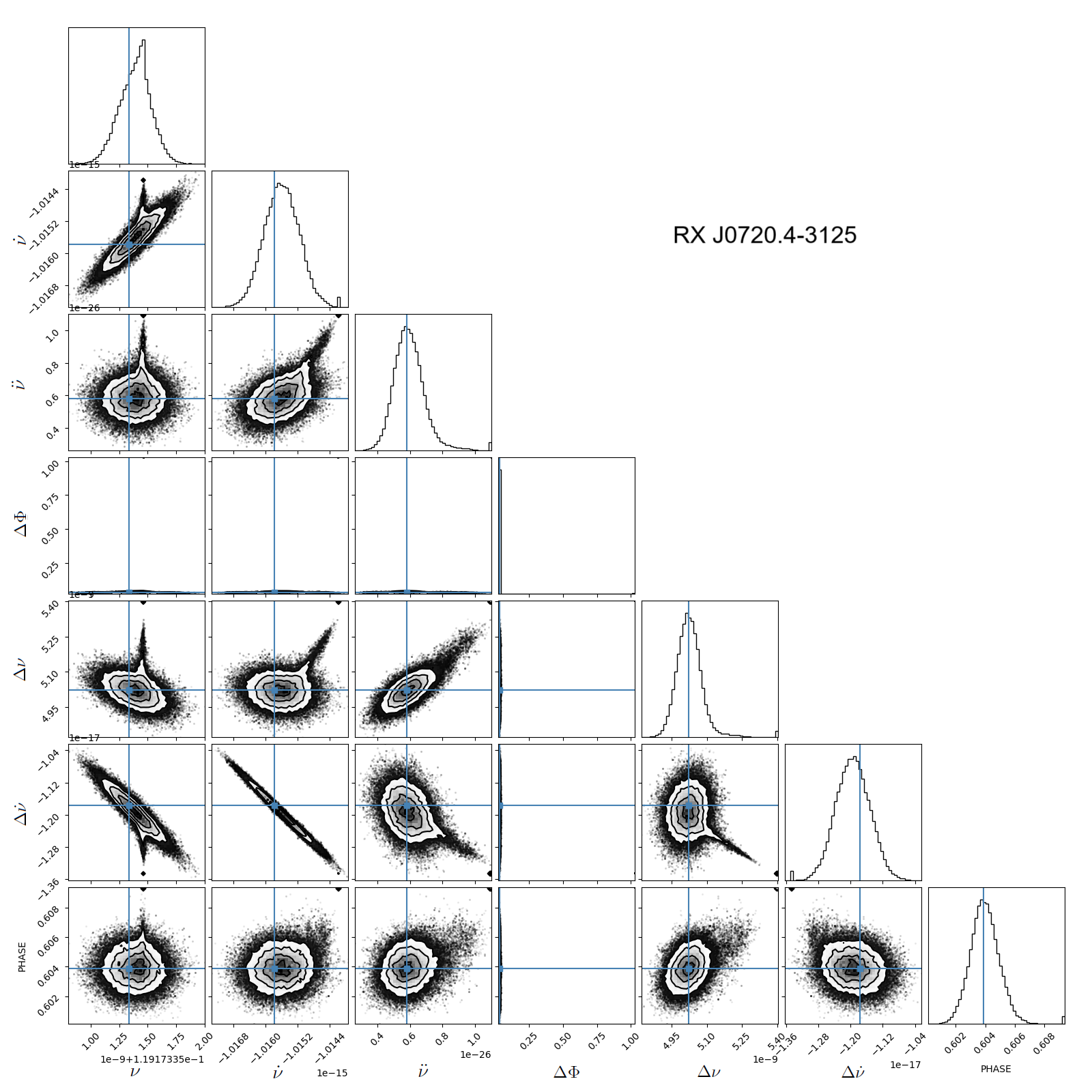}
\caption{Posterior distributions inferred for RX J0720.4$-$3125 for frequency ($\nu$), frequency first ($\dot{\nu}$) and second derivative ($\ddot{\nu}$), the glitch spin phase increment $\Delta\Phi$, the glitch permanent frequency increment $\Delta\nu$, the glitch permanent frequency first derivative increment $\Delta\dot{\nu}$, and the arbitrary pulse phase offset from the event-based likelihood analysis described in Section~\ref{sec:timing}. The blue square marks the sample with the maximum likelihood that was used to fold the event data and produce the pulse profiles shown in Figure~\ref{fig:profiles}. The contours correspond to the 0.5, 1, 1.5, and 2$\sigma$ credible interval levels The median value and credible intervals are reported in Table~\ref{tab:j0720_timing}.}
 \label{fig:j0720_triangle}
\end{figure*}

In Figure~\ref{fig:triangle} and \ref{fig:j0720_triangle} we present the one and two-dimensional posterior distributions of $\nu$ and $\dot{\nu}$ from the timing inference analyses in the form of corner plots. For RX J0720.4$-$3125, the corner plot also shows $\ddot{\nu}$ and the glitch parameters listed in Table~\ref{tab:j0720_timing}. In each figure, the square marks the highest likelihood solution found. In the timing solution parameter (.par) files provided as supplementary material to this paper, the values for the fitted parameters correspond to these highest likelihood points.


\clearpage
\startlongtable
\begin{deluxetable}{lcclR|lcclR}
\tabletypesize{\tiny}
\tablewidth{0pt}
\tablecolumns{10}
\tablecaption{Log of X-ray observations of RX\,J0420.0$-$5022 \label{table:j0420_obs}}
\tablehead{\colhead{ObsID} & \colhead{Date/Start Time} & \colhead{Instrument} & \colhead{Mode/Filter} &  \colhead{$t_{\rm exp}$} &  \colhead{ObsID} & \colhead{Date/Start Time} & \colhead{Instrument} & \colhead{Mode/Filter} &  \colhead{$t_{\rm exp}$}  \\
            \colhead{}     & \colhead{(UTC)} & \colhead{}           &    \colhead{}         & \colhead{(ks)}    & \colhead{}     & \colhead{(UTC)} & \colhead{}           &    \colhead{}         & \colhead{(ks)}   } 
            \startdata
    \multicolumn{10}{c}{\textit{XMM-Newton}\tablenotemark{a}} \\
\hline
0141750101	&	2002-12-30	04:01:38	&	pn	&	FW	/	Thin	&	18.0	&	0651470901	&	2010-10-03	19:23:39	&	pn	&	FW	/	Thin	&	9.1	\\
0141751001	&	2002-12-31	22:17:39	&	pn	&	FW	/	Thin	&	15.1	&		&	2010-10-03	19:18:30	&	MOS1	&	SW	/	Thin	&	12.7	\\
0141751101	&	2003-01-19	17:05:12	&	pn	&	FW	/	Thin	&	18.2	&		&	2010-10-03	19:18:31	&	MOS2	&	SW	/	Thin	&	12.7	\\
0141751201	&	2003-07-25	21:44:58	&	pn	&	FW	/	Thin	&	18	&	0651471001	&	2010-10-04	05:18:10	&	pn	&	FW	/	Thin	&	10.9	\\
0651470201	&	2010-03-30	12:01:11	&	pn	&	FW	/	Thin	&	4.6	&		&	2010-10-04	05:13:02	&	MOS1	&	SW	/	Thin	&	11.5	\\
	&	2010-03-30	11:56:01	&	MOS1	&	SW	/	Thin	&	4.6	&		&	2010-10-04	05:13:03	&	MOS2	&	SW	/	Thin	&	12.5	\\
	&	2010-03-30	11:56:01	&	MOS2	&	SW	/	Thin	&	4.5	&	0651471101	&	2010-10-06	23:03:10	&	pn	&	FW	/	Thin	&	7.3	\\
0651470301	&	2010-04-04	19:02:31	&	pn	&	FW	/	Thin	&	5.7	&		&	2010-10-06	22:58:00	&	MOS1	&	SW	/	Thin	&	9.1	\\
	&	2010-04-04	18:57:20	&	MOS1	&	SW	/	Thin	&	6.3	&		&	2010-10-06	22:58:00	&	MOS2	&	SW	/	Thin	&	9.5	\\
	&	2010-04-04	18:57:23	&	MOS2	&	SW	/	Thin	&	6.3	&	0651471201	&	2010-11-26	09:34:53	&	pn	&	FW	/	Thin	&	3.8	\\
0651470401	&	2010-04-09	08:40:40	&	pn	&	FW	/	Thin	&	5.4	&		&	2010-11-26	09:29:44	&	MOS1	&	SW	/	Thin	&	5.4	\\
	&	2010-04-09	08:35:30	&	MOS1	&	SW	/	Thin	&	7.2	&		&	2010-11-26	09:29:42	&	MOS2	&	SW	/	Thin	&	5.5	\\
	&	2010-04-09	08:35:32	&	MOS2	&	SW	/	Thin	&	7.3	&	0651471301	&	2011-01-13	22:29:23	&	pn	&	FW	/	Thin	&	7.1	\\
0651470501	&	2010-05-21	05:56:19	&	pn	&	FW	/	Thin	&	3.8	&		&	2011-01-13	22:24:16	&	MOS1	&	SW	/	Thin	&	4.5	\\
	&	2010-05-21	05:51:10	&	MOS1	&	SW	/	Thin	&	5.2	&		&	2011-01-13	22:24:14	&	MOS2	&	SW	/	Thin	&	4.9	\\
	&	2010-05-21	05:51:10	&	MOS2	&	SW	/	Thin	&	5.2	&	0651471401	&	2011-03-31	21:19:15	&	pn	&	FW	/	Thin	&	4.9	\\
0651470601	&	2010-07-29	14:26:48	&	pn	&	FW	/	Thin	&	4.5	&		&	2011-03-31	20:16:35	&	MOS1	&	SW	/	Thin	&	10.3	\\
	&	2010-07-29	14:21:38	&	MOS1	&	SW	/	Thin	&	6.4	&		&	2011-03-31	20:16:33	&	MOS1	&	SW	/	Thin	&	10.3	\\
	&	2010-07-29	14:21:38	&	MOS2	&	SW	/	Thin	&	6.4	&	0651471501	&	2011-04-11	07:19:25	&	pn	&	FW	/	Thin	&	3.8	\\
0651470701	&	2010-09-21	08:46:35	&	pn	&	FW	/	Thin	&	6.9	&		&	2011-04-11	07:14:17	&	MOS1	&	SW	/	Thin	&	5.4	\\
	&	2010-09-21	08:41:27	&	MOS1	&	SW	/	Thin	&	9.7	&		&	2011-04-11	07:14:14	&	MOS2	&	SW	/	Thin	&	5.5	\\
	&	2010-09-21	08:41:27	&	MOS2	&	SW	/	Thin	&	9.7	&	0844140401	&	2019-05-22	08:19:03	&	pn	&	FW	/	Thin	&	16.7	\\
0651470801	&	2010-10-02	23:11:59	&	pn	&	FW	/	Thin	&	8.2	&		&			&		&				&		\\
	&	2010-10-02	23:06:50	&	MOS1	&	SW	/	Thin	&	11.5	&		&			&		&				&		\\
	&	2010-10-02	23:06:52	&	MOS2	&	SW	/	Thin	&	11.5	&		&			&		&				&		\\
\enddata
\tablenotetext{a}{Listed exposure times are values after filtering for instances of strong background flaring. EPIC MOS1/2 observations acquired in Full Window mode are not used due to the coarse 2.6\,s time resolution. The imaging mode abbreviations correspond to Full Window (FW) and Small Window (SW).}
\end{deluxetable}

\vspace{-0.3cm}

\startlongtable
\begin{deluxetable}{lcclR|lcclR}
\tabletypesize{\tiny}
\tablewidth{0pt}
\tablecolumns{10}
\tablecaption{Log of X-ray observations of RX\,J0720.4$-$3125 \label{table:j0720_obs}}
\tablehead{\colhead{ObsID} & \colhead{Date/Start Time} & \colhead{Instrument} & \colhead{Mode/Filter} &  \colhead{$t_{\rm exp}$} &  \colhead{ObsID} & \colhead{Date/Start Time} & \colhead{Instrument} & \colhead{Mode/Filter} &  \colhead{$t_{\rm exp}$}  \\
            \colhead{}     & \colhead{(UTC)} & \colhead{}           &    \colhead{}         & \colhead{(ks)}    & \colhead{}     & \colhead{(UTC)} & \colhead{}           &    \colhead{}         & \colhead{(ks)}   } 
            \startdata
\hline
    \multicolumn{10}{c}{\textit{XMM-Newton}\tablenotemark{a}} \\
\hline
0124100101	&	2000-05-13	02:30:54	&	pn	&	FW	/	Thin	&	32.4	&	0400140301	&	2006-05-22	05:07:43	&	pn	&	FW	/	Thin	&	15.5	\\
	&	2000-05-13	02:43:47	&	MOS1	&	FW	/	Thin	&	42.6	&		&	2006-05-22	04:45:36	&	MOS1	&	SW	/	Thin	&	20.7	\\
	&	2000-05-13	02:43:46	&	MOS2	&	SW	/	Thin	&	43.1	&		&	2006-05-22	04:45:36	&	MOS2	&	SW	/	Thin	&	20.7	\\
0132520301	&	2000-11-21	19:20:53	&	pn	&	FW	/	Medium	&	22.8	&	0400140401	&	2006-11-05	11:42:25	&	pn	&	FW	/	Thin	&	17.5	\\
	&	2000-11-21	21:19:16	&	MOS1	&	LW	/	Medium	&	17.8	&		&	2006-11-05	11:20:18	&	MOS1	&	SW	/	Thin	&	21.0	\\
	&	2000-11-21	21:19:12	&	MOS2	&	LW	/	Medium	&	17.8	&		&	2006-11-05	11:20:17	&	MOS2	&	SW	/	Thin	&	21.0	\\
	&	2000-11-21	19:13:54	&	MOS1	&	SW	/	Open	&	6.6	&	0502710201	&	2007-05-05	17:24:22	&	pn	&	FW	/	Thin	&	7.4	\\
	&	2000-11-21	19:08:01	&	MOS2	&	SW	/	Open	&	7.0	&		&	2007-05-05	17:02:02	&	MOS1	&	FW	/	Thin	&	17.4	\\
0156960201	&	2002-11-06	18:14:39	&	pn	&	FW	/	Thin	&	15.0	&		&	2007-05-05 17:02:02		&	MOS2	&	FW	/	Thin	&	17.9	\\
0156960401	&	2002-11-08	19:47:54	&	pn	&	FW	/	Thin	&	24.1	&	0502710301	&	2007-11-17	05:37:12	&	pn	&	FW	/	Thin	&	19.5	\\
	&	2002-11-08	19:25:45	&	MOS1	&	FW	/	Thin	&	31.4	&		&	2007-11-17	05:15:14	&	MOS1	&	FW	/	Thin	&	24.3	\\
	&	2002-11-08T19:25:45		&	MOS2	&	FW	/	Thin	&	31.4	&		&	2007-11-17T05:15:14		&	MOS2	&	FW	/	Thin	&	24.3	\\
0158360201	&	2003-05-02	13:46:27	&	pn	&	SW	/	Thick	&	51.0	&	0554510101	&	2009-03-21	14:37:30	&	pn	&	FW	/	Thin	&	9.1	\\
	&	2003-05-02	13:53:11	&	MOS1	&	SW	/	Open	&	7.3	&	0601170301	&	2009-09-22	04:50:39	&	pn	&	FW	/	Thin	&	2.0	\\
	&	2003-05-02	13:54:04	&	MOS2	&	SW	/	Open	&	7.3	&		&	2009-09-22	04:28:26	&	MOS1	&	SW	/	Thin	&	13.8	\\
0161960201	&	2003-10-27	17:52:19	&	pn	&	SW	/	Thin	&	12.7	&		&	2009-09-22	04:28:26	&	MOS2	&	SW	/	Thin	&	13.4	\\
	&	2003-10-27	17:48:12	&	MOS1	&	SW	/	Open	&	13.4	&	0650920101	&	2011-04-11	00:22:30	&	pn	&	FW	/	Thin	&	11.5	\\
	&	2003-10-27	17:48:09	&	MOS2	&	SW	/	Open	&	13.4	&		&	2011-04-11	00:00:18	&	MOS1	&	SW	/	Thin	&	19.7	\\
	&	2003-10-27	23:18:32	&	pn	&	SW	/	Medium	&	17.4	&		&	2011-04-11	00:00:18	&	MOS2	&	SW	/	Thin	&	20.3	\\
	&	2003-10-28	01:49:26	&	MOS1	&	LW	/	Thin	&	15.5	&	0670700201	&	2011-05-02	23:48:24	&	pn	&	FW	/	Thin	&	10.8	\\
	&	2003-10-28	01:49:26	&	MOS2	&	LW	/	Thin	&	15.5	&		&	2011-05-02	23:26:13	&	MOS1	&	SW	/	Thin	&	19.7	\\
0164560501	&	2004-05-22	10:38:32	&	pn	&	FW	/	Thin	&	20.8	&		&	2011-05-02	23:26:12	&	MOS2	&	SW	/	Thin	&	20.8	\\
	&	2004-05-22	10:16:1	&	MOS1	&	FW	/	Thin	&	25.5	&	0670700301	&	2011-10-01	04:10:31	&	pn	&	FW	/	Thin	&	22.2	\\
0300520201	&	2005-04-28	09:04:02	&	pn	&	FW	/	Thin	&	25.1	&		&	2011-10-01	03:48:17	&	MOS1	&	SW	/	Thin	&	25.8	\\
	&	2005-04-28	08:41:51	&	MOS1	&	SW	/	Thin	&	37.5	&		&	2011-10-01	03:48:19	&	MOS2	&	SW	/	Thin	&	25.8	\\
	&	2005-04-28	08:41:53	&	MOS2	&	SW	/	Thin	&	37.6	&	0690070201	&	2012-09-18	08:53:40	&	pn	&	FW	/	Thin	&	22.3	\\
0300520301	&	2005-09-23	00:07:26	&	pn	&	FW	/	Thin	&	19.9	&		&	2012-09-18	08:31:37	&	MOS1	&	SW	/	Thin	&	25.8	\\
	&	2005-09-22	23:45:17	&	MOS1	&	SW	/	Thin	&	35.3	&		&	2012-09-18	08:31:39	&	MOS2	&	SW	/	Thin	&	25.8	\\
	&	2005-09-22	23:45:17	&	MOS2	&	SW	/	Thin	&	35.1	&	0852980201	&	2019-10-16	23:09:53	&	pn	&	SW	/	Thin	&	65.1	\\
0311590101	&	2005-11-12	22:49:15	&	pn	&	FW	/	Thin	&	32.3	&		&	2019-10-16	23:01:33	&	MOS1	&	SW	/	Thin	&	88.0	\\
	&	2005-11-12	22:27:05	&	MOS1	&	SW	/	Thin	&	38.3	&		&	2019-10-16	23:01:53	&	MOS2	&	SW	/	Thin	&	87.6	\\
	&	2005-11-12	22:27:06	&	MOS2	&	SW	/	Thin	&	38.3	&		&			&		&				&		\\
\hline
    \multicolumn{10}{c}{\textit{Chandra}\tablenotemark{b}} \\                
\hline
368	&	2000-02-01	06:10:30	&	HRC-S	&	LETG	&	5.4	&	5582	&	2005-06-01	12:30:34	&	HRC-S	&	LETG	&	69.7	\\
369	&	2000-02-04	17:16:35	&	HRC-S	&	LETG	&	6.1	&	5584	&	2005-12-17	19:50:45	&	HRC-S	&	LETG	&	14.1	\\
745	&	2000-02-02	02:52:33	&	HRC-S	&	LETG	&	26.1	&	6364	&	2005-08-27	20:34:17	&	HRC-S	&	LETG	&	38.6	\\
2772	&	2001-12-06	14:13:22	&	ACIS-S	&	CC	&	4.0	&	6369	&	2005-10-08	22:10:18	&	HRC-S	&	LETG	&	26.0	\\
2773	&	2001-12-05	05:07:08	&	ACIS-S	&	CC	&	10.6	&	7177	&	2005-10-09	11:21:49	&	HRC-S	&	LETG	&	8.0	\\
2774	&	2001-12-04	16:36:09	&	ACIS-S	&	CC	&	15.0	&	7243	&	2005-12-14	00:52:14	&	HRC-S	&	LETG	&	17.1	\\
4666	&	2004-01-06	04:34:23	&	ACIS-S	&	CC	&	10.1	&	7244	&	2005-12-15	16:50:57	&	HRC-S	&	LETG	&	16.2	\\
4667	&	2004-01-07	05:04:55	&	ACIS-S	&	CC	&	4.8	&	7245	&	2005-12-16	15:50:35	&	HRC-S	&	LETG	&	17.1	\\
4668	&	2004-01-11	11:51:18	&	ACIS-S	&	CC	&	5.1	&	7251	&	2006-02-09	06:41:30	&	HRC-S	&	LETG	&	10.6	\\
4669	&	2004-01-19	02:01:26	&	ACIS-S	&	CC	&	5.2	&	10700	&	2009-02-14	03:18:00	&	HRC-S	&	LETG	&	21.6	\\
4670	&	2004-08-03	22:18:44	&	ACIS-S	&	CC	&	10.1	&	10701	&	2009-09-11	11:12:30	&	HRC-S	&	LETG	&	32.8	\\
4671	&	2004-08-05	04:05:48	&	ACIS-S	&	CC	&	5.1	&	10861	&	2009-01-20	03:55:21	&	HRC-S	&	LETG	&	11.1	\\
4672	&	2004-08-09	04:44:11	&	ACIS-S	&	CC	&	5.1	&	11820	&	2010-06-19	03:00:21	&	HRC-S	&	LETG	&	34.8	\\
4673	&	2004-08-23	03:25:18	&	ACIS-S	&	CC	&	5.1	&	13181	&	2010-11-18	16:36:17	&	HRC-S	&	LETG	&	19.8	\\
5305	&	2004-02-27	04:41:34	&	HRC-S	&	LETG	&	35.5	&	13188	&	2010-11-19	16:08:17	&	HRC-S	&	LETG	&	14.0	\\
5581	&	2005-01-23	06:15:08	&	HRC-S	&	LETG	&	67.7	&	13294	&	2012-11-28	12:57:36	&	HRC-S	&	LETG	&	32.0	\\
\hline
    \multicolumn{10}{c}{\textit{NICER}} \\                
\hline
0020140101	&	2017-06-30	04:09:51	&	XTI	&	\nodata	&	0.46	&	1020140149	&	2018-06-10	03:01:57	&	XTI	&	\nodata	&	0.67	\\
1020140105	&	2017-07-29	19:54:04	&	XTI	&	\nodata	&	0.17	&	1020140150	&	2018-06-11	09:54:55	&	XTI	&	\nodata	&	0.16	\\
1020140107	&	2017-07-31	19:47:34	&	XTI	&	\nodata	&	0.24	&	1020140151	&	2018-12-04	11:31:21	&	XTI	&	\nodata	&	3.1	\\
1020140108	&	2017-08-01	17:24:54	&	XTI	&	\nodata	&	0.27	&	1020140152	&	2018-12-05	21:21:49	&	XTI	&	\nodata	&	0.74	\\
1020140110	&	2017-08-03	18:49:32	&	XTI	&	\nodata	&	0.34	&	1020140153	&	2018-12-06	02:05:12	&	XTI	&	\nodata	&	0.87	\\
1020140121	&	2017-08-16	11:04:01	&	XTI	&	\nodata	&	0.26	&	1020140154	&	2018-12-07	02:48:20	&	XTI	&	\nodata	&	5.1	\\
1020140123	&	2017-10-25	04:27:04	&	XTI	&	\nodata	&	0.49	&	1020140155	&	2018-12-10	01:48:15	&	XTI	&	\nodata	&	2.8	\\
1020140124	&	2017-12-14	00:10:00	&	XTI	&	\nodata	&	4.6	&	1020140156	&	2018-12-11	14:56:28	&	XTI	&	\nodata	&	0.23	\\
1020140125	&	2017-12-15	02:13:31	&	XTI	&	\nodata	&	3.2	&	1020140157	&	2018-12-13	17:57:34	&	XTI	&	\nodata	&	1.7	\\
1020140126	&	2018-05-11	21:49:13	&	XTI	&	\nodata	&	0.69	&	1020140158	&	2018-12-14	00:09:12	&	XTI	&	\nodata	&	1.5	\\
1020140127	&	2018-05-12	00:54:34	&	XTI	&	\nodata	&	0.88	&	1020140159	&	2018-12-15	16:14:41	&	XTI	&	\nodata	&	2.4	\\
1020140128	&	2018-05-15	06:07:01	&	XTI	&	\nodata	&	2.6	&	1020140160	&	2018-12-16	03:14:39	&	XTI	&	\nodata	&	0.55	\\
1020140131	&	2018-05-19	01:19:11	&	XTI	&	\nodata	&	0.095	&	1020140161	&	2018-12-17	08:38:10	&	XTI	&	\nodata	&	0.91	\\
1020140132	&	2018-05-24	01:59:37	&	XTI	&	\nodata	&	0.98	&	1020140162	&	2018-12-19	11:32:13	&	XTI	&	\nodata	&	1.2	\\
1020140134	&	2018-05-26	00:01:32	&	XTI	&	\nodata	&	2.4	&	1020140163	&	2018-12-20	04:32:36	&	XTI	&	\nodata	&	2.3	\\
1020140135	&	2018-05-27	20:42:05	&	XTI	&	\nodata	&	0.80	&	5020140101	&	2022-04-24	22:49:41	&	XTI	&	\nodata	&	0.62	\\
1020140136	&	2018-05-28	21:25:59	&	XTI	&	\nodata	&	0.50	&	5020140102	&	2022-04-25	14:18:04	&	XTI	&	\nodata	&	0.72	\\
1020140137	&	2018-05-29	19:22:44	&	XTI	&	\nodata	&	1.1	&	5020140103	&	2022-04-26	22:55:09	&	XTI	&	\nodata	&	0.45	\\
1020140138	&	2018-05-30	00:05:21	&	XTI	&	\nodata	&	10.0	&	5020140104	&	2022-04-27	00:27:48	&	XTI	&	\nodata	&	1.7	\\
1020140139	&	2018-05-31	00:46:57	&	XTI	&	\nodata	&	4.8	&	5020140105	&	2022-05-03	14:34:21	&	XTI	&	\nodata	&	2.8	\\
1020140140	&	2018-05-31	23:56:35	&	XTI	&	\nodata	&	6.0	&	5020140106	&	2022-05-04	02:49:23	&	XTI	&	\nodata	&	1.9	\\
1020140141	&	2018-06-02	00:26:50	&	XTI	&	\nodata	&	1.2	&	5020140107	&	2022-05-05	05:02:54	&	XTI	&	\nodata	&	1.7	\\
1020140142	&	2018-06-02	23:47:04	&	XTI	&	\nodata	&	5.3	&	5020140108	&	2022-05-05	23:43:16	&	XTI	&	\nodata	&	5.9	\\
1020140143	&	2018-06-04	00:28:36	&	XTI	&	\nodata	&	2.6	&	5020140109	&	2022-05-07	05:09:35	&	XTI	&	\nodata	&	0.19	\\
1020140144	&	2018-06-05	13:36:22	&	XTI	&	\nodata	&	0.55	&	5020140110	&	2022-05-08	18:23:44	&	XTI	&	\nodata	&	0.22	\\
1020140145	&	2018-06-06	23:31:11	&	XTI	&	\nodata	&	0.41	&	5020140112	&	2022-05-11	00:24:03	&	XTI	&	\nodata	&	0.78	\\
1020140146	&	2018-06-07	04:10:26	&	XTI	&	\nodata	&	2.0	&	5020140113	&	2022-05-14	18:23:18	&	XTI	&	\nodata	&	1.0	\\
1020140147	&	2018-06-08	01:48:37	&	XTI	&	\nodata	&	2.3	&	5020140114	&	2022-05-20	21:27:01	&	XTI	&	\nodata	&	0.043	\\
1020140148	&	2018-06-09	02:24:49	&	XTI	&	\nodata	&	7.0	&	5020140115	&	2022-05-21	00:33:46	&	XTI	&	\nodata	&	0.071	\\
\enddata
\tablenotetext{a}{Listed exposure times are values after filtering for instances of strong background flaring. The imaging mode abbreviations correspond to Full Window (FW), Large Window (LW) and Small Window (SW).}
\tablenotetext{b}{For the Chandra HRC-S LETG observation, only the zeroth order (undispersed) point source emission was extracted for the timing analysis.}
\end{deluxetable}

\vspace{-0.3cm}

\startlongtable
\begin{deluxetable}{lcclR|lcclR}
\tabletypesize{\tiny}
\tablewidth{0pt}
\tablecolumns{10}
\tablecaption{Log of X-ray observations of RX\,J0806.4$-$4123 \label{table:j0806_obs}}
\tablehead{\colhead{ObsID} & \colhead{Date/Start Time} & \colhead{Instrument} & \colhead{Mode/Filter} &  \colhead{$t_{\rm exp}$} &  \colhead{ObsID} & \colhead{Date/Start Time} & \colhead{Instrument} & \colhead{Mode/Filter} &  \colhead{$t_{\rm exp}$}  \\
            \colhead{}     & \colhead{(UTC)} & \colhead{}           &    \colhead{}         & \colhead{(ks)}    & \colhead{}     & \colhead{(UTC)} & \colhead{}           &    \colhead{}         & \colhead{(ks)}   } 
            \startdata
\multicolumn{10}{C}{\textit{XMM-Newton}\tablenotemark{a}} \\
\hline
0106260201	&	2000-11-08	14:24:52	&	pn	&	FW	/	Thin	&	6.58	&	0552211101	&	2009-03-31	20:37:10	&	pn	&	SW	/	Thin	&	5.86	\\
0141750501	&	2003-04-24	14:44:30	&	pn	&	FW	/	Thin	&	15.8	&		&	2009-03-31	20:32:00	&	MOS1	&	SW	/	Thin	&	8.1	\\
0552210201	&	2008-05-11	10:51:41	&	pn	&	SW	/	Thin	&	6.05	&		&	2009-03-31	20:32:00	&	MOS2	&	SW	/	Thin	&	8.2	\\
	&	2008-05-11	10:46:30	&	MOS1	&	SW	/	Thin	&	8.5	&	0552211201	&	2009-04-01	18:07:12	&	pn	&	SW	/	Thin	&	6.16	\\
	&	2008-05-11	10:46:29	&	MOS2	&	SW	/	Thin	&	8.5	&		&	2009-04-01	18:02:02	&	MOS1	&	SW	/	Thin	&	0.97	\\
0552210301	&	2008-05-15	06:05:40	&	pn	&	SW	/	Thin	&	6.84	&		&	2009-04-01	18:02:02	&	MOS2	&	SW	/	Thin	&	0.29	\\
	&	2008-05-15	06:00:30	&	MOS1	&	SW	/	Thin	&	9.6	&	0552211501	&	2008-11-09	06:00:56	&	pn	&	SW	/	Thin	&	2.82	\\
	&	2008-05-15	06:00:30	&	MOS2	&	SW	/	Thin	&	9.6	&		&	2008-11-09	05:55:48	&	MOS1	&	SW	/	Thin	&	2.5	\\
0552210401	&	2008-05-29	05:52:26	&	pn	&	SW	/	Thin	&	3.83	&		&	2008-11-09	05:55:50	&	MOS2	&	SW	/	Thin	&	2.4	\\
	&	2008-05-29	05:47:17	&	MOS1	&	SW	/	Thin	&	5.4	&	0552211601	&	2009-04-11	00:12:00	&	pn	&	SW	/	Thin	&	5.93	\\
	&	2008-05-29	05:47:15	&	MOS2	&	SW	/	Thin	&	5.4	&		&	2009-04-11	00:06:49	&	MOS1	&	SW	/	Thin	&	6.9	\\
0552210501	&	2008-06-20	12:33:56	&	pn	&	SW	/	Thin	&	2.96	&		&	2009-04-11	00:06:49	&	MOS2	&	SW	/	Thin	&	6.7	\\
	&	2008-06-20	12:28:45	&	MOS1	&	SW	/	Thin	&	2.5	&	0672980201	&	2011-05-02	19:53:58	&	pn	&	SW	/	Thin	&	6.64	\\
	&	2008-06-20	12:28:45	&	MOS2	&	SW	/	Thin	&	2.1	&		&	2011-05-02	19:48:50	&	MOS1	&	SW	/	Thin	&	8.7	\\
0552210601	&	2008-10-15	10:28:44	&	pn	&	SW	/	Thin	&	6.55	&		&	2011-05-02	19:48:49	&	MOS2	&	SW	/	Thin	&	8.8	\\
	&	2008-10-15	10:23:35	&	MOS1	&	SW	/	Thin	&	9.2	&	0672980301	&	2012-04-20	07:51:15	&	pn	&	SW	/	Thin	&	3.84	\\
	&	2008-10-15	10:23:35	&	MOS2	&	SW	/	Thin	&	9.2	&		&	2012-04-20	07:46:13	&	MOS1	&	SW	/	Thin	&	5.4	\\
0552210901	&	2008-11-04	03:44:15	&	pn	&	SW	/	Thin	&	3.83	&		&	2012-04-20	07:46:11	&	MOS2	&	SW	/	Thin	&	5.5	\\
	&	2008-11-04	03:39:07	&	MOS1	&	SW	/	Thin	&	5.4	&	0844140301	&	2019-05-16	16:31:24	&	pn	&	FW	/	Thin	&	14.4	\\
	&	2008-11-04	03:39:06	&	MOS2	&	SW	/	Thin	&	5.4	&		&			&		                  &				               &		\\
\hline
    \multicolumn{10}{c}{\textit{Chandra}} \\                
\hline
11825	& 2010-03-21 02:26:45	 &  ACIS-S & CC &  18.1	& 20743   & 2018-09-01 17:37:54   & HRC-I   &   LETG     & 29.8     \\
\hline
    \multicolumn{10}{c}{\textit{NICER}} \\     
\hline
2552010101	&	2019-03-11	11:54:20	&	XTI	&	\nodata	&	9.6	&	3553010102	&	2021-02-20	00:30:59	&	XTI	&	\nodata	&	19.8	\\
2552010102	&	2019-03-12	00:16:17	&	XTI	&	\nodata	&	18	&	3553010103	&	2021-02-20	23:45:38	&	XTI	&	\nodata	&	7.1	\\
2552010103	&	2019-03-13	01:00:44	&	XTI	&	\nodata	&	15.3	&	5630010101	&	2023-01-26	04:10:20	&	XTI	&	\nodata	&	15.4	\\
2552010104	&	2019-03-14	00:16:58	&	XTI	&	\nodata	&	10.5	&	5630010102	&	2023-01-27	00:03:40	&	XTI	&	\nodata	&	9.1	\\
2552010201	&	2019-03-14	19:58:41	&	XTI	&	\nodata	&	4.2	&	5630010103	&	2023-01-28	02:41:20	&	XTI	&	\nodata	&	0.76	\\
2552010202	&	2019-03-15	00:35:55	&	XTI	&	\nodata	&	13.6	&	5630010104	&	2023-02-03	20:43:29	&	XTI	&	\nodata	&	1.3	\\
2552010401	&	2019-05-28	01:09:21	&	XTI	&	\nodata	&	19.1	&	5630010105	&	2023-02-04	01:23:07	&	XTI	&	\nodata	&	5	\\
2552010501	&	2020-02-23	21:00:40	&	XTI	&	\nodata	&	0.97	&	5630010106	&	2023-02-06	21:33:22	&	XTI	&	\nodata	&	0.48	\\
2552010502	&	2020-02-24	00:06:40	&	XTI	&	\nodata	&	16.9	&	5630010107	&	2023-02-07	02:12:20	&	XTI	&	\nodata	&	0.81	\\
3553010101	&	2021-02-19	19:52:00	&	XTI	&	\nodata	&	3.9	&		&			&		&		&		\\
\enddata
\tablenotetext{a}{Listed exposure times are values after filtering for instances of strong background flaring. The imaging mode abbreviations correspond to Full Window (FW) and Small Window (SW).}
\end{deluxetable}

\startlongtable
\begin{deluxetable}{lcclR|lcclR}
\tabletypesize{\tiny}
\tablewidth{0pt}
\tablecolumns{10}
\tablecaption{Log of X-ray observations of RX\,J1308.6+2127 \label{table:j1308_obs}}
\tablehead{\colhead{ObsID} & \colhead{Date/Start Time} & \colhead{Instrument} & \colhead{Mode/Filter} &  \colhead{$t_{\rm exp}$} &  \colhead{ObsID} & \colhead{Date/Start Time} & \colhead{Instrument} & \colhead{Mode/Filter} &  \colhead{$t_{\rm exp}$}  \\
            \colhead{}     & \colhead{(UTC)} & \colhead{}           &    \colhead{}         & \colhead{(ks)}    & \colhead{}     & \colhead{(UTC)} & \colhead{}           &    \colhead{}         & \colhead{(ks)}   } 
            \startdata
\multicolumn{10}{C}{\textit{XMM-Newton}\tablenotemark{a}} \\
\hline
0090010101	&	2001-12-31	03:31:51	&	pn	&	SW	/	Thin	&	11.0	&	0402850301	&	2006-06-08	22:44:26	&	pn	&	LW	/	Thin	&	3.9	\\
	&	2001-12-31	03:08:35	&	MOS1	&	FW	/	Thin	&	5.5	&		&	2006-06-08	22:16:12	&	MOS1	&	SW	/	Thin	&	5.3	\\
	&	2001-12-31	03:08:37	&	MOS2	&	FW	/	Thin	&	5.6	&		&	2006-06-08	22:16:13	&	MOS2	&	SW	/	Thin	&	5.3	\\
0157360101	&	2003-01-01	06:36:04	&	pn	&	FW	/	Thin	&	24.2	&	0402850401	&	2006-06-16	21:57:40	&	pn	&	LW	/	Thin	&	5.7	\\
	&	2003-01-01	06:13:59	&	MOS1	&	LW	/	Thin	&	28.3	&		&	2006-06-16	21:29:26	&	MOS1	&	SW	/	Thin	&	7.5	\\
	&	2003-01-01	06:13:58	&	MOS2	&	LW	/	Thin	&	28.3	&		&	2006-06-16	21:29:26	&	MOS2	&	SW	/	Thin	&	7.6	\\
0163560101	&	2003-12-30	06:59:04	&	pn	&	FW	/	Thin	&	23.8	&	0402850501	&	2006-06-27	03:02:26	&	pn	&	LW	/	Thin	&	5.7	\\
	&	2003-12-30	06:36:45	&	MOS1	&	LW	/	Thin	&	31.2	&		&	2006-06-27	02:34:12	&	MOS1	&	SW	/	Thin	&	11.9	\\
	&	2003-12-30	06:36:39	&	MOS2	&	LW	/	Thin	&	31.4	&		&	2006-06-27	02:34:13	&	MOS2	&	SW	/	Thin	&	11.9	\\
0305900201	&	2005-06-25	00:45:02	&	pn	&	FW	/	Thin	&	4.5	&	0402850701	&	2006-12-27	15:08:40	&	pn	&	LW	/	Thin	&	7.6	\\
	&	2005-06-25	00:22:52	&	MOS1	&	SW	/	Thin	&	13.4	&		&	2006-12-27	14:40:27	&	MOS1	&	SW	/	Thin	&	9.8	\\
	&	2005-06-25	00:22:53	&	MOS2	&	SW	/	Thin	&	13.5	&		&	2006-12-27	14:40:25	&	MOS2	&	SW	/	Thin	&	9.8	\\
0305900301	&	2005-06-27	00:25:38	&	pn	&	FW	/	Thin	&	11.4	&	0402850901	&	2006-07-05	05:57:22	&	pn	&	LW	/	Thin	&	3.4	\\
	&	2005-06-27	00:03:28	&	MOS1	&	SW	/	Thin	&	14.1	&		&	2006-07-05	05:29:09	&	MOS1	&	SW	/	Thin	&	7.5	\\
	&	2005-06-27	00:03:29	&	MOS2	&	SW	/	Thin	&	14.0	&		&	2006-07-05	05:29:06	&	MOS2	&	SW	/	Thin	&	7.3	\\
0305900401	&	2005-07-15	08:14:56	&	pn	&	FW	/	Thin	&	10.0	&	0402851001	&	2007-06-11	13:53:13	&	MOS1	&	SW	/	Thin	&	2.6	\\
	&	2005-07-15	07:52:46	&	MOS1	&	SW	/	Thin	&	13.9	&		&	2007-06-11	13:53:12	&	MOS2	&	SW	/	Thin	&	3.2	\\
	&	2005-07-15	07:52:48	&	MOS2	&	SW	/	Thin	&	13.9	&	0844140101	&	2019-12-16	07:13:08	&	pn	&	FW	/	Thin	&	6.3	\\
0305900601	&	2006-01-10	19:00:19	&	pn	&	FW	/	Thin	&	13.2	&		&	2019-12-16	06:47:26	&	MOS1	&	SW	/	Thin	&	13.5	\\
	&	2006-01-10	18:38:12	&	MOS1	&	SW	/	Thin	&	16.1	&		&	2019-12-16	06:47:47	&	MOS2	&	SW	/	Thin	&	13.6	\\
	&	2006-01-10	18:38:12	&	MOS2	&	SW	/	Thin	&	16.1	&		&			&		&				&		\\
\hline
    \multicolumn{10}{c}{\textit{Chandra}} \\                
\hline
4595	&	2005-04-12 20:16:33	&   HRC-S	&	LETG	    &   87.2 & 5526	&	2006-07-11 11:10:16	&	ACIS-S	&	CC	&	15.1 \\
5522	&	2006-02-16 22:02:25	&	ACIS-S	&	CC	&	15.9  & 5527	&	2006-07-12 09:50:18	&	ACIS-S	&	CC	&	5.1 \\
5523	&	2006-02-18 10:12:27	&	ACIS-S	&	CC	&	5.7 & 5528	&	2006-07-15 17:00:25	&	ACIS-S	&	CC	&	5.2 \\
5524	&	2006-02-20 10:19:24	&	ACIS-S	&	CC	&	5.1 & 5529	&	2006-08-03 14:59:40	&	ACIS-S	&	CC	&	5.2 \\
5525	&	2006-03-18 16:19:25	&	ACIS-S	&	CC	&	5.6 &      &                       &           &               &           \\
\hline
    \multicolumn{10}{c}{\textit{NICER}} \\                
\hline
1020280101	&	2017-11-14	03:36:03	&	XTI	&	\nodata	&	0.068	&	1020280120	&	2018-01-23	00:36:52	&	XTI	&	\nodata	&	5.0	\\
1020280104	&	2017-11-18	02:50:49	&	XTI	&	\nodata	&	0.56	&	1020280121	&	2018-01-24	01:11:15	&	XTI	&	\nodata	&	10.9	\\
1020280105	&	2017-11-19	00:27:53	&	XTI	&	\nodata	&	1.0	&	1020280122	&	2018-01-25	00:31:40	&	XTI	&	\nodata	&	2.6	\\
1020280106	&	2017-12-05	02:55:00	&	XTI	&	\nodata	&	0.052	&	1020280123	&	2018-01-26	08:49:28	&	XTI	&	\nodata	&	0.99	\\
1020280108	&	2017-12-06	13:00:13	&	XTI	&	\nodata	&	0.10	&	1020280124	&	2018-01-27	18:53:38	&	XTI	&	\nodata	&	0.11	\\
1020280109	&	2017-12-07	21:13:40	&	XTI	&	\nodata	&	0.78	&	1020280125	&	2018-01-28	07:16:14	&	XTI	&	\nodata	&	0.73	\\
1020280110	&	2017-12-08	06:39:40	&	XTI	&	\nodata	&	1.2	&	1020280126	&	2018-01-29	04:41:29	&	XTI	&	\nodata	&	1.3	\\
1020280111	&	2017-12-09	06:01:40	&	XTI	&	\nodata	&	5.4	&	1020280127	&	2018-01-30	13:18:31	&	XTI	&	\nodata	&	0.12	\\
1020280112	&	2017-12-10	03:49:42	&	XTI	&	\nodata	&	3.5	&	1020280128	&	2018-06-20	02:37:39	&	XTI	&	\nodata	&	8.9	\\
1020280113	&	2017-12-11	06:05:40	&	XTI	&	\nodata	&	1.3	&	1020280129	&	2018-06-21	00:14:19	&	XTI	&	\nodata	&	4.7	\\
1020280114	&	2017-12-16	00:05:01	&	XTI	&	\nodata	&	21.1	&	1020280130	&	2018-06-22	00:58:19	&	XTI	&	\nodata	&	5.4	\\
1020280115	&	2017-12-17	02:07:43	&	XTI	&	\nodata	&	17.7	&	1020280131	&	2018-06-23	09:25:20	&	XTI	&	\nodata	&	1.1	\\
1020280116	&	2017-12-17	23:42:11	&	XTI	&	\nodata	&	17.9	&	1020280132	&	2018-06-24	19:20:26	&	XTI	&	\nodata	&	0.25	\\
1020280117	&	2018-01-20	03:05:24	&	XTI	&	\nodata	&	2.0	&	1020280133	&	2018-06-24	23:58:26	&	XTI	&	\nodata	&	1.9	\\
1020280118	&	2018-01-21	02:10:00	&	XTI	&	\nodata	&	3.6	&	1020280134	&	2018-06-26	02:14:06	&	XTI	&	\nodata	&	0.62	\\
\enddata
\tablenotetext{a}{Listed exposure times are values after filtering for instances of strong background flaring. The imaging mode abbreviations correspond to Full Window (FW) and Small Window (SW).}
\tablenotetext{b}{For the Chandra HRC LETG observation, only the zeroth order (undispersed) point source emission was extracted for the timing analysis.}
\end{deluxetable}

\vspace{-0.5cm}
\startlongtable
\begin{deluxetable}{lcclR|lcclR}
\tabletypesize{\tiny}
\tablewidth{0pt}
\tablecolumns{10}
\tablecaption{Log of X-ray observations of RX\,J1856.5$-$3754 \label{table:j1856_obs}}
\tablehead{\colhead{ObsID} & \colhead{Date/Start Time} & \colhead{Instrument} & \colhead{Mode/Filter} &  \colhead{$t_{\rm exp}$} &  \colhead{ObsID} & \colhead{Date/Start Time} & \colhead{Instrument} & \colhead{Mode/Filter} &  \colhead{$t_{\rm exp}$}  \\
            \colhead{}     & \colhead{(UTC)} & \colhead{}           &    \colhead{}         & \colhead{(ks)}    & \colhead{}     & \colhead{(UTC)} & \colhead{}           &    \colhead{}         & \colhead{(ks)}   } 
\startdata
\multicolumn{10}{C}{\textit{XMM-Newton}\tablenotemark{a}} \\
\hline
0106260101	&	2002-04-08	16:21:27	&	pn	&	SW	/	Thin	&	40.0	&	0727760301	&	2014-09-18	11:03:44	&	pn	&	SW	/	Thin	&	55.0	\\
	&	2002-04-08	16:05:50	&	MOS2	&	SW	/	Thin	&	56.2	&		&	2014-09-18	10:57:56	&	MOS1	&	SW	/	Thin	&	74.6	\\
0165971601	&	2004-09-24	01:42:13	&	pn	&	SW	/	Thin	&	23.1	&		&	2014-09-18	10:58:41	&	MOS2	&	SW	/	Thin	&	74.3	\\
	&	2004-09-24	01:37:02	&	MOS1	&	SW	/	Thin	&	32.1	&	0727760401	&	2015-03-12	11:46:22	&	pn	&	SW	/	Thin	&	51.3	\\
	&	2004-09-24	01:37:01	&	MOS2	&	SW	/	Thin	&	32.1	&		&	2015-03-12	11:40:32	&	MOS1	&	LW	/	Thin	&	68.5	\\
0165971701	&	2004-09-24	11:12:04	&	MOS1	&	SW	/	Medium	&	26.3	&		&	2015-03-12	11:41:04	&	MOS2	&	LW	/	Thin	&	68.1	\\
	&	2004-09-24	11:12:03	&	MOS2	&	SW	/	Medium	&	28.7	&	0727760501	&	2015-10-03	15:56:32	&	pn	&	SW	/	Thin	&	51.7	\\
0165971901	&	2005-03-23	08:34:42	&	pn	&	SW	/	Thin	&	22.8	&		&	2015-10-03	16:57:11	&	MOS2	&	SW	/	Thin	&	65.1	\\
	&	2005-03-23	08:29:16	&	MOS1	&	LW	/	Thin	&	23.7	&	0727760601	&	2016-03-11	21:51:55	&	pn	&	SW	/	Thin	&	42.7	\\
	&	2005-03-23	08:29:32	&	MOS2	&	SW	/	Thin	&	25.5	&		&	2016-03-11	21:46:07	&	MOS1	&	LW	/	Thin	&	46.5	\\
0165972001	&	2005-09-24	07:58:13	&	pn	&	SW	/	Thin	&	23.3	&		&	2016-03-11	21:46:36	&	MOS2	&	LW	/	Thin	&	45.4	\\
	&	2005-09-24	07:53:03	&	MOS1	&	SW	/	Thin	&	32.3	&	0727761001	&	2016-09-23	00:14:46	&	pn	&	SW	/	Thin	&	49.2	\\
	&	2005-09-24	07:53:01	&	MOS2	&	SW	/	Thin	&	32.2	&		&	2016-09-23	00:09:18	&	MOS1	&	SW	/	Thin	&	68.3	\\
0165972101	&	2006-03-26	15:40:29	&	pn	&	SW	/	Thin	&	48.7	&		&	2016-09-23	00:10:03	&	MOS2	&	SW	/	Thin	&	68.3	\\
	&	2006-03-26	15:35:20	&	MOS1	&	SW	/	Thin	&	67.3	&	0727761101	&	2017-03-15	06:45:19	&	pn	&	SW	/	Thin	&	48.3	\\
	&	2006-03-26	15:35:20	&	MOS2	&	SW	/	Thin	&	67.4	&		&	2017-03-15	06:39:50	&	MOS1	&	LW	/	Thin	&	66.2	\\
0201590101	&	2004-04-17	22:17:17	&	pn	&	FT	/	Thin	&	44.3	&		&	2017-03-15	06:40:21	&	MOS2	&	LW	/	Thin	&	66.5	\\
	&	2004-04-17	21:56:28	&	MOS1	&	FW	/	Thin	&	43.0	&	0727761201	&	2017-09-16	17:42:56	&	pn	&	SW	/	Thin	&	51.8	\\
	&	2004-04-17	21:56:28	&	MOS2	&	FW	/	Thin	&	44.4	&		&	2017-09-16	18:54:56	&	MOS2	&	SW	/	Thin	&	55.7	\\
0213080101	&	2005-04-15	17:45:11	&	MOS1	&	LW	/	Thin	&	4.8	&	0727761301	&	2018-04-10	12:18:35	&	pn	&	SW	/	Thin	&	47.4	\\
	&	2005-04-15	18:03:47	&	MOS2	&	LW	/	Thin	&	3.7	&		&	2018-04-10	12:10:15	&	MOS1	&	LW	/	Thin	&	54.0	\\
0412600101	&	2006-10-24	00:33:51	&	pn	&	SW	/	Thin	&	50.8	&		&	2018-04-10	12:10:35	&	MOS2	&	LW	/	Thin	&	50.2	\\
	&	2006-10-24	00:28:42	&	MOS1	&	SW	/	Thin	&	70.2	&	0791580101	&	2016-04-16	17:06:04	&	pn	&	SW	/	Thick	&	12.6	\\
	&	2006-10-24	00:28:42	&	MOS2	&	SW	/	Thin	&	70.2	&		&	2016-04-16	17:00:14	&	MOS1	&	SW	/	Thick	&	17.6	\\
0412600201	&	2007-03-14	20:50:01	&	pn	&	SW	/	Thin	&	42.2	&		&	2016-04-16	17:00:59	&	MOS2	&	SW	/	Thick	&	17.6	\\
	&	2007-03-14	20:01:47	&	MOS1	&	SW	/	Thin	&	43.7	&	0791580201	&	2016-04-16	22:31:04	&	pn	&	SW	/	Medium	&	6.6	\\
	&	2007-03-14	20:01:43	&	MOS2	&	SW	/	Thin	&	34.0	&		&	2016-04-16	23:26:10	&	MOS2	&	SW	/	Medium	&	5.7	\\
0412600301	&	2007-10-04	05:48:49	&	pn	&	SW	/	Thin	&	26.7	&	0791580301	&	2016-04-17	01:32:43	&	pn	&	SW	/	Thin	&	41.9	\\
	&	2007-10-04	05:43:39	&	MOS1	&	SW	/	Thin	&	21.0	&		&	2016-04-17	01:26:54	&	MOS1	&	SW	/	Thin	&	6.0	\\
	&	2007-10-04	05:43:39	&	MOS2	&	SW	/	Thin	&	37.2	&		&	2016-04-17	01:27:40	&	MOS2	&	SW	/	Thin	&	5.9	\\
0412600401	&	2008-03-13	18:49:45	&	pn	&	SW	/	Thin	&	42.7	&	0791580401	&	2016-04-17	03:37:44	&	pn	&	SW	/	Thin	&	13.2	\\
	&	2008-03-13	19:01:02	&	MOS1	&	SW	/	Thin	&	48.6	&		&	2016-04-17	03:31:55	&	MOS1	&	SW	/	Thick	&	18.5	\\
	&	2008-03-13	19:01:32	&	MOS2	&	SW	/	Thin	&	48.9	&		&	2016-04-17	03:32:41	&	MOS2	&	SW	/	Thick	&	18.4	\\
0412600601	&	2008-10-05	01:00:58	&	pn	&	SW	/	Thin	&	43.8	&	0791580501	&	2016-04-17	09:17:43	&	pn	&	SW	/	Thin	&	5.9	\\
	&	2008-10-05	00:49:38	&	MOS1	&	SW	/	Thin	&	61.2	&		&	2016-04-17	09:11:53	&	MOS1	&	SW	/	Medium	&	8.4	\\
	&	2008-10-05	00:49:44	&	MOS2	&	SW	/	Thin	&	61.1	&		&	2016-04-17	09:12:38	&	MOS2	&	SW	/	Medium	&	8.4	\\
0412600701	&	2009-03-19	21:30:04	&	pn	&	SW	/	Thin	&	47.9	&	0791580601	&	2016-04-17	12:04:22	&	pn	&	SW	/	Thin	&	10.8	\\
	&	2009-03-19	21:24:54	&	MOS1	&	SW	/	Thin	&	66.3	&		&	2016-04-17	11:58:33	&	MOS1	&	SW	/	Thin	&	14.8	\\
	&	2009-03-19	21:24:54	&	MOS2	&	SW	/	Thin	&	66.5	&		&	2016-04-17	11:59:19	&	MOS2	&	SW	/	Thin	&	14.9	\\
0412600801	&	2009-10-07	12:07:09	&	pn	&	SW	/	Thin	&	52.9	&	0810840101	&	2018-10-19	07:01:24	&	pn	&	SW	/	Thin	&	48.1	\\
	&	2009-10-07	12:01:59	&	MOS1	&	SW	/	Thin	&	6.1	&		&	2018-10-19	06:53:04	&	MOS1	&	SW	/	Thin	&	67.2	\\
	&	2009-10-07	12:01:59	&	MOS2	&	SW	/	Thin	&	61.5	&		&	2018-10-19	06:53:25	&	MOS2	&	SW	/	Thin	&	67.2	\\
0412600901	&	2010-03-22	02:50:25	&	pn	&	SW	/	Thin	&	51.4	&	0810840201	&	2019-04-12	16:29:45	&	pn	&	SW	/	Thin	&	51.7	\\
	&	2010-03-22	02:45:16	&	MOS1	&	SW	/	Thin	&	70.0	&		&	2019-04-12	16:21:25	&	MOS1	&	LW	/	Thin	&	40.6	\\
	&	2010-03-22	02:45:16	&	MOS2	&	SW	/	Thin	&	70.2	&		&	2019-04-12	16:21:45	&	MOS2	&	LW	/	Thin	&	40.6	\\
0412601101	&	2010-09-28	23:09:10	&	pn	&	SW	/	Thin	&	48.2	&	0810841401	&	2019-09-18	21:58:10	&	pn	&	SW	/	Thin	&	47.7	\\
	&	2010-09-28	23:07:03	&	MOS1	&	SW	/	Thin	&	68.0	&		&	2019-09-18	21:49:49	&	MOS1	&	SW	/	Thin	&	66.6	\\
	&	2010-09-28	23:07:10	&	MOS2	&	SW	/	Thin	&	68.0	&		&	2019-09-18	21:50:12	&	MOS2	&	SW	/	Thin	&	66.6	\\
0412601301	&	2011-03-14	00:47:48	&	pn	&	SW	/	Thin	&	55.9	&	0810841501	&	2020-03-31	23:16:42	&	pn	&	SW	/	Thin	&	50.5	\\
	&	2011-03-14	00:42:25	&	MOS1	&	SW	/	Thin	&	70.2	&		&	2020-03-31	23:08:22	&	MOS1	&	LW	/	Thin	&	71.0	\\
	&	2011-03-14	00:42:22	&	MOS2	&	SW	/	Thin	&	71.1	&		&	2020-03-31	23:08:43	&	MOS2	&	LW	/	Thin	&	70.8	\\
0412601401	&	2012-04-13	07:12:25	&	pn	&	SW	/	Thin	&	46.3	&	0810841601	&	2020-09-15 14:46:00		&	pn	&	SW	/	Thin	&	48.2	\\
	&	2012-04-13	07:07:05	&	MOS1	&	LW	/	Thin	&	44.4	&		&	2020-09-15 14:37:39		&	MOS1	&	SW	/	Thin	&	64.2	\\
	&	2012-04-13	07:07:05	&	MOS2	&	LW	/	Thin	&	46.0	&		&	2020-09-15 14:38:00		&	MOS2	&	SW	/	Thin	&	63.7	\\
0412601501	&	2011-10-05	02:02:47	&	pn	&	SW	/	Thin	&	73.9	&	0810841701	&	2021-04-01 08:36:25		&	pn	&	SW	/	Thin	&	48.8	\\
	&	2011-10-05	01:57:22	&	MOS1	&	LW	/	Thin	&	108.1	&		&	2021-04-01 08:28:05		&	MOS1	&	LW	/	Thin	&	69.2	\\
	&	2011-10-05	01:57:22	&	MOS1	&	LW	/	Thin	&	107.8	&		&	2021-04-01 08:28:26		&	MOS2	&	LW	/	Thin	&	69.1	\\
0412602201	&	2013-03-14	08:29:03	&	pn	&	SW	/	Thin	&	51.5	&	0810841901	&	2021-10-11 12:59:23		&	pn	&	SW	/	Thin	&	48.6	\\
	&	2013-03-14	08:23:43	&	MOS1	&	LW	/	Thin	&	72.8	&		&	2021-10-11 12:51:02		&	MOS1	&	SW	/	Thin	&	66	\\
	&	2013-03-14	08:23:43	&	MOS2	&	LW	/	Thin	&	72.8	&		&	2021-10-11 12:51:23		&	MOS2	&	SW	/	Thin	&	64.2	\\
0412602301	&	2012-09-20	11:23:08	&	pn	&	SW	/	Thin	&	50.0	&	0810842001	&	2022-04-03 00:05:28		&	pn	&	SW	/	Thin	&	49.8	\\
	&	2012-09-20	11:18:05	&	MOS1	&	SW	/	Thin	&	70.4	&		&	2022-04-02 23:57:28		&	MOS2	&	SW	/	Thin	&	65.7	\\
	&	2012-09-20	11:18:05	&	MOS2	&	SW	/	Thin	&	71.7	&	0810842201	&	2022-09-24 03:50:15		&	pn	&	SW	/	Thin	&	46.5	\\
0415180101	&	2007-03-25	05:36:47	&	pn	&	SW	/	Thin	&	25.7	&		&	2022-09-24 03:41:55		&	MOS1	&	SW	/	Thin	&	63.2	\\
	&	2007-03-25	05:31:24	&	MOS1	&	LW	/	Thin	&	25.8	&		&	2022-09-24 03:42:17		&	MOS2	&	SW	/	Thin	&	61.9	\\
	&	2007-03-25	05:31:24	&	MOS2	&	LW	/	Thin	&	25.9	&	0810842301	&	2023-04-01 10:49:51		&	pn	&	SW	/	Thin	&	47.8	\\
0727760101	&	2013-09-14	14:20:54	&	pn	&	SW	/	Thin	&	49.0	&		&	2023-04-01 10:41:30		&	MOS1	&	LW	/	Thin	&	67.9	\\
	&	2013-09-14	14:15:03	&	MOS1	&	SW	/	Thin	&	67.7	&		&	2023-04-01 10:41:51		&	MOS2	&	LW	/	Thin	&	67.9	\\
	&	2013-09-14	14:15:48	&	MOS2	&	SW	/	Thin	&	67.4	&		&			&		&				&		\\
0727760201	&	2014-03-26	06:00:22	&	pn	&	SW	/	Thin	&	49.0	&		&			&		&				&		\\
	&	2014-03-26	05:54:34	&	MOS1	&	LW	/	Thin	&	69.0	&		&			&		&				&		\\
	&	2014-03-26	05:55:04	&	MOS2	&	LW	/	Thin	&	66.8	&		&			&		&				&		\\
\hline
    \multicolumn{10}{c}{\textit{Chandra}} \\     
\hline
113	&	2000-03-10	07:54:08	&	HRC-S	&	LETG	&	55.5	&	6095	&	2005-11-13	03:46:28	&	HRC-I	&	\nodata	&	47.9	\\
3382	&	2001-10-08	08:17:45	&	HRC-S	&	LETG	&	101.9	&	7052	&	2006-06-01	18:08:02	&	HRC-I	&	\nodata	&	46.3	\\
3380	&	2001-10-10	05:05:24	&	HRC-S	&	LETG	&	167.5	&	7053	&	2006-11-18	06:51:52	&	HRC-I	&	\nodata	&	49.6	\\
3381	&	2001-10-12	19:18:22	&	HRC-S	&	LETG	&	171.1	&	15293	&	2013-06-12	14:28:42	&	HRC-S	&	LETG	&	91.8	\\
3399	&	2001-10-15	11:46:02	&	HRC-S	&	LETG	&	9.3	&	14418	&	2013-10-01	05:02:27	&	HRC-S	&	LETG	&	30.2	\\
4286	&	2002-08-06	13:57:40	&	HRC-I	&	\nodata	&	9.9	&	21693	&	2019-06-13	23:54:57	&	HRC-S	&	LETG	&	81.7	\\
4287	&	2002-09-03	18:47:08	&	HRC-I	&	\nodata	&	48.1	&	21896	&	2019-07-15	06:14:19	&	HRC-S	&	LETG	&	22.3	\\
4288	&	2002-09-23	12:21:28	&	HRC-I	&	\nodata	&	8.5	&	22282	&	2019-07-18	15:22:44	&	HRC-S	&	LETG	&	16.3	\\
4356	&	2003-05-04	09:44:49	&	HRC-I	&	\nodata	&	49.7	&	22283	&	2019-07-20	14:35:04	&	HRC-S	&	LETG	&	31.9	\\
5174	&	2004-11-04	12:35:37	&	HRC-I	&	\nodata	&	48.8	&	22284	&	2019-07-21	16:23:44	&	HRC-S	&	LETG	&	15.4	\\
6094	&	2005-06-10	01:09:20	&	HRC-I	&	\nodata	&	49.5	&		&			&		&		&		\\
\hline
    \multicolumn{10}{c}{\textit{NICER}\tablenotemark{b}} \\     
\hline
1020520101	&	2018-04-11	15:18:22	&	XTI	&	\nodata	&	2.1	&	2012130104*	&	2019-07-02	01:11:05	&	XTI	&	\nodata	&	2.8	\\
1020520102	&	2018-04-12	00:19:50	&	XTI	&	\nodata	&	6.0	&	2012130105*	&	2019-07-09	13:54:37	&	XTI	&	\nodata	&	0.67	\\
1020520103	&	2018-04-12	23:45:02	&	XTI	&	\nodata	&	5.4	&	2012130107*	&	2019-07-14	02:38:51	&	XTI	&	\nodata	&	0.19	\\
1020520104	&	2018-04-20	02:41:56	&	XTI	&	\nodata	&	0.33	&	2012130108*	&	2019-08-03	14:59:01	&	XTI	&	\nodata	&	1.5	\\
1020520105	&	2018-04-21	05:03:46	&	XTI	&	\nodata	&	1.2	&	2012130109*	&	2019-08-06	12:48:21	&	XTI	&	\nodata	&	0.31	\\
1020520106	&	2018-04-22	07:17:29	&	XTI	&	\nodata	&	0.96	&	2012130110*	&	2019-08-07	02:42:58	&	XTI	&	\nodata	&	0.37	\\
1020520107	&	2019-02-16	06:09:08	&	XTI	&	\nodata	&	1.0	&	2012130111*	&	2019-08-08	12:36:43	&	XTI	&	\nodata	&	0.76	\\
1020520108	&	2019-02-17	00:51:39	&	XTI	&	\nodata	&	1.1	&	2012130112*	&	2019-08-09	19:32:56	&	XTI	&	\nodata	&	0.30	\\
1020520109	&	2019-02-18	01:38:30	&	XTI	&	\nodata	&	3.0	&	2012130113*	&	2019-08-18	15:43:28	&	XTI	&	\nodata	&	0.45	\\
1020520110	&	2019-02-19	00:46:41	&	XTI	&	\nodata	&	6.3	&	2012130115*	&	2019-08-23	08:25:32	&	XTI	&	\nodata	&	0.10	\\
1020520111	&	2019-02-20	01:31:26	&	XTI	&	\nodata	&	0.65	&	2012130116*	&	2019-08-24	01:27:47	&	XTI	&	\nodata	&	0.072	\\
1020520112	&	2019-02-21	07:01:16	&	XTI	&	\nodata	&	1.6	&	2012130117*	&	2019-09-03	22:56:38	&	XTI	&	\nodata	&	0.48	\\
2020520101	&	2019-03-21	12:10:44	&	XTI	&	\nodata	&	1.2	&	2012130118*	&	2019-09-04	00:29:44	&	XTI	&	\nodata	&	0.66	\\
2020520102	&	2019-03-27	22:50:01	&	XTI	&	\nodata	&	0.43	&	2012130121*	&	2019-09-13	01:41:54	&	XTI	&	\nodata	&	0.34	\\
2020520103	&	2019-03-28	00:33:37	&	XTI	&	\nodata	&	6.9	&	2012130122*	&	2019-09-17	22:50:42	&	XTI	&	\nodata	&	0.26	\\
2614010101	&	2019-03-30	20:43:52	&	XTI	&	\nodata	&	0.85	&	2012130123*	&	2019-09-23	07:14:02	&	XTI	&	\nodata	&	0.50	\\
2614010102	&	2019-03-31	02:42:21	&	XTI	&	\nodata	&	6.2	&	2012130125*	&	2019-09-26	11:10:19	&	XTI	&	\nodata	&	0.044	\\
2614010103	&	2019-04-01	06:37:10	&	XTI	&	\nodata	&	9.4	&	2012130126*	&	2019-11-03	07:18:17	&	XTI	&	\nodata	&	1.1	\\
2614010104	&	2019-04-02	01:20:17	&	XTI	&	\nodata	&	5.4	&	2012130127*	&	2019-11-05	05:44:03	&	XTI	&	\nodata	&	2.9	\\
2614010105	&	2019-04-03	01:33:07	&	XTI	&	\nodata	&	5.5	&	2012130128*	&	2019-11-06	17:22:03	&	XTI	&	\nodata	&	1.3	\\
2614010106	&	2019-04-04	01:15:05	&	XTI	&	\nodata	&	1.2	&	2012130129*	&	2019-11-07	02:38:10	&	XTI	&	\nodata	&	0.47	\\
2614010107	&	2019-04-05	01:59:16	&	XTI	&	\nodata	&	11.3	&	3020520101	&	2020-05-22	07:25:04	&	XTI	&	\nodata	&	2.0	\\
2614010108	&	2019-04-06	00:39:18	&	XTI	&	\nodata	&	15.6	&	3012130101*	&	2020-06-02	22:46:06	&	XTI	&	\nodata	&	0.036	\\
2614010109	&	2019-04-07	00:17:31	&	XTI	&	\nodata	&	13.3	&	3012130103*	&	2020-06-04	02:36:11	&	XTI	&	\nodata	&	0.81	\\
2614010110	&	2019-04-08	02:20:30	&	XTI	&	\nodata	&	1.8	&	3012130104*	&	2020-06-05	03:19:32	&	XTI	&	\nodata	&	1.3	\\
2614010111	&	2019-04-08	23:57:28	&	XTI	&	\nodata	&	3.1	&	3012130105*	&	2020-06-09	10:43:21	&	XTI	&	\nodata	&	1.4	\\
2614010112	&	2019-04-10	00:41:29	&	XTI	&	\nodata	&	3.9	&	3012130106*	&	2020-06-10	01:09:07	&	XTI	&	\nodata	&	1.6	\\
2614010113	&	2019-04-11	02:59:38	&	XTI	&	\nodata	&	4.9	&	3012130107*	&	2020-06-13	06:10:54	&	XTI	&	\nodata	&	0.22	\\
2614010114	&	2019-04-13	12:11:57	&	XTI	&	\nodata	&	0.43	&	3012130108*	&	2020-06-15	14:06:52	&	XTI	&	\nodata	&	1.2	\\
2614010115	&	2019-04-14	12:55:20	&	XTI	&	\nodata	&	0.88	&	3012130109*	&	2020-06-16	05:26:08	&	XTI	&	\nodata	&	0.83	\\
2614010116	&	2019-04-16	17:19:00	&	XTI	&	\nodata	&	0.27	&	3012130110*	&	2020-06-18	06:49:28	&	XTI	&	\nodata	&	1.0	\\
2614010117	&	2019-04-19	10:10:22	&	XTI	&	\nodata	&	1.7	&	3020520103	&	2020-06-25	02:59:46	&	XTI	&	\nodata	&	0.43	\\
2614010118	&	2019-04-21	16:30:21	&	XTI	&	\nodata	&	2.3	&	3020520105	&	2020-06-28	00:47:10	&	XTI	&	\nodata	&	0.66	\\
2614010119	&	2019-04-22	00:16:38	&	XTI	&	\nodata	&	8.0	&	3020520107	&	2020-06-30	00:48:38	&	XTI	&	\nodata	&	3.1	\\
2614010120	&	2019-04-27	02:40:44	&	XTI	&	\nodata	&	2.2	&	3020520108	&	2020-07-01	01:37:37	&	XTI	&	\nodata	&	1.2	\\
2614010121	&	2019-04-28	01:52:06	&	XTI	&	\nodata	&	2.6	&	3020520109	&	2020-07-02	01:08:02	&	XTI	&	\nodata	&	3.3	\\
2614010122	&	2019-04-29	00:45:52	&	XTI	&	\nodata	&	1.3	&	3020520110	&	2020-07-03	03:32:24	&	XTI	&	\nodata	&	2.9	\\
2614010123	&	2019-05-05	11:05:12	&	XTI	&	\nodata	&	0.25	&	3020520111	&	2020-07-04	05:25:24	&	XTI	&	\nodata	&	2.5	\\
2614010124	&	2019-05-12	15:35:18	&	XTI	&	\nodata	&	0.51	&	3020520112	&	2020-07-05	17:08:39	&	XTI	&	\nodata	&	0.13	\\
2614010125	&	2019-05-13	00:50:22	&	XTI	&	\nodata	&	6.4	&	3020520114	&	2020-08-31	15:54:42	&	XTI	&	\nodata	&	0.24	\\
2614010126	&	2019-05-14	01:36:13	&	XTI	&	\nodata	&	1.3	&	4020520101	&	2021-06-08	05:46:28	&	XTI	&	\nodata	&	0.23	\\
2614010127	&	2019-05-17	19:26:09	&	XTI	&	\nodata	&	0.29	&	4020520102	&	2021-06-10	13:08:09	&	XTI	&	\nodata	&	1.3	\\
2614010128	&	2019-05-22	16:21:46	&	XTI	&	\nodata	&	1.8	&	5020520101	&	2022-04-24	11:13:02	&	XTI	&	\nodata	&	0.16	\\
2614010129	&	2019-05-23	03:06:32	&	XTI	&	\nodata	&	1.0	&	5020520103	&	2022-05-11	04:58:06	&	XTI	&	\nodata	&	0.44	\\
2614010130	&	2019-05-25	01:46:40	&	XTI	&	\nodata	&	2.3	&	5020520104	&	2022-05-20	16:08:35	&	XTI	&	\nodata	&	0.12	\\
2614010131	&	2019-05-26	01:02:03	&	XTI	&	\nodata	&	0.80	&	5020520106	&	2022-06-05	17:45:17	&	XTI	&	\nodata	&	0.17	\\
2614010132	&	2019-05-27	07:43:54	&	XTI	&	\nodata	&	2.8	&	5020520107	&	2022-06-07	10:01:23	&	XTI	&	\nodata	&	0.18	\\
2614010133	&	2019-05-28	00:45:10	&	XTI	&	\nodata	&	0.19	&	5020520108	&	2022-06-14	07:55:10	&	XTI	&	\nodata	&	0.28	\\
2614010134	&	2019-06-12	01:41:58	&	XTI	&	\nodata	&	0.45	&	5020520109	&	2022-06-23	06:57:18	&	XTI	&	\nodata	&	0.41	\\
2614010135	&	2019-06-14	06:23:16	&	XTI	&	\nodata	&	0.57	&	5020520110	&	2022-06-24	07:48:00	&	XTI	&	\nodata	&	0.14	\\
2614010136	&	2019-06-15	00:52:43	&	XTI	&	\nodata	&	9.4	&	5020520111	&	2022-06-26	10:57:38	&	XTI	&	\nodata	&	0.055	\\
2614010137	&	2019-06-16	04:48:05	&	XTI	&	\nodata	&	1.6	&	5020520112	&	2022-07-04	04:47:17	&	XTI	&	\nodata	&	0.21	\\
2614010138	&	2019-06-17	03:22:05	&	XTI	&	\nodata	&	10.2	&	5020520113	&	2022-07-23	23:10:22	&	XTI	&	\nodata	&	0.66	\\
2614010139	&	2019-06-18	01:18:55	&	XTI	&	\nodata	&	5.1	&	5020520114	&	2022-07-24	00:43:16	&	XTI	&	\nodata	&	4.7	\\
2012130101*	&	2019-06-27	03:10:45	&	XTI	&	\nodata	&	0.23	&	5020520115	&	2022-08-28	04:12:21	&	XTI	&	\nodata	&	0.46	\\
2012130102*	&	2019-06-28	05:28:28	&	XTI	&	\nodata	&	0.15	&	5020520117	&	2022-10-10	00:47:34	&	XTI	&	\nodata	&	0.18	\\
2012130103*	&	2019-07-01	23:37:45	&	XTI	&	\nodata	&	0.20	&		&			&		&		&		\\
\enddata
\tablenotetext{a}{Listed exposure times are values after filtering for instances of strong background flaring. The imaging mode abbreviations correspond to Full Window (FW), Large Window (LW), Small Window (SW), and Fast Timing (FT).}
\tablenotetext{b}{NICER observations marked with an asterisk were acquired with a $1\farcm 47$ pointing offset from RX J1856.5$-$3754 for instrument calibration purposes.}
\end{deluxetable}

\clearpage

\startlongtable
\begin{deluxetable}{lcclR|lcclR}
\tabletypesize{\tiny}
\tablewidth{0pt}
\tablecolumns{10}
\tablecaption{Log of X-ray observations of RX\,J2143.0+0654 \label{table:j2143_obs}}
\tablehead{\colhead{ObsID} & \colhead{Date/Start Time} & \colhead{Instrument} & \colhead{Mode/Filter} &  \colhead{$t_{\rm exp}$} &  \colhead{ObsID} & \colhead{Date/Start Time} & \colhead{Instrument} & \colhead{Mode/Filter} &  \colhead{$t_{\rm exp}$}  \\
            \colhead{}     & \colhead{(UTC)} & \colhead{}           &    \colhead{}         & \colhead{(ks)}    & \colhead{}     & \colhead{(UTC)} & \colhead{}           &    \colhead{}         & \colhead{(ks)}   } 
            \startdata
\multicolumn{10}{c}{\textit{XMM-Newton}\tablenotemark{a}} \\
\hline
0201150101	&	2004-05-31	15:33:53	&	pn	&	SW	/	Thin	&	19.4	&	0502041201	&	2007-11-08	03:47:33	&	pn	&	SW	/	Thin	&	6.6	\\
	&	2004-05-31	15:28:45	&	MOS1	&	SW	/	Thin	&	23.2	&		&	2007-11-08	03:42:25	&	MOS1	&	SW	/	Thin	&	9.3	\\
	&	2004-05-31	15:28:45	&	MOS2	&	SW	/	Thin	&	23.3	&		&	2007-11-08	03:42:23	&	MOS2	&	SW	/	Thin	&	9.3	\\
0502040601	&	2007-05-13	16:33:06	&	pn	&	SW	/	Thin	&	9.0	&	0502041301	&	2007-11-23	10:25:18	&	pn	&	SW	/	Thin	&	3.7	\\
	&	2007-05-13	16:27:57	&	MOS1	&	SW	/	Thin	&	12.5	&		&	2007-11-23	10:20:10	&	MOS1	&	SW	/	Thin	&	4.5	\\
	&	2007-05-13	16:27:56	&	MOS2	&	SW	/	Thin	&	12.6	&		&	2007-11-23	10:20:08	&	MOS2	&	SW	/	Thin	&	4.5	\\
0502040701	&	2007-05-17	21:02:54	&	pn	&	SW	/	Thin	&	9.1	&	0502041401	&	2007-12-10	00:47:48	&	pn	&	SW	/	Thin	&	5.4	\\
	&	2007-05-17	20:57:42	&	MOS1	&	SW	/	Thin	&	12.7	&		&	2007-12-10	00:42:40	&	MOS1	&	SW	/	Thin	&	7.6	\\
	&	2007-05-17	20:57:45	&	MOS2	&	SW	/	Thin	&	12.7	&		&	2007-12-10	00:42:39	&	MOS2	&	SW	/	Thin	&	7.6	\\
0502040801	&	2007-05-30	14:13:22	&	pn	&	SW	/	Thin	&	4.9	&	0502041501	&	2008-05-03	03:14:02	&	pn	&	SW	/	Thin	&	7.3	\\
	&	2007-05-30	14:08:14	&	MOS1	&	SW	/	Thin	&	6.0	&		&	2008-05-03	03:08:51	&	MOS1	&	SW	/	Thin	&	6.2	\\
	&	2007-05-30	14:08:12	&	MOS2	&	SW	/	Thin	&	6.2	&		&	2008-05-03	03:08:55	&	MOS2	&	SW	/	Thin	&	6.6	\\
0502040901	&	2007-06-12	20:50:17	&	pn	&	SW	/	Thin	&	5.8	&	0502041801	&	2008-05-19	04:17:53	&	pn	&	SW	/	Thin	&	5.4	\\
	&	2007-06-12	20:45:07	&	MOS1	&	SW	/	Thin	&	8.2	&		&	2008-05-19	04:12:43	&	MOS1	&	SW	/	Thin	&	7.6	\\
	&	2007-06-12	20:45:08	&	MOS2	&	SW	/	Thin	&	8.2	&		&	2008-05-19	04:12:43	&	MOS2	&	SW	/	Thin	&	7.6	\\
0502041001	&	2007-11-03	09:40:58	&	pn	&	SW	/	Thin	&	5.9	&	0844140201	&	2019-05-22	16:07:03	&	MOS1	&	SW	/	Thin	&	19.0	\\
	&	2007-11-03	09:35:49	&	MOS1	&	SW	/	Thin	&	8.4	&		&	2019-05-22	16:07:25	&	MOS2	&	SW	/	Thin	&	19.0	\\
	&	2007-11-03	09:35:48	&	MOS2	&	SW	/	Thin	&	8.4	&		&			&		&				&		\\
0502041101	&	2007-11-07	04:22:07	&	pn	&	SW	/	Thin	&	7.7	&		&			&		&				&		\\
	&	2007-11-07	04:17:00	&	MOS1	&	SW	/	Thin	&	10.9	&		&			&		&				&		\\
	&	2007-11-07	04:16:58	&	MOS2	&	SW	/	Thin	&	10.9	&		&			&		&				&		\\
 \hline
    \multicolumn{10}{c}{\textit{Chandra}\tablenotemark{b}} \\                
\hline
 11824     & 2009-12-24 05:41:01 &  ACIS-S &   CC &   12.1 & 20748     & 2018-09-12 2:51:17 &	HRC-S &	LETG  & 90.5  \\
 21669     & 2018-08-18 14:28:11 &	HRC-S &	LETG  &  15.2 & 21843     & 2018-09-14 18:11:37 &	HRC-S &	LETG  & 49.2  \\
 21670     & 2018-08-19 4:40:16 &	HRC-S &	LETG  &  20.2 & & & & & \\
\hline
    \multicolumn{10}{c}{\textit{NICER}} \\
\hline
5617010101	&	2022-07-26	22:09:00	&	XTI	&	\nodata	&	0.52	&	6713040101	&	2023-05-25	06:31:40	&	XTI	&	\nodata	&	9.5	\\
5617010102	&	2022-07-27	00:08:20	&	XTI	&	\nodata	&	13.3	&	6713040102	&	2023-05-26	00:48:20	&	XTI	&	\nodata	&	9.6	\\
5617010103	&	2022-07-28	00:35:40	&	XTI	&	\nodata	&	14.6	&	6713040103	&	2023-05-27	00:17:00	&	XTI	&	\nodata	&	7.2	\\
5617010104	&	2022-07-29	01:27:40	&	XTI	&	\nodata	&	14.5	&	6713040104	&	2023-05-28	00:53:20	&	XTI	&	\nodata	&	3.4	\\
5617010105	&	2022-07-30	00:22:07	&	XTI	&	\nodata	&	20.4	&	6713040105	&	2023-05-29	01:38:48	&	XTI	&	\nodata	&	7.0	\\
5617010106	&	2022-07-30	23:35:27	&	XTI	&	\nodata	&	7.1	&	6713040106	&	2023-05-30	01:03:40	&	XTI	&	\nodata	&	9.6	\\
5617010107	&	2022-08-01	05:05:53	&	XTI	&	\nodata	&	4.6	&	6713040201	&	2023-08-21	11:27:57	&	XTI	&	\nodata	&	4.7	\\
5617010201	&	2022-08-03	00:35:20	&	XTI	&	\nodata	&	18.5	&	6713040202	&	2023-08-21	23:59:00	&	XTI	&	\nodata	&	9.6	\\
5617010202	&	2022-08-04	01:07:48	&	XTI	&	\nodata	&	16.6	&	6713040203	&	2023-08-23	00:47:20	&	XTI	&	\nodata	&	3.9	\\
5617010203	&	2022-08-05	00:21:29	&	XTI	&	\nodata	&	10.9	&	6713040204	&	2023-08-24	00:05:40	&	XTI	&	\nodata	&	3.7	\\
5617010301	&	2022-09-22	13:08:00	&	XTI	&	\nodata	&	6.4	&	6713040205	&	2023-08-25	02:26:20	&	XTI	&	\nodata	&	6.4	\\
5617010302	&	2022-09-22	23:58:00	&	XTI	&	\nodata	&	12.8	&	6713040206	&	2023-08-26	00:15:40	&	XTI	&	\nodata	&	6.3	\\
5617010303	&	2022-09-24	00:43:07	&	XTI	&	\nodata	&	12.4	&	6713040207	&	2023-08-27	08:44:40	&	XTI	&	\nodata	&	6.0	\\
5617010304	&	2022-09-25	04:33:43	&	XTI	&	\nodata	&	11.8	&	6713040208	&	2023-08-28	00:14:20	&	XTI	&	\nodata	&	8.5	\\
5617010305	&	2022-09-26	00:41:14	&	XTI	&	\nodata	&	12.5	&		&			&		&		&		\\
\enddata
\tablenotetext{a}{Listed exposure times are values after filtering for instances of strong background flaring. The imaging mode abbreviation SW is for Small Window.}
\tablenotetext{b}{For the Chandra HRC-S LETG observation, only the zeroth order (undispersed) point source emission was extracted for the timing analysis.}
\end{deluxetable}

\clearpage

\bibliography{sample631}{}

\begin{thebibliography}{}
\expandafter\ifx\csname natexlab\endcsname\relax\def\natexlab#1{#1}\fi
\providecommand{\url}[1]{\href{#1}{#1}}
\providecommand{\dodoi}[1]{doi:~\href{http://doi.org/#1}{\nolinkurl{#1}}}
\providecommand{\doeprint}[1]{\href{http://ascl.net/#1}{\nolinkurl{http://ascl.net/#1}}}
\providecommand{\doarXiv}[1]{\href{https://arxiv.org/abs/#1}{\nolinkurl{https://arxiv.org/abs/#1}}}

\bibitem[{Abbott {et~al.}(2018)Abbott, Abbott, Abbott, Acernese, Ackley, Adams, Adams, Addesso, Adhikari, Adya, Affeldt, Agarwal, Agathos, Agatsuma, Aggarwal, Aguiar, Aiello, Ain, Ajith, Allen, Allen, Allocca, Aloy, Altin, Amato, Ananyeva, Anderson, Anderson, Angelova, Antier, Appert, Arai, Araya, Areeda, Ar\`ene, Arnaud, Arun, Ascenzi, Ashton, Ast, Aston, Astone, Atallah, Aubin, Aufmuth, Aulbert, AultONeal, Austin, Avila-Alvarez, Babak, Bacon, Badaracco, Bader, Bae, Baker, Baldaccini, Ballardin, Ballmer, Banagiri, Barayoga, Barclay, Barish, Barker, Barkett, Barnum, Barone, Barr, Barsotti, Barsuglia, Barta, Bartlett, Bartos, Bassiri, Basti, Batch, Bawaj, Bayley, Bazzan, B\'ecsy, Beer, Bejger, Belahcene, Bell, Beniwal, Bensch, Berger, Bergmann, Bernuzzi, Bero, Berry, Bersanetti, Bertolini, Betzwieser, Bhandare, Bilenko, Bilgili, Billingsley, Billman, Birch, Birney, Birnholtz, Biscans, Biscoveanu, Bisht, Bitossi, Bizouard, Blackburn, Blackman, Blair, Blair, Blair, Bloemen, Bock, Bode, Boer, Boetzel, Bogaert,
  Bohe, Bondu, Bonilla, Bonnand, Booker, Boom, Booth, Bork, Boschi, Bose, Bossie, Bossilkov, Bosveld, Bouffanais, Bozzi, Bradaschia, Brady, Bramley, Branchesi, Brau, Briant, Brighenti, Brillet, Brinkmann, Brisson, Brockill, Brooks, Brown, Brunett, Buchanan, Buikema, Bulik, Bulten, Buonanno, Buskulic, Buy, Byer, Cabero, Cadonati, Cagnoli, Cahillane, Calder\'on~Bustillo, Callister, Calloni, Camp, Canepa, Canizares, Cannon, Cao, Cao, Capano, Capocasa, Carbognani, Caride, Carney, Carullo, Casanueva~Diaz, Casentini, Caudill, Cavagli\`a, Cavalier, Cavalieri, Cella, Cepeda, Cerd\'a-Dur\'an, Cerretani, Cesarini, Chaibi, Chamberlin, Chan, Chao, Charlton, Chase, Chassande-Mottin, Chatterjee, Chatziioannou, Cheeseboro, Chen, Chen, Chen, Cheng, Chia, Chincarini, Chiummo, Chmiel, Cho, Cho, Chow, Christensen, Chu, Chua, Chua, Chung, Chung, Ciani, Ciobanu, Ciolfi, Cipriano, Cirelli, Cirone, Clara, Clark, Clearwater, Cleva, Cocchieri, Coccia, Cohadon, Cohen, Colla, Collette, Collins, Cominsky, Constancio, Conti, Cooper,
  Corban, Corbitt, Cordero-Carri\'on, Corley, Cornish, Corsi, Cortese, Costa, Cotesta, Coughlin, Coughlin, Coulon, Countryman, Couvares, Covas, Cowan, Coward, Cowart, Coyne, Coyne, Creighton, Creighton, Cripe, Crowder, Cullen, Cumming, Cunningham, Cuoco, Canton, D\'alya, Danilishin, D'Antonio, Danzmann, Dasgupta, Da~Silva~Costa, Dattilo, Dave, Davier, Davis, Daw, Day, DeBra, Deenadayalan, Degallaix, De~Laurentis, Del\'eglise, Del~Pozzo, Demos, Denker, Dent, De~Pietri, Derby, Dergachev, De~Rosa, De~Rossi, DeSalvo, de~Varona, Dhurandhar, D\'{\i}az, Dietrich, Di~Fiore, Di~Giovanni, Di~Girolamo, Di~Lieto, Ding, Di~Pace, Di~Palma, Di~Renzo, Dmitriev, Doctor, Dolique, Donovan, Dooley, Doravari, Dorrington, Dovale~\'Alvarez, Downes, Drago, Dreissigacker, Driggers, Du, Dupej, Dwyer, Easter, Edo, Edwards, Effler, Eggenstein, Ehrens, Eichholz, Eikenberry, Eisenmann, Eisenstein, Essick, Estelles, Estevez, Etienne, Etzel, Evans, Evans, Fafone, Fair, Fairhurst, Fan, Farinon, Farr, Farr, Fauchon-Jones, Favata, Fays, Fee,
  Fehrmann, Feicht, Fejer, Feng, Fernandez-Galiana, Ferrante, Ferreira, Ferrini, Fidecaro, Fiori, Fiorucci, Fishbach, Fisher, Fishner, Fitz-Axen, Flaminio, Fletcher, Fong, Font, Forsyth, Forsyth, Fournier, Frasca, Frasconi, Frei, Freise, Frey, Frey, Fritschel, Frolov, Fulda, Fyffe, Gabbard, Gadre, Gaebel, Gair, Gammaitoni, Ganija, Gaonkar, Garcia, Garc\'{\i}a-Quir\'os, Garufi, Gateley, Gaudio, Gaur, Gayathri, Gemme, Genin, Gennai, George, George, Gergely, Germain, Ghonge, Ghosh, Ghosh, Ghosh, Giacomazzo, Giaime, Giardina, Giazotto, Gill, Giordano, Glover, Goetz, Goetz, Goncharov, Gonz\'alez, Gonzalez~Castro, Gopakumar, Gorodetsky, Gossan, Gosselin, Gouaty, Grado, Graef, Granata, Grant, Gras, Gray, Greco, Green, Green, Gretarsson, Groot, Grote, Grunewald, Gruning, Guidi, Gulati, Guo, Gupta, Gupta, Gushwa, Gustafson, Gustafson, Halim, Hall, Hall, Hamilton, Hamilton, Hammond, Haney, Hanke, Hanks, Hanna, Hannam, Hannuksela, Hanson, Hardwick, Harms, Harry, Harry, Hart, Haster, Haughian, Healy, Heidmann, Heintze,
  Heitmann, Hello, Hemming, Hendry, Heng, Hennig, Heptonstall, Hernandez, Heurs, Hild, Hinderer, Ho, Hoak, Hochheim, Hofman, Holland, Holt, Holz, Hopkins, Horst, Hough, Houston, Howell, Hreibi, Huerta, Huet, Hughey, Hulko, Husa, Huttner, Huynh-Dinh, Iess, Indik, Ingram, Inta, Intini, Irwin, Isa, Isac, Isi, Iyer, Izumi, Jacqmin, Jani, Jaranowski, Johnson, Johnson, Jones, Jones, Jonker, Ju, Junker, Kalaghatgi, Kalogera, Kamai, Kandhasamy, Kang, Kanner, Kapadia, Karki, Karvinen, Kasprzack, Katolik, Katsanevas, Katsavounidis, Katzman, Kaufer, Kawabe, Keerthana, K\'ef\'elian, Keitel, Kemball, Kennedy, Key, Khalili, Khamesra, Khan, Khan, Khan, Khan, Khazanov, Kijbunchoo, Kim, Kim, Kim, Kim, Kim, Kim, King, King, Kinley-Hanlon, Kirchhoff, Kissel, Kleybolte, Klimenko, Knowles, Koch, Koehlenbeck, Koley, Kondrashov, Kontos, Korobko, Korth, Kowalska, Kozak, Kr\"amer, Kringel, Krishnan, Kr\'olak, Kuehn, Kumar, Kumar, Kumar, Kuo, Kutynia, Kwang, Lackey, Lai, Landry, Landry, Lang, Lange, Lantz, Lanza, Lartaux-Vollard,
  Lasky, Laxen, Lazzarini, Lazzaro, Leaci, Leavey, Lee, Lee, Lee, Lee, Lee, Lehmann, Lenon, Leonardi, Leroy, Letendre, Levin, Li, Li, Li, Linker, Littenberg, Liu, Liu, Lo, Lockerbie, London, Longo, Lorenzini, Loriette, Lormand, Losurdo, Lough, Lousto, Lovelace, L\"uck, Lumaca, Lundgren, Lynch, Ma, Macas, Macfoy, Machenschalk, MacInnis, Macleod, Maga\~na Hernandez, Maga\~na Sandoval, Maga\~na Zertuche, Magee, Majorana, Maksimovic, Man, Mandic, Mangano, Mansell, Manske, Mantovani, Marchesoni, Marion, M\'arka, M\'arka, Markakis, Markosyan, Markowitz, Maros, Marquina, Martelli, Martellini, Martin, Martin, Martynov, Mason, Massera, Masserot, Massinger, Masso-Reid, Mastrogiovanni, Matas, Matichard, Matone, Mavalvala, Mazumder, McCann, McCarthy, McClelland, McCormick, McCuller, McGuire, McIver, McManus, McRae, McWilliams, Meacher, Meadors, Mehmet, Meidam, Mejuto-Villa, Melatos, Mendell, Mendoza-Gandara, Mercer, Mereni, Merilh, Merzougui, Meshkov, Messenger, Messick, Metzdorff, Meyers, Miao, Michel, Middleton,
  Mikhailov, Milano, Miller, Miller, Miller, Miller, Millhouse, Mills, Milovich-Goff, Minazzoli, Minenkov, Ming, Mishra, Mitra, Mitrofanov, Mitselmakher, Mittleman, Moffa, Mogushi, Mohan, Mohapatra, Montani, Moore, Moraru, Moreno, Morisaki, Mours, Mow-Lowry, Mueller, Muir, Mukherjee, Mukherjee, Mukherjee, Mukund, Mullavey, Munch, Mu\~niz, Muratore, Murray, Nagar, Napier, Nardecchia, Naticchioni, Nayak, Neilson, Nelemans, Nelson, Nery, Neunzert, Nevin, Newport, Ng, Ng, Nguyen, Nguyen, Nichols, Nielsen, Nissanke, Nitz, Nocera, Nolting, North, Nuttall, Obergaulinger, Oberling, O'Brien, O'Dea, Ogin, Oh, Oh, Ohme, Ohta, Okada, Oliver, Oppermann, Oram, O'Reilly, Ormiston, Ortega, O'Shaughnessy, Ossokine, Ottaway, Overmier, Owen, Pace, Pagano, Page, Page, Pai, Pai, Palamos, Palashov, Palomba, Pal-Singh, Pan, Pan, Pang, Pang, Pankow, Pannarale, Pant, Paoletti, Paoli, Papa, Parida, Parker, Pascucci, Pasqualetti, Passaquieti, Passuello, Patil, Patricelli, Pearlstone, Pedersen, Pedraza, Pedurand, Pekowsky, Pele, Penn,
  Perego, Perez, Perreca, Perri, Pfeiffer, Phelps, Phukon, Piccinni, Pichot, Piergiovanni, Pierro, Pillant, Pinard, Pinto, Pirello, Pitkin, Poggiani, Popolizio, Porter, Possenti, Post, Powell, Prasad, Pratt, Pratten, Predoi, Prestegard, Principe, Privitera, Prodi, Prokhorov, Puncken, Punturo, Puppo, P\"urrer, Qi, Quetschke, Quintero, Quitzow-James, Raab, Rabeling, Radkins, Raffai, Raja, Rajan, Rajbhandari, Rakhmanov, Ramirez, Ramos-Buades, Rana, Rapagnani, Raymond, Razzano, Read, Regimbau, Rei, Reid, Reitze, Ren, Ricci, Ricker, Riemenschneider, Riles, Rizzo, Robertson, Robie, Robinet, Robson, Rocchi, Rolland, Rollins, Roma, Romano, Romel, Romie, Rosi\ifmmode~\acute{n}\else \'{n}\fi{}ska, Ross, Rowan, R\"udiger, Ruggi, Rutins, Ryan, Sachdev, Sadecki, Sakellariadou, Salconi, Saleem, Salemi, Samajdar, Sammut, Sampson, Sanchez, Sanchez, Sanchis-Gual, Sandberg, Sanders, Sarin, Sassolas, Sathyaprakash, Saulson, Sauter, Savage, Sawadsky, Schale, Scheel, Scheuer, Schmidt, Schnabel, Schofield, Sch\"onbeck, Schreiber,
  Schuette, Schulte, Schutz, Schwalbe, Scott, Scott, Seidel, Sellers, Sengupta, Sentenac, Sequino, Sergeev, Setyawati, Shaddock, Shaffer, Shah, Shahriar, Shaner, Shao, Shapiro, Shawhan, Shen, Shoemaker, Shoemaker, Siellez, Siemens, Sieniawska, Sigg, Silva, Singer, Singh, Singhal, Sintes, Slagmolen, Slaven-Blair, Smith, Smith, Smith, Somala, Son, Sorazu, Sorrentino, Souradeep, Spencer, Srivastava, Staats, Steinke, Steinlechner, Steinlechner, Steinmeyer, Steltner, Stevenson, Stocks, Stone, Stops, Strain, Stratta, Strigin, Strunk, Sturani, Stuver, Summerscales, Sun, Sunil, Suresh, Sutton, Swinkels, Szczepa\ifmmode~\acute{n}\else \'{n}\fi{}czyk, Tacca, Tait, Talbot, Talukder, Tanner, T\'apai, Taracchini, Tasson, Taylor, Taylor, Tewari, Theeg, Thies, Thomas, Thomas, Thomas, Thorne, Thrane, Tiwari, Tiwari, Tokmakov, Toland, Tonelli, Tornasi, Torres-Forn\'e, Torrie, T\"oyr\"a, Travasso, Traylor, Trinastic, Tringali, Trovato, Trozzo, Tsang, Tse, Tso, Tsuna, Tsukada, Tuyenbayev, Ueno, Ugolini, Urban, Usman, Vahlbruch,
  Vajente, Valdes, van Bakel, van Beuzekom, van~den Brand, Van Den~Broeck, Vander-Hyde, van~der Schaaf, van Heijningen, van Veggel, Vardaro, Varma, Vass, Vas\'uth, Vecchio, Vedovato, Veitch, Veitch, Venkateswara, Venugopalan, Verkindt, Vetrano, Vicer\'e, Viets, Vinciguerra, Vine, Vinet, Vitale, Vo, Vocca, Vorvick, Vyatchanin, Wade, Wade, Wade, Walet, Walker, Wallace, Walsh, Wang, Wang, Wang, Wang, Wang, Ward, Warner, Was, Watchi, Weaver, Wei, Weinert, Weinstein, Weiss, Wellmann, Wen, Wessel, We\ss{}els, Westerweck, Wette, Whelan, Whiting, Whittle, Wilken, Williams, Williams, Williamson, Willis, Willke, Wimmer, Winkler, Wipf, Wittel, Woan, Woehler, Wofford, Wong, Worden, Wright, Wu, Wysocki, Xiao, Yam, Yamamoto, Yancey, Yang, Yap, Yazback, Yu, Yu, Yvert, Zadro\ifmmode~\dot{z}\else \.{z}\fi{}ny, Zanolin, Zelenova, Zendri, Zevin, Zhang, Zhang, Zhang, Zhang, Zhang, Zhao, Zhou, Zhou, Zhu, Zhu, Zimmerman, Zlochower, Zucker, \& Zweizig}]{PhysRevLett.121.161101}
Abbott, B.~P., Abbott, R., Abbott, T.~D., {et~al.} 2018, Phys. Rev. Lett., 121, 161101, \dodoi{10.1103/PhysRevLett.121.161101}

\bibitem[{{AlGendy} \& {Morsink}(2014)}]{2014ApJ...791...78A}
{AlGendy}, M., \& {Morsink}, S.~M. 2014, \apj, 791, 78, \dodoi{10.1088/0004-637X/791/2/78}

\bibitem[{{Astropy Collaboration} {et~al.}(2022){Astropy Collaboration}, {Price-Whelan}, {Lim}, {Earl}, {Starkman}, {Bradley}, {Shupe}, {Patil}, {Corrales}, {Brasseur}, {N{\"o}the}, {Donath}, {Tollerud}, {Morris}, {Ginsburg}, {Vaher}, {Weaver}, {Tocknell}, {Jamieson}, {van Kerkwijk}, {Robitaille}, {Merry}, {Bachetti}, {G{\"u}nther}, {Aldcroft}, {Alvarado-Montes}, {Archibald}, {B{\'o}di}, {Bapat}, {Barentsen}, {Baz{\'a}n}, {Biswas}, {Boquien}, {Burke}, {Cara}, {Cara}, {Conroy}, {Conseil}, {Craig}, {Cross}, {Cruz}, {D'Eugenio}, {Dencheva}, {Devillepoix}, {Dietrich}, {Eigenbrot}, {Erben}, {Ferreira}, {Foreman-Mackey}, {Fox}, {Freij}, {Garg}, {Geda}, {Glattly}, {Gondhalekar}, {Gordon}, {Grant}, {Greenfield}, {Groener}, {Guest}, {Gurovich}, {Handberg}, {Hart}, {Hatfield-Dodds}, {Homeier}, {Hosseinzadeh}, {Jenness}, {Jones}, {Joseph}, {Kalmbach}, {Karamehmetoglu}, {Ka{\l}uszy{\'n}ski}, {Kelley}, {Kern}, {Kerzendorf}, {Koch}, {Kulumani}, {Lee}, {Ly}, {Ma}, {MacBride}, {Maljaars}, {Muna}, {Murphy}, {Norman},
  {O'Steen}, {Oman}, {Pacifici}, {Pascual}, {Pascual-Granado}, {Patil}, {Perren}, {Pickering}, {Rastogi}, {Roulston}, {Ryan}, {Rykoff}, {Sabater}, {Sakurikar}, {Salgado}, {Sanghi}, {Saunders}, {Savchenko}, {Schwardt}, {Seifert-Eckert}, {Shih}, {Jain}, {Shukla}, {Sick}, {Simpson}, {Singanamalla}, {Singer}, {Singhal}, {Sinha}, {Sip{\H{o}}cz}, {Spitler}, {Stansby}, {Streicher}, {{\v{S}}umak}, {Swinbank}, {Taranu}, {Tewary}, {Tremblay}, {de Val-Borro}, {Van Kooten}, {Vasovi{\'c}}, {Verma}, {de Miranda Cardoso}, {Williams}, {Wilson}, {Winkel}, {Wood-Vasey}, {Xue}, {Yoachim}, {Zhang}, {Zonca}, \& {Astropy Project Contributors}}]{2022ApJ...935..167A}
{Astropy Collaboration}, {Price-Whelan}, A.~M., {Lim}, P.~L., {et~al.} 2022, \apj, 935, 167, \dodoi{10.3847/1538-4357/ac7c74}

\bibitem[{{Blandford} \& {Romani}(1988)}]{1988MNRAS.234P..57B}
{Blandford}, R.~D., \& {Romani}, R.~W. 1988, \mnras, 234, 57P, \dodoi{10.1093/mnras/234.1.57P}

\bibitem[{{Bogdanov} {et~al.}(2019){Bogdanov}, {Guillot}, {Ray}, {Wolff}, {Chakrabarty}, {Ho}, {Kerr}, {Lamb}, {Lommen}, {Ludlam}, {Milburn}, {Montano}, {Miller}, {Baub{\"o}ck}, {{\"O}zel}, {Psaltis}, {Remillard}, {Riley}, {Steiner}, {Strohmayer}, {Watts}, {Wood}, {Zeldes}, {Enoto}, {Okajima}, {Kellogg}, {Baker}, {Markwardt}, {Arzoumanian}, \& {Gendreau}}]{2019ApJ...887L..25B}
{Bogdanov}, S., {Guillot}, S., {Ray}, P.~S., {et~al.} 2019, \apjl, 887, L25, \dodoi{10.3847/2041-8213/ab53eb}

\bibitem[{{Borghese} {et~al.}(2015){Borghese}, {Rea}, {Coti Zelati}, {Tiengo}, \& {Turolla}}]{2015ApJ...807L..20B}
{Borghese}, A., {Rea}, N., {Coti Zelati}, F., {Tiengo}, A., \& {Turolla}, R. 2015, \apjl, 807, L20, \dodoi{10.1088/2041-8205/807/1/L20}

\bibitem[{{Borghese} {et~al.}(2017){Borghese}, {Rea}, {Coti Zelati}, {Tiengo}, {Turolla}, \& {Zane}}]{2017MNRAS.468.2975B}
{Borghese}, A., {Rea}, N., {Coti Zelati}, F., {et~al.} 2017, \mnras, 468, 2975, \dodoi{10.1093/mnras/stx632}

\bibitem[{{Buschmann} {et~al.}(2021){Buschmann}, {Co}, {Dessert}, \& {Safdi}}]{2021PhRvL.126b1102B}
{Buschmann}, M., {Co}, R.~T., {Dessert}, C., \& {Safdi}, B.~R. 2021, \prl, 126, 021102, \dodoi{10.1103/PhysRevLett.126.021102}

\bibitem[{{Cropper} {et~al.}(2004){Cropper}, {Haberl}, {Zane}, \& {Zavlin}}]{2004MNRAS.351.1099C}
{Cropper}, M., {Haberl}, F., {Zane}, S., \& {Zavlin}, V.~E. 2004, \mnras, 351, 1099, \dodoi{10.1111/j.1365-2966.2004.07854.x}

\bibitem[{{De Grandis} {et~al.}(2021){De Grandis}, {Taverna}, {Turolla}, {Gnarini}, {Popov}, {Zane}, \& {Wood}}]{2021ApJ...914..118D}
{De Grandis}, D., {Taverna}, R., {Turolla}, R., {et~al.} 2021, \apj, 914, 118, \dodoi{10.3847/1538-4357/abfdac}

\bibitem[{{De Grandis} {et~al.}(2022){De Grandis}, {Rigoselli}, {Mereghetti}, {Younes}, {Pizzochero}, {Taverna}, {Tiengo}, {Turolla}, \& {Zane}}]{2022MNRAS.516.4932D}
{De Grandis}, D., {Rigoselli}, M., {Mereghetti}, S., {et~al.} 2022, \mnras, 516, 4932, \dodoi{10.1093/mnras/stac2587}

\bibitem[{{de Jager} \& {B{\"u}sching}(2010)}]{2010A&A...517L...9D}
{de Jager}, O.~C., \& {B{\"u}sching}, I. 2010, \aap, 517, L9, \dodoi{10.1051/0004-6361/201014362}

\bibitem[{{de Jager} {et~al.}(1989){de Jager}, {Raubenheimer}, \& {Swanepoel}}]{1989A&A...221..180D}
{de Jager}, O.~C., {Raubenheimer}, B.~C., \& {Swanepoel}, J.~W.~H. 1989, \aap, 221, 180

\bibitem[{{De Luca} {et~al.}(2005){De Luca}, {Caraveo}, {Mereghetti}, {Negroni}, \& {Bignami}}]{2005ApJ...623.1051D}
{De Luca}, A., {Caraveo}, P.~A., {Mereghetti}, S., {Negroni}, M., \& {Bignami}, G.~F. 2005, \apj, 623, 1051, \dodoi{10.1086/428567}

\bibitem[{{de Vries} {et~al.}(2004){de Vries}, {Vink}, {M{\'e}ndez}, \& {Verbunt}}]{2004A&A...415L..31D}
{de Vries}, C.~P., {Vink}, J., {M{\'e}ndez}, M., \& {Verbunt}, F. 2004, \aap, 415, L31, \dodoi{10.1051/0004-6361:20040009}

\bibitem[{{Dessert} {et~al.}(2020){Dessert}, {Foster}, \& {Safdi}}]{2020ApJ...904...42D}
{Dessert}, C., {Foster}, J.~W., \& {Safdi}, B.~R. 2020, \apj, 904, 42, \dodoi{10.3847/1538-4357/abb4ea}

\bibitem[{{FERMI-LAT Collaboration} {et~al.}(2022){FERMI-LAT Collaboration}, {Ajello}, {Atwood}, {Baldini}, {Ballet}, {Barbiellini}, {Bastieri}, {Bellazzini}, {Berretta}, {Bhattacharyya}, {Bissaldi}, {Blandford}, {Bloom}, {Bonino}, {Bruel}, {Buehler}, {Burns}, {Buson}, {Cameron}, {Caraveo}, {Cavazzuti}, {Cibrario}, {Ciprini}, {Clark}, {Cognard}, {Coronado-Bl{\'a}zquez}, {Crnogorcevic}, {Cromartie}, {Crowter}, {Cutini}, {D'Ammando}, {De Gaetano}, {de Palma}, {Digel}, {Di Lalla}, {Fana Dirirsa}, {Di Venere}, {Dom{\'\i}nguez}, {Ferrara}, {Fiori}, {Franckowiak}, {Fukazawa}, {Funk}, {Fusco}, {Gammaldi}, {Gargano}, {Gasparrini}, {Giglietto}, {Giordano}, {Giroletti}, {Green}, {Grenier}, {Guillemot}, {Guiriec}, {Gustafsson}, {Harding}, {Hays}, {Hewitt}, {Horan}, {Hou}, {J{\'o}hannesson}, {Keith}, {Kerr}, {Kramer}, {Kuss}, {Larsson}, {Latronico}, {Li}, {Longo}, {Loparco}, {Lovellette}, {Lubrano}, {Maldera}, {Manfreda}, {Mart{\'\i}-Devesa}, {Mazziotta}, {Mereu}, {Michelson}, {Mirabal}, {Mitthumsiri}, {Mizuno},
  {Monzani}, {Morselli}, {Negro}, {Nieder}, {Ojha}, {Omodei}, {Orienti}, {Orlando}, {Ormes}, {Paneque}, {Parthasarathy}, {Pei}, {Persic}, {Pesce-Rollins}, {Pillera}, {Poon}, {Porter}, {Principe}, {Racusin}, {Rain{\`o}}, {Rando}, {Rani}, {Ransom}, {Ray}, {Razzano}, {Razzaque}, {Reimer}, {Reimer}, {Roy}, {S{\'a}nchez-Conde}, {Saz Parkinson}, {Scargle}, {Scotton}, {Serini}, {Sgr{\`o}}, {Siskind}, {Smith}, {Spandre}, {Spiewak}, {Spinelli}, {Stairs}, {Suson}, {Swihart}, {Tabassum}, {Thayer}, {Theureau}, {Torres}, {Troja}, {Valverde}, {Wadiasingh}, {Wood}, \& {Zaharijas}}]{2022Sci...376..521F}
{FERMI-LAT Collaboration}, {Ajello}, M., {Atwood}, W.~B., {et~al.} 2022, Science, 376, 521, \dodoi{10.1126/science.abm3231}

\bibitem[{{Foreman-Mackey} {et~al.}(2013){Foreman-Mackey}, {Hogg}, {Lang}, \& {Goodman}}]{2013PASP..125..306F}
{Foreman-Mackey}, D., {Hogg}, D.~W., {Lang}, D., \& {Goodman}, J. 2013, \pasp, 125, 306, \dodoi{10.1086/670067}

\bibitem[{{Fruscione} {et~al.}(2006){Fruscione}, {McDowell}, {Allen}, {Brickhouse}, {Burke}, {Davis}, {Durham}, {Elvis}, {Galle}, {Harris}, {Huenemoerder}, {Houck}, {Ishibashi}, {Karovska}, {Nicastro}, {Noble}, {Nowak}, {Primini}, {Siemiginowska}, {Smith}, \& {Wise}}]{2006SPIE.6270E..1VF}
{Fruscione}, A., {McDowell}, J.~C., {Allen}, G.~E., {et~al.} 2006, in Society of Photo-Optical Instrumentation Engineers (SPIE) Conference Series, Vol. 6270, Society of Photo-Optical Instrumentation Engineers (SPIE) Conference Series, ed. D.~R. {Silva} \& R.~E. {Doxsey}, 62701V, \dodoi{10.1117/12.671760}

\bibitem[{{Gabriel} {et~al.}(2004){Gabriel}, {Denby}, {Fyfe}, {Hoar}, {Ibarra}, {Ojero}, {Osborne}, {Saxton}, {Lammers}, \& {Vacanti}}]{2004ASPC..314..759G}
{Gabriel}, C., {Denby}, M., {Fyfe}, D.~J., {et~al.} 2004, in Astronomical Society of the Pacific Conference Series, Vol. 314, Astronomical Data Analysis Software and Systems (ADASS) XIII, ed. F.~{Ochsenbein}, M.~G. {Allen}, \& D.~{Egret}, 759

\bibitem[{{Gaensler}(2005)}]{2005AdSpR..35.1116G}
{Gaensler}, B.~M. 2005, Advances in Space Research, 35, 1116, \dodoi{10.1016/j.asr.2005.01.026}

\bibitem[{{Gendreau} {et~al.}(2016){Gendreau}, {Arzoumanian}, {Adkins}, {Albert}, {Anders}, {Aylward}, {Baker}, {Balsamo}, {Bamford}, {Benegalrao}, {Berry}, {Bhalwani}, {Black}, {Blaurock}, {Bronke}, {Brown}, {Budinoff}, {Cantwell}, {Cazeau}, {Chen}, {Clement}, {Colangelo}, {Coleman}, {Coopersmith}, {Dehaven}, {Doty}, {Egan}, {Enoto}, {Fan}, {Ferro}, {Foster}, {Galassi}, {Gallo}, {Green}, {Grosh}, {Ha}, {Hasouneh}, {Heefner}, {Hestnes}, {Hoge}, {Jacobs}, {J{\o}rgensen}, {Kaiser}, {Kellogg}, {Kenyon}, {Koenecke}, {Kozon}, {LaMarr}, {Lambertson}, {Larson}, {Lentine}, {Lewis}, {Lilly}, {Liu}, {Malonis}, {Manthripragada}, {Markwardt}, {Matonak}, {Mcginnis}, {Miller}, {Mitchell}, {Mitchell}, {Mohammed}, {Monroe}, {Montt de Garcia}, {Mul{\'e}}, {Nagao}, {Ngo}, {Norris}, {Norwood}, {Novotka}, {Okajima}, {Olsen}, {Onyeachu}, {Orosco}, {Peterson}, {Pevear}, {Pham}, {Pollard}, {Pope}, {Powers}, {Powers}, {Price}, {Prigozhin}, {Ramirez}, {Reid}, {Remillard}, {Rogstad}, {Rosecrans}, {Rowe}, {Sager}, {Sanders},
  {Savadkin}, {Saylor}, {Schaeffer}, {Schweiss}, {Semper}, {Serlemitsos}, {Shackelford}, {Soong}, {Struebel}, {Vezie}, {Villasenor}, {Winternitz}, {Wofford}, {Wright}, {Yang}, \& {Yu}}]{2016SPIE.9905E..1HG}
{Gendreau}, K.~C., {Arzoumanian}, Z., {Adkins}, P.~W., {et~al.} 2016, in Society of Photo-Optical Instrumentation Engineers (SPIE) Conference Series, Vol. 9905, Space Telescopes and Instrumentation 2016: Ultraviolet to Gamma Ray, ed. J.-W.~A. {den Herder}, T.~{Takahashi}, \& M.~{Bautz}, 99051H, \dodoi{10.1117/12.2231304}

\bibitem[{{Gotthelf} \& {Halpern}(2009)}]{2009ApJ...695L..35G}
{Gotthelf}, E.~V., \& {Halpern}, J.~P. 2009, \apjl, 695, L35, \dodoi{10.1088/0004-637X/695/1/L35}

\bibitem[{{Gunn} \& {Ostriker}(1969)}]{1969Natur.221..454G}
{Gunn}, J.~E., \& {Ostriker}, J.~P. 1969, \nat, 221, 454, \dodoi{10.1038/221454a0}

\bibitem[{{Haberl}(2004)}]{2004AdSpR..33..638H}
{Haberl}, F. 2004, Advances in Space Research, 33, 638, \dodoi{10.1016/j.asr.2003.07.022}

\bibitem[{{Haberl}(2007)}]{2007Ap&SS.308..181H}
---. 2007, \apss, 308, 181, \dodoi{10.1007/s10509-007-9342-x}

\bibitem[{{Haberl} {et~al.}(1997){Haberl}, {Motch}, {Buckley}, {Zickgraf}, \& {Pietsch}}]{1997A&A...326..662H}
{Haberl}, F., {Motch}, C., {Buckley}, D.~A.~H., {Zickgraf}, F.~J., \& {Pietsch}, W. 1997, \aap, 326, 662

\bibitem[{{Haberl} {et~al.}(1998){Haberl}, {Motch}, \& {Pietsch}}]{1998AN....319...97H}
{Haberl}, F., {Motch}, C., \& {Pietsch}, W. 1998, Astronomische Nachrichten, 319, 97, \dodoi{10.1002/asna.2123190145}

\bibitem[{{Haberl} {et~al.}(1999){Haberl}, {Pietsch}, \& {Motch}}]{1999A&A...351L..53H}
{Haberl}, F., {Pietsch}, W., \& {Motch}, C. 1999, \aap, 351, L53, \dodoi{10.48550/arXiv.astro-ph/9911159}

\bibitem[{{Haberl} {et~al.}(2006){Haberl}, {Turolla}, {de Vries}, {Zane}, {Vink}, {M{\'e}ndez}, \& {Verbunt}}]{2006A&A...451L..17H}
{Haberl}, F., {Turolla}, R., {de Vries}, C.~P., {et~al.} 2006, \aap, 451, L17, \dodoi{10.1051/0004-6361:20065093}

\bibitem[{{Haberl} \& {Zavlin}(2002)}]{2002A&A...391..571H}
{Haberl}, F., \& {Zavlin}, V.~E. 2002, \aap, 391, 571, \dodoi{10.1051/0004-6361:20020778}

\bibitem[{{Haberl} {et~al.}(2004){Haberl}, {Motch}, {Zavlin}, {Reinsch}, {G{\"a}nsicke}, {Cropper}, {Schwope}, {Turolla}, \& {Zane}}]{2004AandA...424..635H}
{Haberl}, F., {Motch}, C., {Zavlin}, V.~E., {et~al.} 2004, \aap, 424, 635, \dodoi{10.1051/0004-6361:20040440}

\bibitem[{{Hambaryan} {et~al.}(2002){Hambaryan}, {Hasinger}, {Schwope}, \& {Schulz}}]{2002A&A...381...98H}
{Hambaryan}, V., {Hasinger}, G., {Schwope}, A.~D., \& {Schulz}, N.~S. 2002, \aap, 381, 98, \dodoi{10.1051/0004-6361:20011425}

\bibitem[{{Hambaryan} {et~al.}(2017){Hambaryan}, {Suleimanov}, {Haberl}, {Schwope}, {Neuh{\"a}user}, {Hohle}, \& {Werner}}]{2017A&A...601A.108H}
{Hambaryan}, V., {Suleimanov}, V., {Haberl}, F., {et~al.} 2017, \aap, 601, A108, \dodoi{10.1051/0004-6361/201630368}

\bibitem[{{Hambaryan} {et~al.}(2011){Hambaryan}, {Suleimanov}, {Schwope}, {Neuh{\"a}user}, {Werner}, \& {Potekhin}}]{2011A&A...534A..74H}
{Hambaryan}, V., {Suleimanov}, V., {Schwope}, A.~D., {et~al.} 2011, \aap, 534, A74, \dodoi{10.1051/0004-6361/201117548}

\bibitem[{{Harding} {et~al.}(1999){Harding}, {Contopoulos}, \& {Kazanas}}]{1999ApJ...525L.125H}
{Harding}, A.~K., {Contopoulos}, I., \& {Kazanas}, D. 1999, \apjl, 525, L125, \dodoi{10.1086/312339}

\bibitem[{{Ho}(2007)}]{2007MNRAS.380...71H}
{Ho}, W. C.~G. 2007, \mnras, 380, 71, \dodoi{10.1111/j.1365-2966.2007.12043.x}

\bibitem[{{Ho}(2015)}]{2015MNRAS.452..845H}
---. 2015, \mnras, 452, 845, \dodoi{10.1093/mnras/stv1339}

\bibitem[{{Ho} {et~al.}(2007){Ho}, {Kaplan}, {Chang}, {van Adelsberg}, \& {Potekhin}}]{2007MNRAS.375..821H}
{Ho}, W. C.~G., {Kaplan}, D.~L., {Chang}, P., {van Adelsberg}, M., \& {Potekhin}, A.~Y. 2007, \mnras, 375, 821, \dodoi{10.1111/j.1365-2966.2006.11376.x}

\bibitem[{{Ho} {et~al.}(2008){Ho}, {Potekhin}, \& {Chabrier}}]{2008ApJS..178..102H}
{Ho}, W. C.~G., {Potekhin}, A.~Y., \& {Chabrier}, G. 2008, \apjs, 178, 102, \dodoi{10.1086/589238}

\bibitem[{{Hobbs} {et~al.}(2006){Hobbs}, {Edwards}, \& {Manchester}}]{2006MNRAS.369..655H}
{Hobbs}, G.~B., {Edwards}, R.~T., \& {Manchester}, R.~N. 2006, \mnras, 369, 655, \dodoi{10.1111/j.1365-2966.2006.10302.x}

\bibitem[{{Hohle} {et~al.}(2012){Hohle}, {Haberl}, {Vink}, {de Vries}, {Turolla}, {Zane}, \& {M{\'e}ndez}}]{2012MNRAS.423.1194H}
{Hohle}, M.~M., {Haberl}, F., {Vink}, J., {et~al.} 2012, \mnras, 423, 1194, \dodoi{10.1111/j.1365-2966.2012.20946.x}

\bibitem[{{Hohle} {et~al.}(2009){Hohle}, {Haberl}, {Vink}, {Turolla}, {Hambaryan}, {Zane}, {de Vries}, \& {M{\'e}ndez}}]{2009A&A...498..811H}
---. 2009, \aap, 498, 811, \dodoi{10.1051/0004-6361/200810812}

\bibitem[{{Hohle} {et~al.}(2010){Hohle}, {Haberl}, {Vink}, {Turolla}, {Zane}, {de Vries}, \& {M{\'e}ndez}}]{2010A&A...521A..11H}
---. 2010, \aap, 521, A11, \dodoi{10.1051/0004-6361/200913661}

\bibitem[{Hunter(2007)}]{Hunter:2007}
Hunter, J.~D. 2007, Computing in Science \& Engineering, 9, 90, \dodoi{10.1109/MCSE.2007.55}

\bibitem[{{Kaplan} {et~al.}(2011){Kaplan}, {Kamble}, {van Kerkwijk}, \& {Ho}}]{2011ApJ...736..117K}
{Kaplan}, D.~L., {Kamble}, A., {van Kerkwijk}, M.~H., \& {Ho}, W.~C.~G. 2011, \apj, 736, 117, \dodoi{10.1088/0004-637X/736/2/117}

\bibitem[{{Kaplan} {et~al.}(2002{\natexlab{a}}){Kaplan}, {Kulkarni}, \& {van Kerkwijk}}]{2002ApJ...579L..29K}
{Kaplan}, D.~L., {Kulkarni}, S.~R., \& {van Kerkwijk}, M.~H. 2002{\natexlab{a}}, \apjl, 579, L29, \dodoi{10.1086/344923}

\bibitem[{{Kaplan} {et~al.}(2002{\natexlab{b}}){Kaplan}, {Kulkarni}, {van Kerkwijk}, \& {Marshall}}]{2002ApJ...570L..79K}
{Kaplan}, D.~L., {Kulkarni}, S.~R., {van Kerkwijk}, M.~H., \& {Marshall}, H.~L. 2002{\natexlab{b}}, \apjl, 570, L79, \dodoi{10.1086/341102}

\bibitem[{{Kaplan} \& {van Kerkwijk}(2005{\natexlab{a}})}]{2005ApJ...628L..45K}
{Kaplan}, D.~L., \& {van Kerkwijk}, M.~H. 2005{\natexlab{a}}, \apjl, 628, L45, \dodoi{10.1086/432536}

\bibitem[{{Kaplan} \& {van Kerkwijk}(2005{\natexlab{b}})}]{2005ApJ...635L..65K}
---. 2005{\natexlab{b}}, \apjl, 635, L65, \dodoi{10.1086/499241}

\bibitem[{{Kaplan} \& {van Kerkwijk}(2009{\natexlab{a}})}]{2009ApJ...705..798K}
---. 2009{\natexlab{a}}, \apj, 705, 798, \dodoi{10.1088/0004-637X/705/1/798}

\bibitem[{{Kaplan} \& {van Kerkwijk}(2009{\natexlab{b}})}]{2009ApJ...692L..62K}
---. 2009{\natexlab{b}}, \apjl, 692, L62, \dodoi{10.1088/0004-637X/692/1/L62}

\bibitem[{{Kaplan} \& {van Kerkwijk}(2011)}]{2011ApJ...740L..30K}
---. 2011, \apjl, 740, L30, \dodoi{10.1088/2041-8205/740/1/L30}

\bibitem[{{Kaplan} {et~al.}(2007){Kaplan}, {van Kerkwijk}, \& {Anderson}}]{2007ApJ...660.1428K}
{Kaplan}, D.~L., {van Kerkwijk}, M.~H., \& {Anderson}, J. 2007, \apj, 660, 1428, \dodoi{10.1086/513309}

\bibitem[{{Kargaltsev} {et~al.}(2015){Kargaltsev}, {Cerutti}, {Lyubarsky}, \& {Striani}}]{2015SSRv..191..391K}
{Kargaltsev}, O., {Cerutti}, B., {Lyubarsky}, Y., \& {Striani}, E. 2015, \ssr, 191, 391, \dodoi{10.1007/s11214-015-0171-x}

\bibitem[{{Keane} \& {Kramer}(2008)}]{2008MNRAS.391.2009K}
{Keane}, E.~F., \& {Kramer}, M. 2008, \mnras, 391, 2009, \dodoi{10.1111/j.1365-2966.2008.14045.x}

\bibitem[{{Kondratiev} {et~al.}(2009){Kondratiev}, {McLaughlin}, {Lorimer}, {Burgay}, {Possenti}, {Turolla}, {Popov}, \& {Zane}}]{2009ApJ...702..692K}
{Kondratiev}, V.~I., {McLaughlin}, M.~A., {Lorimer}, D.~R., {et~al.} 2009, \apj, 702, 692, \dodoi{10.1088/0004-637X/702/1/692}

\bibitem[{{Kurpas} {et~al.}(2024){Kurpas}, {Schwope}, {Pires}, \& {Haberl}}]{2024arXiv240117290K}
{Kurpas}, J., {Schwope}, A.~D., {Pires}, A.~M., \& {Haberl}, F. 2024, arXiv e-prints, arXiv:2401.17290, \dodoi{10.48550/arXiv.2401.17290}

\bibitem[{{Kurpas} {et~al.}(2023){Kurpas}, {Schwope}, {Pires}, {Haberl}, \& {Buckley}}]{2023A&A...674A.155K}
{Kurpas}, J., {Schwope}, A.~D., {Pires}, A.~M., {Haberl}, F., \& {Buckley}, D.~A.~H. 2023, \aap, 674, A155, \dodoi{10.1051/0004-6361/202346375}

\bibitem[{{Luo} {et~al.}(2021){Luo}, {Ransom}, {Demorest}, {Ray}, {Archibald}, {Kerr}, {Jennings}, {Bachetti}, {van Haasteren}, {Champagne}, {Colen}, {Phillips}, {Zimmerman}, {Stovall}, {Lam}, \& {Jenet}}]{2021ApJ...911...45L}
{Luo}, J., {Ransom}, S., {Demorest}, P., {et~al.} 2021, \apj, 911, 45, \dodoi{10.3847/1538-4357/abe62f}

\bibitem[{{Malacaria} {et~al.}(2019){Malacaria}, {Bogdanov}, {Ho}, {Enoto}, {Ray}, {Arzoumanian}, {Cazeau}, {Gendreau}, {Guillot}, {G{\"u}ver}, {Jaisawal}, {Wolff}, {NICER Magnetar}, \& {Magnetospheres Team}}]{2019ApJ...880...74M}
{Malacaria}, C., {Bogdanov}, S., {Ho}, W. C.~G., {et~al.} 2019, \apj, 880, 74, \dodoi{10.3847/1538-4357/ab2875}

\bibitem[{{Manchester} {et~al.}(2005){Manchester}, {Hobbs}, {Teoh}, \& {Hobbs}}]{2005AJ....129.1993M}
{Manchester}, R.~N., {Hobbs}, G.~B., {Teoh}, A., \& {Hobbs}, M. 2005, \aj, 129, 1993, \dodoi{10.1086/428488}

\bibitem[{{Melatos}(1997)}]{1997MNRAS.288.1049M}
{Melatos}, A. 1997, \mnras, 288, 1049, \dodoi{10.1093/mnras/288.4.1049}

\bibitem[{{Meszaros}(1992)}]{1992herm.book.....M}
{Meszaros}, P. 1992, {High-energy radiation from magnetized neutron stars}

\bibitem[{{Miller} {et~al.}(2021){Miller}, {Lamb}, {Dittmann}, {Bogdanov}, {Arzoumanian}, {Gendreau}, {Guillot}, {Ho}, {Lattimer}, {Loewenstein}, {Morsink}, {Ray}, {Wolff}, {Baker}, {Cazeau}, {Manthripragada}, {Markwardt}, {Okajima}, {Pollard}, {Cognard}, {Cromartie}, {Fonseca}, {Guillemot}, {Kerr}, {Parthasarathy}, {Pennucci}, {Ransom}, \& {Stairs}}]{2021ApJ...918L..28M}
{Miller}, M.~C., {Lamb}, F.~K., {Dittmann}, A.~J., {et~al.} 2021, \apjl, 918, L28, \dodoi{10.3847/2041-8213/ac089b}

\bibitem[{{Mori} \& {Ho}(2007)}]{2007MNRAS.377..905M}
{Mori}, K., \& {Ho}, W. C.~G. 2007, \mnras, 377, 905, \dodoi{10.1111/j.1365-2966.2007.11663.x}

\bibitem[{{Nasa High Energy Astrophysics Science Archive Research Center (Heasarc)}(2014)}]{2014ascl.soft08004N}
{Nasa High Energy Astrophysics Science Archive Research Center (Heasarc)}. 2014, {HEAsoft: Unified Release of FTOOLS and XANADU}, Astrophysics Source Code Library, record ascl:1408.004.
\newblock \doeprint{1408.004}

\bibitem[{{Nishimura} {et~al.}(1986){Nishimura}, {Mitsuda}, \& {Itoh}}]{1986PASJ...38..819N}
{Nishimura}, J., {Mitsuda}, K., \& {Itoh}, M. 1986, \pasj, 38, 819

\bibitem[{{Parthasarathy} {et~al.}(2020){Parthasarathy}, {Johnston}, {Shannon}, {Lentati}, {Bailes}, {Dai}, {Kerr}, {Manchester}, {Os{\l}owski}, {Sobey}, {van Straten}, \& {Weltevrede}}]{2020MNRAS.494.2012P}
{Parthasarathy}, A., {Johnston}, S., {Shannon}, R.~M., {et~al.} 2020, \mnras, 494, 2012, \dodoi{10.1093/mnras/staa882}

\bibitem[{{Pavlov} {et~al.}(1994){Pavlov}, {Shibanov}, {Ventura}, \& {Zavlin}}]{1994A&A...289..837P}
{Pavlov}, G.~G., {Shibanov}, Y.~A., {Ventura}, J., \& {Zavlin}, V.~E. 1994, \aap, 289, 837

\bibitem[{{Perna} {et~al.}(2013){Perna}, {Vigan{\`o}}, {Pons}, \& {Rea}}]{2013MNRAS.434.2362P}
{Perna}, R., {Vigan{\`o}}, D., {Pons}, J.~A., \& {Rea}, N. 2013, \mnras, 434, 2362, \dodoi{10.1093/mnras/stt1181}

\bibitem[{{Pires} {et~al.}(2014){Pires}, {Haberl}, {Zavlin}, {Motch}, {Zane}, \& {Hohle}}]{2014A&A...563A..50P}
{Pires}, A.~M., {Haberl}, F., {Zavlin}, V.~E., {et~al.} 2014, \aap, 563, A50, \dodoi{10.1051/0004-6361/201423380}

\bibitem[{{Pires} {et~al.}(2019){Pires}, {Schwope}, {Haberl}, {Zavlin}, {Motch}, \& {Zane}}]{2019A&A...623A..73P}
{Pires}, A.~M., {Schwope}, A.~D., {Haberl}, F., {et~al.} 2019, \aap, 623, A73, \dodoi{10.1051/0004-6361/201834801}

\bibitem[{{Posselt} {et~al.}(2024){Posselt}, {Pavlov}, {Ho}, \& {Haberl}}]{Posselt24}
{Posselt}, B., {Pavlov}, G.~G., {Ho}, W.~C.~G., \& {Haberl}, F. 2024, \apj, submitted

\bibitem[{{Potekhin}(2014)}]{2014PhyU...57..735P}
{Potekhin}, A.~Y. 2014, Physics Uspekhi, 57, 735, \dodoi{10.3367/UFNe.0184.201408a.0793}

\bibitem[{{Riley} {et~al.}(2021){Riley}, {Watts}, {Ray}, {Bogdanov}, {Guillot}, {Morsink}, {Bilous}, {Arzoumanian}, {Choudhury}, {Deneva}, {Gendreau}, {Harding}, {Ho}, {Lattimer}, {Loewenstein}, {Ludlam}, {Markwardt}, {Okajima}, {Prescod-Weinstein}, {Remillard}, {Wolff}, {Fonseca}, {Cromartie}, {Kerr}, {Pennucci}, {Parthasarathy}, {Ransom}, {Stairs}, {Guillemot}, \& {Cognard}}]{2021ApJ...918L..27R}
{Riley}, T.~E., {Watts}, A.~L., {Ray}, P.~S., {et~al.} 2021, \apjl, 918, L27, \dodoi{10.3847/2041-8213/ac0a81}

\bibitem[{{Sartore} {et~al.}(2012){Sartore}, {Tiengo}, {Mereghetti}, {De Luca}, {Turolla}, \& {Haberl}}]{2012A&A...541A..66S}
{Sartore}, N., {Tiengo}, A., {Mereghetti}, S., {et~al.} 2012, \aap, 541, A66, \dodoi{10.1051/0004-6361/201118489}

\bibitem[{{Schwope} {et~al.}(2005){Schwope}, {Hambaryan}, {Haberl}, \& {Motch}}]{2005A&A...441..597S}
{Schwope}, A.~D., {Hambaryan}, V., {Haberl}, F., \& {Motch}, C. 2005, \aap, 441, 597, \dodoi{10.1051/0004-6361:20053125}

\bibitem[{{Schwope} {et~al.}(1999){Schwope}, {Hasinger}, {Schwarz}, {Haberl}, \& {Schmidt}}]{1999A&A...341L..51S}
{Schwope}, A.~D., {Hasinger}, G., {Schwarz}, R., {Haberl}, F., \& {Schmidt}, M. 1999, \aap, 341, L51, \dodoi{10.48550/arXiv.astro-ph/9811326}

\bibitem[{{Schwope} {et~al.}(2009){Schwope}, {Erben}, {Kohnert}, {Lamer}, {Steinmetz}, {Strassmeier}, {Zinnecker}, {Bechtold}, {Diolaiti}, {Fontana}, {Gallozzi}, {Giallongo}, {Ragazzoni}, {de Santis}, \& {Testa}}]{2009AandA...499..267S}
{Schwope}, A.~D., {Erben}, T., {Kohnert}, J., {et~al.} 2009, \aap, 499, 267, \dodoi{10.1051/0004-6361/200811041}

\bibitem[{Smith {et~al.}(2023)Smith, Abdollahi, Ajello, Bailes, Baldini, Ballet, Baring, Bassa, Gonzalez, Bellazzini, Berretta, Bhattacharyya, Bissaldi, Bonino, Bottacini, Bregeon, Bruel, Burgay, Burnett, Cameron, Camilo, Caputo, Caraveo, Cavazzuti, Chiaro, Ciprini, Clark, Cognard, Corongiu, Orestano, Crnogorcevic, Cuoco, Cutini, D’Ammando, de~Angelis, DeCesar, Gaetano, de~Menezes, Deneva, de~Palma, Lalla, Dirirsa, Venere, Domínguez, Dumora, Fegan, Ferrara, Fiori, Fleischhack, Flynn, Franckowiak, Freire, Fukazawa, Fusco, Galanti, Gammaldi, Gargano, Gasparrini, Giacchino, Giglietto, Giordano, Giroletti, Green, Grenier, Guillemot, Guiriec, Gustafsson, Harding, Hays, Hewitt, Horan, Hou, Jankowski, Johnson, Johnson, Johnston, Kataoka, Keith, Kerr, Kramer, Kuss, Latronico, Lee, Li, Li, Limyansky, Longo, Loparco, Lorusso, Lovellette, Lower, Lubrano, Lyne, Maan, Maldera, Manchester, Manfreda, Marelli, Martí-Devesa, Mazziotta, McEnery, Mereu, Michelson, Mickaliger, Mitthumsiri, Mizuno, Moiseev, Monzani, Morselli,
  Negro, Nemmen, Nieder, Nuss, Omodei, Orienti, Orlando, Ormes, Palatiello, Paneque, Panzarini, Parthasarathy, Persic, Pesce-Rollins, Pillera, Poon, Porter, Possenti, Principe, Rainò, Rando, Ransom, Ray, Razzano, Razzaque, Reimer, Reimer, Renault-Tinacci, Romani, Sánchez-Conde, Parkinson, Scotton, Serini, Sgrò, Shannon, Sharma, Shen, Siskind, Spandre, Spinelli, Stappers, Stephens, Suson, Tabassum, Tajima, Tak, Theureau, Thompson, Tibolla, Torres, Valverde, Venter, Wadiasingh, Wang, Wang, Wang, Weltevrede, Wood, Yan, Zaharijas, Zhang, \& Zhu}]{Smith_2023}
Smith, D.~A., Abdollahi, S., Ajello, M., {et~al.} 2023, The Astrophysical Journal, 958, 191, \dodoi{10.3847/1538-4357/acee67}

\bibitem[{{Str{\"u}der} {et~al.}(2001){Str{\"u}der}, {Briel}, {Dennerl}, {Hartmann}, {Kendziorra}, {Meidinger}, {Pfeffermann}, {Reppin}, {Aschenbach}, {Bornemann}, {Br{\"a}uninger}, {Burkert}, {Elender}, {Freyberg}, {Haberl}, {Hartner}, {Heuschmann}, {Hippmann}, {Kastelic}, {Kemmer}, {Kettenring}, {Kink}, {Krause}, {M{\"u}ller}, {Oppitz}, {Pietsch}, {Popp}, {Predehl}, {Read}, {Stephan}, {St{\"o}tter}, {Tr{\"u}mper}, {Holl}, {Kemmer}, {Soltau}, {St{\"o}tter}, {Weber}, {Weichert}, {von Zanthier}, {Carathanassis}, {Lutz}, {Richter}, {Solc}, {B{\"o}ttcher}, {Kuster}, {Staubert}, {Abbey}, {Holland}, {Turner}, {Balasini}, {Bignami}, {La Palombara}, {Villa}, {Buttler}, {Gianini}, {Lain{\'e}}, {Lumb}, \& {Dhez}}]{2001A&A...365L..18S}
{Str{\"u}der}, L., {Briel}, U., {Dennerl}, K., {et~al.} 2001, \aap, 365, L18, \dodoi{10.1051/0004-6361:20000066}

\bibitem[{{Suleimanov} {et~al.}(2012){Suleimanov}, {Pavlov}, \& {Werner}}]{2012ApJ...751...15S}
{Suleimanov}, V.~F., {Pavlov}, G.~G., \& {Werner}, K. 2012, \apj, 751, 15, \dodoi{10.1088/0004-637X/751/1/15}

\bibitem[{{Tiengo} \& {Mereghetti}(2007)}]{2007ApJ...657L.101T}
{Tiengo}, A., \& {Mereghetti}, S. 2007, \apjl, 657, L101, \dodoi{10.1086/513143}

\bibitem[{{Turner} {et~al.}(2001){Turner}, {Abbey}, {Arnaud}, {Balasini}, {Barbera}, {Belsole}, {Bennie}, {Bernard}, {Bignami}, {Boer}, {Briel}, {Butler}, {Cara}, {Chabaud}, {Cole}, {Collura}, {Conte}, {Cros}, {Denby}, {Dhez}, {Di Coco}, {Dowson}, {Ferrando}, {Ghizzardi}, {Gianotti}, {Goodall}, {Gretton}, {Griffiths}, {Hainaut}, {Hochedez}, {Holland}, {Jourdain}, {Kendziorra}, {Lagostina}, {Laine}, {La Palombara}, {Lortholary}, {Lumb}, {Marty}, {Molendi}, {Pigot}, {Poindron}, {Pounds}, {Reeves}, {Reppin}, {Rothenflug}, {Salvetat}, {Sauvageot}, {Schmitt}, {Sembay}, {Short}, {Spragg}, {Stephen}, {Str{\"u}der}, {Tiengo}, {Trifoglio}, {Tr{\"u}mper}, {Vercellone}, {Vigroux}, {Villa}, {Ward}, {Whitehead}, \& {Zonca}}]{2001A&A...365L..27T}
{Turner}, M.~J.~L., {Abbey}, A., {Arnaud}, M., {et~al.} 2001, \aap, 365, L27, \dodoi{10.1051/0004-6361:20000087}

\bibitem[{{Turolla}(2009)}]{2009ASSL..357..141T}
{Turolla}, R. 2009, in Astrophysics and Space Science Library, Vol. 357, Astrophysics and Space Science Library, ed. W.~{Becker}, 141, \dodoi{10.1007/978-3-540-76965-1_7}

\bibitem[{{van Kerkwijk} \& {Kaplan}(2008)}]{2008ApJ...673L.163V}
{van Kerkwijk}, M.~H., \& {Kaplan}, D.~L. 2008, \apjl, 673, L163, \dodoi{10.1086/528796}

\bibitem[{{van Kerkwijk} {et~al.}(2007){van Kerkwijk}, {Kaplan}, {Pavlov}, \& {Mori}}]{2007ApJ...659L.149V}
{van Kerkwijk}, M.~H., {Kaplan}, D.~L., {Pavlov}, G.~G., \& {Mori}, K. 2007, \apjl, 659, L149, \dodoi{10.1086/518030}

\bibitem[{{Walter} {et~al.}(2010){Walter}, {Eisenbei{\ss}}, {Lattimer}, {Kim}, {Hambaryan}, \& {Neuh{\"a}user}}]{2010ApJ...724..669W}
{Walter}, F.~M., {Eisenbei{\ss}}, T., {Lattimer}, J.~M., {et~al.} 2010, \apj, 724, 669, \dodoi{10.1088/0004-637X/724/1/669}

\bibitem[{{Walter} {et~al.}(1996){Walter}, {Wolk}, \& {Neuh{\"a}user}}]{1996Natur.379..233W}
{Walter}, F.~M., {Wolk}, S.~J., \& {Neuh{\"a}user}, R. 1996, \nat, 379, 233, \dodoi{10.1038/379233a0}

\bibitem[{{Zampieri} {et~al.}(2001){Zampieri}, {Campana}, {Turolla}, {Chieregato}, {Falomo}, {Fugazza}, {Moretti}, \& {Treves}}]{2001A&A...378L...5Z}
{Zampieri}, L., {Campana}, S., {Turolla}, R., {et~al.} 2001, \aap, 378, L5, \dodoi{10.1051/0004-6361:20011151}

\bibitem[{{Zane} {et~al.}(2005){Zane}, {Cropper}, {Turolla}, {Zampieri}, {Chieregato}, {Drake}, \& {Treves}}]{2005ApJ...627..397Z}
{Zane}, S., {Cropper}, M., {Turolla}, R., {et~al.} 2005, \apj, 627, 397, \dodoi{10.1086/430138}

\bibitem[{{Zane} {et~al.}(2002){Zane}, {Haberl}, {Cropper}, {Zavlin}, {Lumb}, {Sembay}, \& {Motch}}]{2002MNRAS.334..345Z}
{Zane}, S., {Haberl}, F., {Cropper}, M., {et~al.} 2002, \mnras, 334, 345, \dodoi{10.1046/j.1365-8711.2002.05480.x}

\end{thebibliography}
\bibliographystyle{aasjournal}

\end{document}